\documentclass[
  aps,prd,
  preprint,          %  toggle: reprint (2-col)  /  preprint (1-col)
  amsmath,amssymb,
  superscriptaddress,
  nofootinbib,
  floatfix,
]{revtex4-2}

\usepackage[T1]{fontenc}
\usepackage{xcolor}
\newcommand{\blue}{\color{blue}}
\usepackage{graphicx}
\usepackage{hyperref}
\hypersetup{colorlinks=true,linkcolor=blue,citecolor=blue,urlcolor=blue}
\usepackage{amsmath}
\usepackage{mathrsfs}
\usepackage{slashed}
\usepackage{bm}
\usepackage{adjustbox}

\newcommand{\tr}{\mathrm{tr}}

\newcommand{\Id}{\mathbf{1}}

\newcommand{\vev}[1]{\langle #1 \rangle}
\newcommand{\kAB}{\kappa_{AB}}

\newcommand{\RAB}{R_{AB}}
\newcommand{\CR}{C^{R}}
\newcommand{\chiR}{\chi^{R}}
\begin{document}

\title{Renormalized symmetry probes for emergent symmetries and
non-chiral lattice fermions}

\author{Ting-Wai Chiu}
\email{twchiu@phys.ntu.edu.tw}
\affiliation{Department of Physics, National Taiwan Normal University, Taipei
11677, Taiwan}
\affiliation{Institute of Physics, Academia Sinica, Taipei 11529, Taiwan}
\affiliation{Department of Physics, National Taiwan University, Taipei 10617,
Taiwan}
\affiliation{Physics Division, National Center for Theoretical Sciences, Taipei
10617, Taiwan}
\affiliation{Nuclear Science Division, Lawrence Berkeley National Laboratory,
Berkeley, CA 94720, USA}

\begin{abstract}
The degeneracy of hadron correlators in symmetry-related channels signals
symmetry restoration in high-temperature QCD.
Renormalization-group-invariant probes of this degeneracy need care when the
regularization breaks chiral symmetry, or when the symmetry is not exact in the
massless action, as for the emergent chiral-spin symmetry $SU(2)_{CS}$.
We renormalize the thermal and screening meson correlators of every flavor
sector by the thermal correlator of the same channel in the $\bar ud$ sector at
a common reference point.
The symmetry ratio $\kAB$ then depends on the renormalization only through the
ratio $\RAB$.
For exact symmetries, $\RAB=1$ and $\kAB$ is RG invariant and scheme independent
in mass-independent schemes.
For emergent symmetries, $\RAB\neq1$ carries the relative logarithmic cutoff
dependence of the two channels, which cancels in $\kAB$, and the residual
dependence on the reference scale is physical.
The framework applies to Wilson, twisted-mass and staggered fermions, where
$\RAB$ absorbs the channel-dependent renormalization, and the tensor pair
$(T_k,X_k)$ keeps $R_{T_kX_k}=1$ on isotropic Wilson-type lattices.
Lattice artifacts are a small correction to the chiral-spin probes $\kappa_{CS}$
and $\kappa_{fade}$, but of the same order as the signal of the probes of the
exact symmetries.
For free quarks, we derive $\kappa_{fade}=2/3$ for every mass and separation,
and $\kappa_{CS}\le1/3$.
We apply the renormalized probes to $N_f=2+1+1$ lattice QCD with optimal
domain-wall quarks at the physical point, with $a=0.064$~fm and
$T=192$--$1539$~MeV.
The hierarchical restoration of the chiral symmetries from the light to the
charm sectors is confirmed, and it is invariant under renormalization.
For the emergence of chiral-spin symmetry, the $\bar uc$ and $\bar sc$ sectors
have the widest chiral-spin windows, also at a tighter threshold.
Thus the spin-1 mesons with one charm quark are the most attractive channels for
the chiral-spin regime.
\end{abstract}

\maketitle

%======================================================================
\section{Introduction}
\label{sec:intro}
%======================================================================

The restoration of chiral and axial symmetries in high-temperature QCD is
diagnosed
by the degeneracy of hadron correlators.
This degeneracy follows from the symmetry transformation, which relates the
hadron correlation functions
at every space-time point.
The temporal correlator $C(t)$ is a thermal trace over states, while the
screening correlator $C(z)$ is a ground-state matrix element of the transfer
matrix along $z$.
The temporal correlator therefore has the spectral representation
$C(t)=\int_0^\infty
d\omega\,\rho(\omega,T)\,\cosh[\omega(t-\tfrac{1}{2T})]/\sinh(\tfrac{%
\omega}{2T})$,
so $C_A(t)=C_B(t)$ at every $t$ is equivalent to $\rho_A=\rho_B$.
The screening correlator is a sum over screening states, $C(z)=\sum_i\alpha_i
e^{-m_iz}$,
so $C_A(z)=C_B(z)$ implies $\alpha_{A,i}=\alpha_{B,i}$ and $m_{A,i}=m_{B,i}$ for
every $i$.
The degeneracy of the thermal or screening masses alone therefore does not
guarantee the degeneracy
of the correlators.
A thermal mass describes only the lowest peak of $\rho$,
and channels with the same screening mass can differ in their amplitudes.
For free quarks, for example, all screening correlators have the screening mass
$2\pi T$, but some fall as $e^{-2\pi Tz}/z$ and others as $e^{-2\pi Tz}/z^2$, so
a mass comparison would declare degenerate multiplets that are never degenerate
(Appendix~\ref{app:screen}).
Susceptibilities, the sums of the correlation functions over all space-time
points,
combine the amplitudes and the masses into one number per channel, 
so their degeneracy is a necessary condition for the degeneracy of the
correlators and serves as a probe of the exact symmetries.
However, susceptibilities have no dependence on the separation, so they cannot
probe emergent symmetries
such as the approximate chiral-spin symmetry $SU(2)_{CS}$, which is broken at
the short distances
that dominate the sums.

The probes below are therefore built from the full correlators
(Sec.~\ref{subsec:cs}).
A quantitative measure of the residual breaking is the ratio
\begin{equation}
\kAB(T) = \frac{|\mathcal{O}_A(T) - \mathcal{O}_B(T)|}
               {|\mathcal{O}_A(T)| + |\mathcal{O}_B(T)|}.
\label{eq:kappa_generic}
\end{equation}
Here $A$ and $B$ label a pair of symmetry partners.
The quantity $\mathcal{O}_\Gamma$ is a renormalized correlator or
susceptibility.
By the triangle inequality, $0\le\kAB\le1$.
The value $\kAB=0$ occurs only if the partners are degenerate.
The value $\kAB=1$ occurs if they have opposite signs.
For partners of the same sign, $\kAB$ equals
$|(\mathcal{O}_A-\mathcal{O}_B)/(\mathcal{O}_A+\mathcal{O}_B)|$.
The correlators and susceptibilities are those of the vacuum-subtracted
operators $O_\Gamma^a-\vev{O_\Gamma^a}$ defined in Sec.~\ref{subsec:defs}.
The ratio $\kAB$ built from the bare spatial correlators was first employed to
study $U(1)_A$, $SU(2)_L\times SU(2)_R$ and the emergent $SU(2)_{CS}$
in thermal QCD with optimal domain-wall quarks~\cite{Chiu:2024jyz,Chiu:2025sym}.
Built from susceptibilities, it was used to study the restoration of 
$SU(2)_L\times SU(2)_R$ and $U(1)_A$ in 
$N_f=2+1+1$ lattice QCD at the physical point~\cite{Chiu:2026sxy}.
Its renormalization for chiral fermions was established in
Ref.~\cite{Chiu:2026upk}.
The present paper builds $\kAB$ from renormalized correlators and uses it to
study the emergent $SU(2)_{CS}$ for the first time.

The RG invariance of Eq.~(\ref{eq:kappa_generic}) requires that the
renormalization of the
two channels cancels between numerator and denominator.
Suppose $A$ and $B$ are related by an exact symmetry of the regularized massless
action.
Then they share a single constant $Z_A=Z_B$, and the cancellation is automatic.
This holds for solutions of the Ginsparg--Wilson
relation~\cite{Ginsparg:1981bj,Luscher:1998pqa}.
Examples are overlap fermions~\cite{Neuberger:1997fp} and domain-wall fermions
with an infinite fifth
dimension~\cite{Kaplan:1992bt,Shamir:1993zy,Borici:1999zw,Chiu:2002ir,%
Brower:2012vk}.
Two circumstances break the cancellation.
They are the subject of this paper.

The first circumstance is a non-chiral regularization.
Most finite-temperature studies use actions without exact chiral symmetry.
Examples are improved and highly improved staggered
quarks~\cite{Bazavov:2011nk,HotQCD:2019xnw,Dick:2015twa}.
Others are $O(a)$-improved Wilson fermions~\cite{Brandt:2016daq,Aarts:2026} 
and twisted-mass Wilson fermions~\cite{Kotov:2021rah}.
For Wilson-type discretizations, the explicit breaking spoils $Z_A=Z_B$.
For staggered fermions, the renormalization depends on the spin-taste structure
of each channel (Sec.~\ref{subsec:staggered}).
Equation~(\ref{eq:kappa_generic}) is then not automatically RG invariant.

The second circumstance is more fundamental.
It is independent of the regularization and arises for any emergent symmetry.
An example is the approximate emergent chiral-spin symmetry
$SU(2)_{CS}$~\cite{Glozman:2014mka}.
This symmetry was first observed above the crossover temperature $T_c$
in the spatial and temporal meson correlators of $N_f=2$
QCD~\cite{Rohrhofer:2019qwq,Rohrhofer:2019qal}.
Recent studies used $N_f=2+1+1$ QCD at the physical point,
with temporal and spatial correlators of the $\bar u d$
sector~\cite{Chiu:2023hnm},
and with spatial correlators of all six light and heavy sectors
($\bar u d, \bar u s, \bar u c, \bar s s, \bar s c, \bar c
c$)~\cite{Chiu:2024jyz}.
The $b$ quark was added in $N_f=2+1+1+1$ QCD~\cite{Chiu:2025sym}.
An emergent symmetry relates channels that no symmetry of the massless action
connects,
so the corresponding operators have different anomalous dimensions in general.
This holds even in the continuum limit and for chiral fermions.
Hence $Z_A \neq Z_B$ for emergent partners, whatever the regularization.

We treat both cases in one framework.
Each bare correlator, thermal or screening, in every flavor sector, is divided
by the thermal correlator of the same channel in the $\bar ud$ sector at a
single reference point $(t_r,T_r)$ common to all channels.
This gives the renormalized correlators $\CR_\Gamma(t)$ and $\CR_\Gamma(z)$,
from which the symmetry ratio $\kAB$ is built.
We express $\kAB$ through the renormalization ratio $\RAB$, which isolates all
renormalization effects.

Our main results are the following.
\begin{itemize}
\item[(i)] For partners of an exact symmetry of the massless action, $\RAB=1$,
and $\kAB$ is RG invariant and scheme independent in mass-independent schemes.
For emergent partners $\RAB\neq1$, and the residual scheme dependence of $\kAB$
is the scale dependence of the emergent degeneracy (Sec.~\ref{sec:RAB}).
\item[(ii)] The tensor pair $(T_k,X_k)$ consists of the electric and magnetic
components of one tensor. Hence $R_{T_kX_k}=1$ exactly for chiral fermions and,
on isotropic lattices, for Wilson-type fermions. Then $\kappa_{T_kX_k}$ is a
$U(1)_A$ probe that needs no renormalization on Wilson-type actions
(Sec.~\ref{subsec:exact}).
\item[(iii)] Two probes measure the chiral-spin symmetry, $\kappa_{CS}$ for its
emergence and $\kappa_{fade}$ for its fading.
For free quarks, we derive $\kappa_{CS}\le1/3$ and $\kappa_{fade}=2/3$ for every
quark mass, and these values set the reference levels of the thresholds
(Sec.~\ref{subsec:cs} and Appendix~\ref{app:free}).
\item[(iv)] The temperature-subtracted susceptibility gives the same probes with
the same renormalization constants (Sec.~\ref{sec:susceptibility}).
\item[(v)] In $N_f=2+1+1$ QCD at the physical point, the renormalized
$\kappa_{PS}$, $\kappa_{V_kA_k}$, and $\kappa_{T_kX_k}$ of the exact symmetries
are RG invariant, and so are the restoration temperatures.
They confirm the scenario of Ref.~\cite{Chiu:2024jyz}, the hierarchical
restoration of the chiral symmetries from the light to the charm sectors
(Sec.~\ref{subsec:num_hier}).
At $zT=1$ and $\epsilon=0.05$, $T_c^{ud}<192$~MeV, $T_c^{uds}=214(2)(11)$~MeV
and $T_c^{udsc}=733(1)(9)$~MeV, and $T_1^{ud}<192$~MeV,
$T_1^{uds}=217(3)(12)$~MeV and $T_1^{udsc}=742(1)(9)$~MeV.
These are the new results based on the connected $z$-correlators of the group
generators.
\item[(vi)] The $\bar uc$ and $\bar sc$ sectors have the widest chiral-spin
windows.
At $zT=1$, with the $N_t=4$ reference, $\epsilon_{CS}=0.2$ and
$\epsilon_{fade}=0.35$, they are $<192$--$689(4)$~MeV and
$241(1)(5)$--$692(4)$~MeV, against $364(5)$--$691(4)$~MeV for $\bar ud$
(Table~\ref{tab:cswin}).
At $\epsilon_{CS}=0.15$ they are $393$--$689$~MeV and $410$--$692$~MeV, against
$490$--$691$~MeV for $\bar ud$ (Table~\ref{tab:cswin_eps}).
\end{itemize}

The paper is organized as follows.
Section~\ref{sec:correlators} defines the renormalized correlators and the
choice of the reference point.
It also gives the bookkeeping form of $\kAB$ in terms of $\RAB$.
Section~\ref{sec:RAB} distinguishes exact from emergent symmetries through
$\RAB$, and it builds the chiral-spin probes $\kappa_{CS}$ and $\kappa_{fade}$
for thermal and screening correlators.
Section~\ref{sec:susceptibility} extends the construction to the
temperature-subtracted susceptibility.
Section~\ref{sec:nonchiral} discusses the construction for Wilson, twisted-mass
and staggered fermions.
Section~\ref{sec:numerics} applies the probes to $N_f=2+1+1$ ensembles with
optimal domain-wall quarks at $a=0.064$~fm.
Section~\ref{sec:summary} gives a summary and outlook.
Appendix~\ref{app:flavor} collects the flavor conventions.
It expresses the singlet and the diagonal nonsinglet correlators through
connected and disconnected quark-line contractions.
Appendix~\ref{app:free} derives the free-fermion values of $\kappa_{CS}$ and
$\kappa_{fade}$.
Appendix~\ref{app:tables} tabulates the bare correlators that enter the main
tables.

%======================================================================
\section{Renormalized meson correlators}
\label{sec:correlators}
%======================================================================

\subsection{Definitions}
\label{subsec:defs}

The meson operators are the quark bilinears
\begin{equation}
O_\Gamma^a(x)=\bar q(x)\,\Gamma\,t^a\,q(x).
\label{eq:Oa_def}
\end{equation}
Here $\Gamma$ is the Dirac structure and $t^a$ the flavor structure.
The $t^a$ ($a=1,\ldots,N_f^2-1$) are the generators of $SU(N_f)$ for the flavor
nonsinglets, with $t^a=\tau^a/2$ for $N_f=2$.
The flavor singlet is $t^0=\Id/\sqrt{2N_f}$, and the normalization is
$\tr_F(t^at^b)=\delta^{ab}/2$ for $a,b=0,\ldots,N_f^2-1$, where $\tr_F$ is the
trace over flavor indices.
The zero-momentum temporal correlator and the spatial (screening) correlator are
\begin{align}
C^a_\Gamma(t,T) &= \sum_{\vec x}\,\big[\vev{O_\Gamma^a(\vec
x,t)\,O_\Gamma^{a\dagger}(0)}
                 -|\vev{O_\Gamma^a}|^2\big],
\nonumber\\
C^a_\Gamma(z,T) &=
\sum_{t,x,y}\,\big[\vev{O_\Gamma^a(x,y,z,t)\,O_\Gamma^{a\dagger}(0)}
                 -|\vev{O_\Gamma^a}|^2\big].
\label{eq:corr_def}
\end{align}
The temperature is $T=1/(aN_t)$, and there is no sum over the flavor index $a$.
The subtraction of $\vev{O_\Gamma^a}$ removes the factorized constant piece of
the correlator.
It is needed for the flavor singlet and for the nonsinglets with diagonal
generators, where $\vev{O_\Gamma^a}$ can be nonzero, in practice in the scalar
channel.
It is unnecessary for the off-diagonal nonsinglets $\bar q_1\Gamma q_2$ with
$q_1\neq q_2$, since $\vev{O_\Gamma^a}$ vanishes identically there.
Appendix~\ref{app:flavor} collects the generators and decomposes the singlet and
the diagonal nonsinglets into connected and disconnected parts.
Below, we suppress the flavor index $a$ of $C^a_\Gamma$.

The spatial correlator keeps the full spatial extent $N_s$ of the lattice at
every temperature, so it stays resolved in the hot phase where $N_t$ shrinks.
We write the expressions below for the temporal correlator.
The spatial forms follow by $t\to z$, with the reference point $(t_r,T_r)$ of
Sec.~\ref{subsec:common} unchanged.
We use the manifestly positive normalization of
Refs.~\cite{Chiu:2026sxy,Chiu:2026upk} for $C(t)$ and extend it to $C(z)$.
The conjugate operator is $O_\Gamma^{a\dagger}=\bar q\,\bar\Gamma\,t^a q$ with
$\bar\Gamma=\gamma_4\Gamma^\dagger\gamma_4=\pm\Gamma$.
The sign of each channel is fixed by reflection positivity along the propagation
axis, that is, by $\gamma_4\Gamma^\dagger\gamma_4$ for $C(t)$ and by
$\gamma_3\Gamma^\dagger\gamma_3$ for $C(z)$, with $z$ as the direction of the
transfer matrix, so every channel is positive.

Table~\ref{tab:bilinear} classifies the meson interpolators $\bar q_1\Gamma
q_2$.
They are the scalar $S$, the pseudoscalar $P$, the vector $V_k$, the axial
vector $A_k$, the tensor vector $T_k$ and the axial-tensor vector $X_k$.
The last two obey $X_k=\gamma_5\gamma_4\gamma_k=\gamma_5\,T_k$, so $X_k$ is the
$\gamma_5$ partner of $T_k$ and both are built on the same $\gamma_4\gamma_k$
structure.
For the $z$-correlators, the temporal index $4$ is replaced by the direction $3$
along the separation, and then $k$ runs over $1,2,4$.
Without the index $k$, $V$, $A$, $T$ and $X$ denote the averages over $k=1,2,3$
for the $t$-correlators and over $k=1,2$ for the $z$-correlators, while the
components $V_4$, $A_4$, $T_4$ and $X_4$ of the $z$-correlators are kept
separate.
Otherwise the index $k$ is written.

\begin{table}[ht]
\caption{The classification of meson interpolators $\bar q_1 \Gamma q_2 $, and
their names and notations. The $\Gamma$ matrices in the second column are for
the $t$-correlators. Those in the third column are for the $z$-correlators.}
\label{tab:bilinear}
\begin{ruledtabular}
\begin{tabular}{lcc}
Name and notation & $\Gamma$ (for $t$-correlators) &  $\Gamma$ (for
$z$-correlators)  \\
\hline
Scalar ($S$)
&  $\Id$           &  $\Id$                         \\
Pseudoscalar ($P$)
&  $\gamma_5$      &  $\gamma_5$                    \\
Vector ($V_k$)
&  $\gamma_k \ (k=1,2,3)$  &  $\gamma_k \ (k=1,2,4)$                    \\
Axial vector ($A_k$)
&  $\gamma_5 \gamma_k \ (k=1,2,3)$   &  $\gamma_5 \gamma_k \ (k=1,2,4)$
\\
Tensor vector ($T_k$)
&  $\gamma_4 \gamma_k \ (k=1,2,3)$   &  $\gamma_3 \gamma_k \ (k=1,2,4)$
\\
Axial-tensor vector ($X_k$)
&  $\gamma_5 \gamma_4 \gamma_k \ (k=1,2,3)$  &  $\gamma_5 \gamma_3 \gamma_k \
(k=1,2,4)$  \\
\end{tabular}
\end{ruledtabular}
\end{table}

The bare correlator carries the multiplicative renormalization of the operator,
\begin{equation}
C_\Gamma(t,T) = Z_\Gamma^2\, \CR_\Gamma(t,T).
\label{eq:Zdef}
\end{equation}
The constant $Z_\Gamma$ is a property of the operator alone, independent of the
separation and the temperature.
At fixed $t>0$ the two operators never coincide, so the correlator has no
additive divergence for any lattice discretization, Wilson
included~\cite{Chiu:2026upk}.
The coincident-point singularity at $t\to0$ produces only the additive
divergence of the susceptibility (Sec.~\ref{sec:susceptibility}).
The only divergence to remove at fixed $t>0$ is therefore the multiplicative
$Z_\Gamma^2$.

\subsection{Common normalization}
\label{subsec:common}

We define $Z_\Gamma^2$ as the value $C_\Gamma(t_r,T_r)$ at a reference point
$(t_r,T_r)$ common to all channels:
\begin{equation}
\CR_\Gamma(t,T) = \frac{C_\Gamma(t,T)}{C_\Gamma(t_r,T_r)} =
\frac{C_\Gamma(t,T)}{Z_\Gamma^2},
\label{eq:CR}
\end{equation}
and the spatial form is $\CR_\Gamma(z,T)=C_\Gamma(z,T)/C_\Gamma(t_r,T_r)$, with
the same thermal reference point.
The reference value is that of the $\bar ud$ sector, and it is used for every
flavor sector, which makes the scheme mass independent
(Sec.~\ref{subsec:num_setup}).
Since $Z_\Gamma^2$ multiplies the correlator at every separation and temperature
alike, the logarithmic divergence cancels in the ratio.
Thus our renormalization scheme is $\CR_\Gamma(t_r,T_r) = 1$ for all channels.
For partners of an exact symmetry, we use instead $Z_A=Z_B$, which differs from
this definition only by the residual breaking at $T_r$
(Sec.~\ref{subsec:exact}).

The cancellation of the multiplicative logarithmic divergence in $C_\Gamma(t,T)$
holds for any reference point,
but the choice of $(t_r,T_r)$ nevertheless matters.
The ratio of two reference correlators, $C_A(t_r,T_r)/C_B(t_r,T_r)$, 
enters the symmetry ratio below as $\RAB$.
Three aspects fix the choice.

\emph{The reference temperature.}
The reference temperature is admissible under the same conditions as for the
regularized susceptibility
of Ref.~\cite{Chiu:2026upk}.
First, it lies far enough above the restoration temperatures $T_c^{\bar qQ}$ of
$SU(2)_L\times SU(2)_R$ and $T_1^{\bar qQ}$ of $U(1)_A$ in the sector $\bar qQ$
(Sec.~\ref{subsec:num_hier}), so that the exact partners are degenerate
within errors, $C_A(t_r,T_r)=C_B(t_r,T_r)$.
Second, it has $N_t^r\ge4$ time slices, so that the thermal midpoint lies at
least two lattice units from the source.
The thermal midpoint is the middle time slice $t_r=N_t^r/2$ of the reference
lattice, or $\tau=1/(2T_r)$ in the continuum.
Throughout, ``midpoint'' means this thermal midpoint, never a midpoint in a
spatial direction.
At $N_t^r=2$ or equivalently $T_r = 1/(2a)$ the midpoint is the first lattice
unit,
where cutoff effects are largest, and no other slice exists to check the
degeneracy of the exact partners.
The reference $T_r = 1/(2a)$ is then inadmissible.
Third, for the continuum limit, $T_r$ is held fixed in physical units as
$a\to0$.
In practice, $T_r$ is the highest admissible temperature of the set
(Sec.~\ref{subsec:cs}).

\emph{The reference separation.}
For the thermal correlators we take the thermal midpoint for two reasons.
First, the midpoint is the separation at which excited-state and discretization
contamination is smallest.
Second, for free fermions all six channels are degenerate there
(Appendix~\ref{app:free}),
because the terms that distinguish them vanish by the antiperiodicity of the
quark propagator.
At the high-$T$ midpoint the reference ratios are therefore
$C_A/C_B=1+O(\alpha_s(T_r))$ for every pair,
so the normalization is unity at tree level and the noninteracting values of
Sec.~\ref{subsec:cs}
and Appendix~\ref{app:free} apply up to $O(\alpha_s(T_r))$ corrections.
These corrections are not small for partners of the $SU(2)_{CS}$ emergent
symmetry at the temperatures of present simulations.
At $T_r=770$~MeV the midpoint ratios $C_V/C_T$ and $C_V/C_S$ are $1.10$ and
$0.85$, and at $T_r=513$~MeV they are $1.19$ and $0.67$
(Sec.~\ref{sec:numerics}).
The scheme dependence of the crossing temperatures of the emergent probes is
correspondingly $12$--$22\%$, and we quote it with the results.
For free quarks, only a reference at the midpoint gives $R_{VT}=R_{VS}=1$,
whereas a reference at $tT=1/4$ gives $R_{VT}=1.55$ and $R_{VS}=0.74$ on the
free $N_t=16$ lattice, and $1.23$ and $0.84$ on the free $N_t=4$ lattice
(Appendix~\ref{app:freelat}).

\emph{The screening correlators.}
On a finite lattice the screening correlator also has a middle slice, $z=N_s/2$.
For antiperiodic boundary conditions in $z$, the free-quark correlators of all
channels have the same value there, as at the thermal midpoint.
The average over periodic and antiperiodic boundary conditions in the $z$
direction removes this equality (Appendix~\ref{app:screen}),
so the averaged screening correlators have no separation at which all channels
coincide.
The $z$-correlators and the $t$-correlators of the same operator renormalize
with the same constants,
because they are different sums over the same correlation function
$\vev{O_\Gamma(x,y,z,t)\,O_\Gamma^\dagger(0)}$.
We therefore use the same $Z_\Gamma$ and the same ratios $\RAB$, fixed at the
thermal midpoint, for the screening probes.

By construction, every $\kAB$ built from Eq.~(\ref{eq:CR}) vanishes at the
reference point,
so the probes measure the departure from the reference degeneracy.

A different normalization has been used for $U(1)_A$ with Wilson-clover
fermions~\cite{Aarts:2026}.
There each correlator is divided by its own value at the midpoint of the same
ensemble, $\tilde C_\Gamma(t,T)=C_\Gamma(t,T)/C_\Gamma(N_t/2,T)$, and the pair
$(P,S)$ is compared through the ratio of $\tilde C_P-\tilde C_S$ to $\tilde
C_P+\tilde C_S$, either at each $t$ or summed from a short-distance cutoff to
the midpoint.
This normalization is free of $Z_P$ and $Z_S$, but it removes the amplitude of
each channel at every temperature, and the ratio vanishes at the midpoint by
construction.
It therefore tests only whether the two correlators are proportional,
$C_P(t,T)\propto C_S(t,T)$, i.e., whether the spectral functions satisfy
$\rho_P=c\,\rho_S$ for some constant $c$.
But the constant $c$, which contains $Z_P^2/Z_S^2$ together with the ratio of
the physical amplitudes, drops out, so the vanishing of the ratio
$(\tilde C_P-\tilde C_S)/(\tilde C_P+\tilde C_S)$ is necessary but not
sufficient for degeneracy~\cite{Chiu:2026upk}.
On the other hand, Eq.~(\ref{eq:CR}) uses one reference point for all channels
and all temperatures, so it retains the amplitudes.
Ref.~\cite{Chiu:2023hnm} normalized the correlators of $V,A,T_4,X_4$ to unity at
the first lattice unit, which discards the amplitude of each channel.
Ref.~\cite{Chiu:2024jyz} used the correlators without normalization, and thus
retains the amplitudes.
For the exact symmetries its probes coincide with the renormalized ones of the
present paper, since $\RAB=1$ for exact partners.
For the emergent pairs they are the bare probes, with $\RAB=1$.

\subsection{The symmetry ratio and the bookkeeping form}
\label{subsec:kappa_corr}

The correlator-level symmetry ratio is Eq.~(\ref{eq:kappa_generic}) with
$\mathcal{O}_\Gamma=\CR_\Gamma$:
\begin{equation}
\kAB(t,T) = \frac{|\CR_A(t,T) - \CR_B(t,T)|}
                 {|\CR_A(t,T)| + |\CR_B(t,T)|}.
\label{eq:kappa_CR}
\end{equation}
We avoid forming each $\CR_\Gamma$ explicitly by keeping the bare correlators
and folding the renormalization into a single ratio.
We multiply numerator and denominator of Eq.~(\ref{eq:kappa_CR}) by $Z_A^2>0$
and use Eq.~(\ref{eq:Zdef}).
This gives
\begin{equation}
\kAB(t,T) = \frac{|C_A(t,T) - \RAB\, C_B(t,T)|}
                 {|C_A(t,T)| + \RAB\, |C_B(t,T)|},
\qquad
\RAB \equiv \frac{Z_A^2}{Z_B^2}.
\label{eq:kappa_RAB}
\end{equation}
The two forms are algebraically identical.
In Eq.~(\ref{eq:kappa_RAB}) the bare correlators enter directly, and the single
ratio $\RAB$ carries
the entire renormalization, so no individual $Z_\Gamma^2$ is ever applied.
The ratio $\RAB$ is cutoff independent only when $A$ and $B$ share an anomalous
dimension,
and otherwise it carries their relative logarithmic cutoff dependence
(Sec.~\ref{subsec:emergent}).
For partners of an exact symmetry of the regularized massless action it is unity
by symmetry.
For all other pairs, the reference convention of Sec.~\ref{subsec:common} fixes
it nonperturbatively
as the ratio of the reference correlators at $(t_r,T_r)$, which is equivalent to
choosing the normalization scheme of Eq.~(\ref{eq:CR}).

%======================================================================
\section{Exact symmetries versus emergent symmetries}
\label{sec:RAB}
%======================================================================

\subsection{Exact symmetry of the massless action: $\RAB=1$}
\label{subsec:exact}

Suppose $A$ and $B$ are related by an exact symmetry of the massless action.
That symmetry rotates the operators into each other, so they share one
renormalization constant
and one anomalous dimension.
Hence $Z_A=Z_B$ and $\RAB=1$, and Eq.~(\ref{eq:kappa_RAB}) reduces to
\begin{equation}
\kAB(t,T) = \frac{|C_A(t,T) - C_B(t,T)|}{|C_A(t,T)| + |C_B(t,T)|}.
\label{eq:kappa_exact}
\end{equation}
\emph{The bare correlators may then be used directly, with no renormalization at
all.}
The scalar and pseudoscalar form such a pair under $U(1)_A$.
The vector and axial vector form one under $SU(2)_L\times SU(2)_R$.
For two flavors, $SU(2)_L\times SU(2)_R$ does not relate $S$ to $P$, so $(S,P)$
is a genuine $U(1)_A$ pair, and so is the tensor pair below.
For $N_f\ge3$ massless flavors, $SU(N_f)_L\times SU(N_f)_R$ alone equates the
nonsinglet $S$ and $P$, and likewise $T$ and $X$, so these pairs probe $U(1)_A$
only through the nonzero quark masses~\cite{Chiu:2026upk}.
In a mass-independent scheme $\kAB$ is then RG invariant and scheme independent,
since the common anomalous dimension cancels and no scale enters.
Chiral lattice fermions (overlap fermions and domain-wall fermions with infinite
fifth dimension)
realize this because their lattice chiral symmetry enforces
$Z_A=Z_B$~\cite{Chiu:2026upk}.
For domain-wall fermions at finite $L_s$ the enforcement holds up to the small
residual chiral breaking.

\paragraph*{The tensor pair: a $U(1)_A$ probe for non-chiral fermions.}
The tensor and axial-tensor channels $T_k=\gamma_4\gamma_k$ and
$X_k=\gamma_5\gamma_4\gamma_k=\gamma_5 T_k$ are a third such pair, related by
the $U(1)_A$ transformation like $S$ and $P$.
They have a special property.
Since $\gamma_5\gamma_4\gamma_k\propto\gamma_i\gamma_j$ ($i,j,k$ cyclic), $T_k$
and $X_k$ are the electric ($\sigma_{4k}$) and magnetic ($\sigma_{ij}$)
components of one tensor $\bar q\sigma_{\mu\nu}q$.
They renormalize with a single $Z_T$, as is standard for the tensor
current~\cite{Chimirri:2024tns}.
For chiral fermions this follows from the $U(1)_A$ rotation, which relates
$T_k$ and $X_k$~\cite{Chiu:2026upk}.
For Wilson-type fermions on isotropic lattices, including clover and
twisted-mass, it follows from the hypercubic group.
A rotation by $90^\circ$ in the $(4,i)$ plane turns $\sigma_{4j}$ into
$\sigma_{ij}$, and with the reflections the six components form one irreducible
representation of the hypercubic group.
The tensor channels of the $t$- and $z$-correlators (Table~\ref{tab:bilinear})
are all among these six components, so they share one $Z_T$.
Hence \emph{$R_{T_kX_k}=1$ exactly, for chiral and isotropic Wilson-type actions
alike}, whereas $R_{SP}=1$ and $R_{VA}=1$ hold only for chiral fermions.
For staggered fermions the rotations act on spin and taste together, and
$Z_{T_k}=Z_{X_k}$ holds only for spin-taste pairs connected by the staggered
symmetry group (Sec.~\ref{subsec:staggered}).
On anisotropic lattices, and for actions that single out one direction such as
minimally doubled
fermions~\cite{Karsten:1981gd,Wilczek:1987kw,Creutz:2007af,Borici:2007kz}, the
rotations in the $(4,i)$ planes are not symmetries, and $T_k$ and $X_k$ need
separate renormalization constants.

The pair $(T_k,X_k)$ is therefore a $U(1)_A$ probe whose renormalization is
immune to the explicit chiral breaking of isotropic Wilson-type actions.
On these actions $\kappa_{T_kX_k}$ is RG invariant and scheme independent in
mass-independent schemes, without any renormalization constant. Only the
$O(a^n)$ breaking in the correlators themselves remains
(Sec.~\ref{sec:nonchiral}).
To our knowledge, this property has not been used before.

\subsection{Emergent symmetry: $\RAB\neq1$}
\label{subsec:emergent}

An emergent symmetry relates channels that no symmetry of the massless action
connects. The vector and the tensor are an example, related by the chiral-spin
symmetry of Sec.~\ref{subsec:cs}.
Two things set them apart from an exact-partner pair.

First, no symmetry relates them, and their correlators differ already for free
quarks.
For free quarks, $C_V\ne C_T$ at every separation except the thermal midpoint
(Appendix~\ref{app:free}), in the continuum and for chiral lattice fermions.

Second, the two carry different anomalous dimensions.
The vector current is conserved, so $\gamma_V=0$, while the tensor has
$\gamma_T\ne0$.
Hence $R_{VT}$ depends on the scale at which it is fixed:
\begin{equation}
R_{VT}=\frac{Z_V^2}{Z_T^2}\neq1,
\qquad
\frac{d\,R_{VT}}{d\ln\mu}\neq0.
\label{eq:RVT}
\end{equation}
Here $Z_\Gamma^2=C_\Gamma(t_r,T_r)$ is the reference normalization of
Eq.~(\ref{eq:CR}), and the scale is that of the reference, $\mu\sim T_r$. At the
midpoint reference the free correlators coincide, so
$R_{VT}=1+O(\alpha_s(T_r))$, and $R_{VT}\neq1$ is an interaction effect.
The running is set by the anomalous dimension $\gamma_T$ of the tensor. In the
data, $R_{VT}=1.105$ at $T_r=770$~MeV and $1.193$ at $T_r=513$~MeV
(Sec.~\ref{sec:numerics}). This change contains the running, but also the
temperature dependence of the midpoint correlators.
The inequality $R_{VT}\ne1$ survives $a\to0$, so it is a genuine feature of the
emergent pair and not a discretization artifact.

Scale independence alone does not imply $\RAB=1$.
A shared anomalous dimension, for instance $\gamma_A=\gamma_B=0$, removes only
the running of $\RAB$ and leaves a finite cutoff-independent constant.
The value $\RAB=1$ requires in addition a symmetry that equates the two
normalizations, $Z_A=Z_B$.

The cutoff dependence of $\RAB$ does not make $\kAB$ ill defined.
At fixed lattice spacing and fixed scale $\RAB$ is a number with no continuum
limit by itself, since $Z_T$ depends logarithmically on the cutoff.
But $\RAB$ enters only through $\RAB\,C_B=Z_A^2\,\CR_B$, which carries the same
factor $Z_A^2$ as $C_A=Z_A^2\,\CR_A$.
The factor cancels between numerator and denominator, so $\kAB$ has a continuum
limit.

What survives is a residual dependence on the scale at which $\RAB$ is fixed.
\emph{An emergent symmetry is not a symmetry of the action but an approximate,
scale-dependent degeneracy.}
Whether the vector and tensor correlators coincide is itself a scale-dependent
statement, and the scheme dependence of $\kAB$ expresses this.
Fixing $\RAB$ at the common reference point of Sec.~\ref{subsec:common} sets the
scheme, and the probe then measures the emergent degeneracy relative to that
scale.

\subsection{Probing the chiral-spin symmetry}
\label{subsec:cs}

\paragraph*{The symmetry.}
The chiral-spin group $SU(2)_{CS}$~\cite{Glozman:2014mka,Glozman:2015qva}
rotates the quark field
with the generators
\begin{equation}
\vec{\Sigma}_\mu=\{\gamma_\mu,\;i\gamma_5\gamma_\mu,\;-\gamma_5\},\qquad
q\to\exp\!\big(\tfrac{i}{2}\,\vec{\varepsilon}\cdot\vec{\Sigma}_\mu\big)\,q.
\label{eq:CS_gen}
\end{equation}
The Euclidean direction $\mu$ is fixed, and $\Sigma^3_\mu$ is $-\gamma_5$, the
generator of the
$U(1)_A$ rotation. Hence $U(1)_A$ is a subgroup of $SU(2)_{CS}$, while the axial
rotations of $SU(2)_L\times SU(2)_R$ are not contained in it.
The $SU(2)_{CS}$ transformation mixes left- and right-handed quarks, so it is
not a symmetry of the massless action.
The spatial part of the kinetic term and the chromomagnetic part of the
quark--gluon interaction break it.
What is invariant, for every $\mu$, is the chromoelectric interaction $\bar
q\gamma_4A_4q=q^\dagger A_4q$ together with the color charge $Q^a=\int
d^3x\,q^\dagger T^aq$,
since $\bar q=q^\dagger\gamma_4$ transforms as $\bar q\to\bar
q\gamma_4U^{-1}\gamma_4$ with $U$ unitary.
Suppose the symmetry holds approximately in a window of temperature.
Then the quarks in that window are not free, since the free theory breaks
$SU(2)_{CS}$
at any temperature (Appendix~\ref{app:free}).
Thus the degrees of freedom are hadron-like objects bound predominantly by the
chromoelectric interaction.

\paragraph*{Multiplets.}
The direction $\mu$ is fixed by the requirement that the rotation must not mix
operators of different spin
along the correlator~\cite{Rohrhofer:2019qwq,Chiu:2023hnm}.
This gives $\mu=4$ for the temporal correlators and $\mu=1,2$ for the spatial
correlators along $z$.
For the temporal correlators, the nonsinglet bilinears of
Table~\ref{tab:bilinear} fall
into two kinds of multiplets.
The $SU(2)_{CS}$ singlets are $A_k$, and the triplets are $\{V_k,T_k,X_k\}$ with
$k=1,2,3$.
Here $A_k$, $V_k$, $T_k$ and $X_k$ denote the renormalized correlators
$C^R_\Gamma$ of the vector mesons.
The multiplicity in $k$ is the $S_3$ symmetry of the temporal correlators,
which enlarges the group to $SU(2)_{CS}\times S_3$.
For the spatial correlators, the two directions $\mu=1,2$ together give the
singlets
$V_1,V_2,V_4$ and the triplets $\{A_1,T_4,X_4\}$, $\{A_2,T_4,X_4\}$,
$\{A_4,T_1,X_1\}$
and $\{A_4,T_2,X_2\}$.
With the $S_2$ symmetry in $k=1,2$, the $SU(2)_{CS}\times S_2$ transformations
with $\mu=1$ and $\mu=2$
together give the multiplets
\begin{align}
&(V_1, V_2); (A_1, A_2, T_4, X_4);\notag\\
&V_4; (A_4, T_1, T_2, X_1, X_2).
\end{align}
Suppose $SU(2)_L\times SU(2)_R$ is restored, as it is in the $ud$ sector above a
temperature $T_c^{ud}$ (Sec.~\ref{subsec:num_hier}).
Then the singlet joins the triplet through $V=A$ and $V_4=A_4$.
The multiplets become the chirally completed sets
\begin{align}
\text{temporal:}\quad &M_1=\{V,A,T,X\},\notag\\
\text{spatial:}\quad &M_2=\{V,A,T_4,X_4\},\qquad M_4=\{V_4,A_4,T,X\}.
\label{eq:multiplets}
\end{align}
Here $M_1$, $M_2$ and $M_4$ are the flavor-nonsinglet content of the $SU(4)$
multiplets~\cite{Glozman:2014mka} [Eqs.~(25)--(27) of
Ref.~\cite{Chiu:2023hnm}].
In the manifestly positive normalization all channels are positive,
while the raw free Dirac traces of $M_4$ carry the opposite sign
(Appendix~\ref{app:screen}),
which is absorbed into the sign fixed by reflection positivity along $z$.

\paragraph*{Which splittings probe the symmetry.}
Not every splitting within a multiplet measures $SU(2)_{CS}$ itself.
The degeneracies $T-X$ and $T_4-X_4$ are those of the subgroup $U(1)_A$, and
$V-A$ is that of $SU(2)_L\times SU(2)_R$.
Since $U(1)_A\subset SU(2)_{CS}$, a breaking of $U(1)_A$ breaks $SU(2)_{CS}$
with it.
The splittings intrinsic to $SU(2)_{CS}$ are those between members of a triplet
that are not $U(1)_A$ partners, namely $V-T$ or $V-X$ for the temporal
correlators and $A-T_4$ or $A-X_4$ for the spatial ones.
From each pair we take the one whose free-fermion quark-mass terms cancel, $V-T$
and $A-X_4$ (Appendix~\ref{app:free}), so that the probes carry no explicit
quark-mass dependence at tree level.
Because $SU(2)_L\times SU(2)_R$ is restored at $T_c^{ud}\lesssim T_1^{ud}$ for
the light quarks, where $T_1^{ud}$ is the restoration temperature of $U(1)_A$
(Sec.~\ref{subsec:num_hier}), $A-X_4$ carries the same information as $V-X_4$ in
the window of interest, and likewise $V-T$ carries the same information as
$A-X$.

Within a chiral-spin triplet, only the pairs between a vector-type channel
($V$ or $A$) and a tensor-type channel ($T$ or $X$) have $\RAB\neq1$.
Since all tensor channels share one $Z_T$ (Sec.~\ref{subsec:exact}), $R_{VX}$,
$R_{VT_4}$ and $R_{VX_4}$ all equal $R_{VT}$ of Eq.~(\ref{eq:RVT}).

Two tests follow, in an order fixed by the group structure for light and heavy
quarks alike.
The first is the $U(1)_A$ test within the triplet, $\kappa_{TX}$ or
$\kappa_{T_4X_4}$.
If it fails, $SU(2)_{CS}$ is broken whatever the intrinsic splitting.
Only if it passes does the intrinsic splitting decide.
For light quarks with domain-wall valence quarks, the $U(1)_A$ and
$SU(2)_L\times SU(2)_R$ tests are passed well below the temperature where
$SU(2)_{CS}$ emerges (Sec.~\ref{subsec:num_subgroup}). In the light sectors near
$T_c$ these tests are the most sensitive to the residual chiral breaking of the
domain-wall valence quarks (Sec.~\ref{subsec:outlook}).
For heavy quarks, $U(1)_A$ is broken over a wide range and $\kappa_{TX}$ can
exceed $\kappa_{VT}$, so a small $\kappa_{VT}$ then says nothing about
$SU(2)_{CS}$.

\paragraph*{Emergence: $\kappa_{CS}$.}
The renormalized probe $\kappa_{AB}$ of Eq.~(\ref{eq:kappa_RAB}) for the
intrinsic pair measures the emergence of $SU(2)_{CS}$:
\begin{align}
\kappa_{CS}(t,T)=\kappa_{VT}(t,T),\qquad
\kappa_{CS}(z,T)=\kappa_{AX_4}(z,T).
\label{eq:kappa_CS}
\end{align}
The emergence of $SU(2)_{CS}$ is signalled by two criteria, applied in this
order:
\begin{align}
\kappa_{TX}(t,T)\le\epsilon_{CS},\quad \kappa_{CS}(t,T)\le\epsilon_{CS},
\qquad
\kappa_{T_4X_4}(z,T)\le\epsilon_{CS},\quad \kappa_{CS}(z,T)\le\epsilon_{CS}.
\label{eq:CS_criterion}
\end{align}
We take $\epsilon_{CS}\le0.2$ as a reasonable upper bound.
The first criterion is the effective restoration of the subgroup $U(1)_A$ within
the triplet, which holds above the temperature $T_1^{\bar qQ}$ of the sector.
The second is the intrinsic chiral-spin degeneracy, which holds above $T_{CS}$,
where $\kappa_{CS}$ falls through $\epsilon_{CS}$.
The second triplet of the spatial correlators, $\{A_4,T,X\}\subset M_4$, would
give the analogous probe $\kappa_{A_4X}$ for the chiral-spin emergence, with
$\kappa_{TX}$ as its $U(1)_A$ test.
But $M_4$ does not merge with $M_2$, nor with the multiplet $M_0=\{S,P\}$
introduced below, even at infinite temperature in the free-fermion limit
(Appendix~\ref{app:screen}), so it plays no role in the fading of
$SU(2)_{CS}$.
We therefore study both the emergence and the fading of $SU(2)_{CS}$ with $M_2$,
with $\kappa_{T_4X_4}$ as the $U(1)_A$ test and $\kappa_{CS}=\kappa_{AX_4}$.

\paragraph*{Fading: $\kappa_{fade}$.}
Genuine chiral-spin symmetry has another feature: the chirally completed
multiplet $M_1$ ($M_2$) stays separated from the $U(1)_A$ multiplet
$M_0=\{S,P\}$ of the scalar and pseudoscalar mesons.
Let the temperature increase beyond $T_{CS}$.
The splitting within $M_1$ ($M_2$) stays small while its separation from $M_0$
decreases, and for $T>T_{fade}$ the multiplet $M_1$ ($M_2$) is no longer cleanly
separated from $M_0$.
The chiral-spin symmetry is then washed out, and what remains is the
$SU(2)_L\times SU(2)_R\times U(1)_A$ chiral symmetry of free quarks.
The signal is not a growth of the intrinsic splitting but the shrinking of the
separation $|M_0-M_1|$ to the scale of that splitting.
The renormalized probe for it is the ratio of the two:
\begin{align}
\kappa_{fade}(t,T)=\frac{|V-T|}{|M_0-M_1|},\qquad
\kappa_{fade}(z,T)=\frac{|A-X_4|}{|M_0-M_2|}.
\label{eq:kappa_fade}
\end{align}
In the bookkeeping form of Sec.~\ref{subsec:kappa_corr}, the ingredients are
\begin{align}
&V-T = C_V^R-C_T^R \rightarrow C_V-R_{VT}C_T, \qquad
A-X_4 = C_A^R-C_{X_4}^R \rightarrow R_{VA}C_A-R_{VX_4}C_{X_4}, \notag\\
&M_0 = \tfrac{1}{2}(S+P)=\tfrac{1}{2}(C_S^R+C_P^R)
  \rightarrow \tfrac{1}{2}(R_{VS}C_S+R_{VP}C_P), \notag\\
&M_1 = \tfrac{1}{4}(V+A+T+X)
  \rightarrow \tfrac{1}{4}(C_V+R_{VA}C_A+R_{VT}C_T+R_{VX}C_X), \notag\\
&M_2 = \tfrac{1}{4}(V+A+T_4+X_4)
  \rightarrow \tfrac{1}{4}(C_V+R_{VA}C_A+R_{VT_4}C_{T_4}+R_{VX_4}C_{X_4}).
\label{eq:M1M2}
\end{align}
The last expressions follow by cancelling the common factor $Z_V^2$ from
numerator and denominator.
The chiral-spin symmetry remains a distinct emergent symmetry as long as
\begin{align}
\kappa_{fade}\le\epsilon_{fade}.
\label{eq:fade_criterion}
\end{align}
We take $\epsilon_{fade}\le0.35$ as a reasonable upper bound, which is about
half the free value.
As $T$ increases, $\kappa_{CS}$ decreases while $\kappa_{fade}$ rises toward its
noninteracting value $2/3$.
The temperature $T_{fade}$ is where $\kappa_{fade}$ rises through
$\epsilon_{fade}$. By the order of the criteria, the window of the emergent
$SU(2)_{CS}$ is $\max(T_1^{\bar qQ},T_{CS})<T<T_{fade}$.
It depends on $\epsilon_{CS}$ and $\epsilon_{fade}$.
The thermal and the screening probes weight the correlation function
differently, so they need not give the same window away from exact restoration
(Sec.~\ref{subsec:num_cs}). They must agree only in the limit of exact
restoration.

\paragraph*{Free-fermion values.}
The noninteracting theory fixes the endpoints of these probes, which makes the
definition of the chiral-spin window quantitative.
For free quarks of any masses $m_1,m_2$ (Appendix~\ref{app:free}),
\begin{align}
\kappa_{fade}(t)=\kappa_{fade}(z)=\frac23,\qquad
0\le\kappa_{CS}\le\frac13 .
\label{eq:free_values}
\end{align}
The value of $\kappa_{fade}$ holds at any lattice spacing, because the common
cutoff-dependent factor cancels. For the thermal probe it holds away from the
midpoint, where numerator and denominator both vanish. The bound on
$\kappa_{CS}$ is proven for the thermal probe and checked numerically for the
screening probe (Appendix~\ref{app:screen}), and $\kappa_{CS}$ decreases with
the separation, to zero at the thermal midpoint and slowly in $z$.
The multiplets $M_1$ and $M_2$ contain each channel together with its chiral
partner, so every quark-mass term cancels in $M_0-M_1$ and $M_0-M_2$.
The intrinsic splittings $V-T$ and $A-X_4$ are mass free, and \emph{the value
$\kappa_{fade}=2/3$ therefore holds from light to heavy quarks}.
These values make precise the statement that free quarks possess no chiral-spin
symmetry.
Their $\kappa_{CS}$ is small, yet the intrinsic splitting inside $M_1$ ($M_2$)
is a fixed fraction $2/3$ of its separation from $M_0$, so $SU(2)_{CS}$ has
faded completely.
A small $\kappa_{CS}$ is therefore necessary but not sufficient for the
chiral-spin symmetry.

\paragraph*{Separations and temperatures.}
At the thermal midpoint of any $N_t$, all channels coincide for free quarks, so
$\kappa_{CS}$ vanishes, and both the numerator and the denominator of
$\kappa_{fade}$ vanish as well.
We therefore evaluate the thermal probes at $tT=1/4$ and $3/8$, where the
channels are resolved and the short-distance region $t\to0$ is avoided.
We evaluate the screening probes at $zT=1/2$, $1$ and $2$.

The range of temperatures depends on the probe.
The probes of the exact symmetries have $\RAB=1$, so they do not depend on the
reference point, and we evaluate them at all temperatures.
The probes $\kappa_{CS}$ and $\kappa_{fade}$ contain $\RAB\neq1$ from the
reference point, and we evaluate them only for $T<T_r$, since the renormalized
$\kappa_{fade}$ has a pole above $T_r$ (Sec.~\ref{subsec:num_cs}).
This is why $T_r$ is the highest admissible temperature of the set
(Sec.~\ref{subsec:common}).
At $T=T_r$ itself, the renormalized $M_0$ and $M_1$ coincide at $t_r$ by
construction, so we do not evaluate the thermal probes there.
The screening probes at $zT\ge1$ lie at $z\ge2t_r$, and we evaluate them at
$T=T_r$ as well.
For them the normalization at $t_r$ raises $\kappa_{fade}$ by about $5\%$ at
$zT=1$, which we quote as a systematic uncertainty of $+20$~MeV on the upper end
of the chiral-spin window (Sec.~\ref{subsec:num_cs}).

Two remarks close this section.
First, $\kappa_{fade}$ cannot be reconstructed from thermal or screening masses
alone.
For free quarks, all channels in $M_0$ and $M_1$ share the same leading thermal
mass, and the screening mass is $2\pi T$, so a comparison of masses gives a
vanishing numerator and denominator and no ratio.
The value $\kappa_{fade}=2/3$ comes entirely from the correlator amplitudes,
which is why the probe is built from the full renormalized correlators.
Second, the probes of Eqs.~(\ref{eq:kappa_CS}) and (\ref{eq:kappa_fade}) involve
only the correlators and the ratios $\RAB$, so they apply to continuum QCD and
to lattice QCD with chiral or non-chiral fermions alike
(Sec.~\ref{sec:nonchiral}).

%======================================================================
\section{Renormalized susceptibility}
\label{sec:susceptibility}
%======================================================================

The susceptibility is the space-time sum of the correlation function.
For a given operator, $\chi_\Gamma(T)=\sum_t C_\Gamma(t,T)=\sum_z
C_\Gamma(z,T)$.
It carries the \emph{same} $Z_\Gamma^2$ as the correlation function it sums,
because $Z_\Gamma$ renormalizes the operator and does not depend on the
summation.

Unlike the correlator at fixed $t>0$, the susceptibility has an additive
divergence, because the sum runs through $t\to0$.
This part is temperature independent, being a short-distance property of the
operator product, so it cancels in $\chi_\Gamma(T)-\chi_\Gamma(T_r)$ for any
lattice fermion.
The same common reference normalization removes the multiplicative divergence:
\begin{equation}
\chiR_\Gamma(T) = \frac{\chi_\Gamma(T)-\chi_\Gamma(T_r)}{C_\Gamma(t_r,T_r)}.
\label{eq:chiR}
\end{equation}
The associated ratio is Eq.~(\ref{eq:kappa_generic}) with
$\mathcal{O}_\Gamma=\chiR_\Gamma$, and in bookkeeping form it reads
\begin{equation}
\kAB(T) = \frac{|\chi^{\rm reg}_A(T) - \RAB\,\chi^{\rm reg}_B(T)|}
               {|\chi^{\rm reg}_A(T)| + \RAB\,|\chi^{\rm reg}_B(T)|},
\qquad
\chi^{\rm reg}_\Gamma(T)\equiv\chi_\Gamma(T)-\chi_\Gamma(T_r).
\label{eq:kappa_chi_RAB}
\end{equation}
The ratio $\RAB=Z_A^2/Z_B^2$ is the same as before, and the classification of
Sec.~\ref{sec:RAB} carries over.
For exact partners $\RAB=1$, and Eq.~(\ref{eq:kappa_chi_RAB}) is then RG
invariant. It still depends on $T_r$ through the subtracted $\chi_\Gamma(T_r)$,
which drops out of the numerator for partners degenerate at $T_r$ but not out of
the denominator.
For emergent partners $\RAB\neq1$, and Eq.~(\ref{eq:kappa_chi_RAB}) is then
scheme dependent.
Both $\chi^{\rm reg}_A$ and $\RAB\,\chi^{\rm reg}_B$ still carry the same
logarithmically divergent factor $Z_A^2$, in the numerator and in the
denominator, so only the ratio $\kAB$ is cutoff independent.

The susceptibility is a weaker degeneracy test than the correlation function,
but a stronger one than the screening mass.
It integrates the correlation function over all separations, so it carries the
amplitudes and the masses of a channel in one integrated number, not
individually.
The screening mass carries only the exponent of the ground state, with no
information on the amplitudes.
The susceptibility remains the most robust single number because it requires no
fit, and it complements the correlator probes for partners of an exact symmetry
of the massless action, whose degeneracy holds at every separation.

For an emergent symmetry, the susceptibility probes are insensitive to the
symmetry.
The sum in the susceptibility is dominated by the shortest separations, where
the kinetic term breaks $SU(2)_{CS}$ maximally.
The kinetic term respects $SU(2)_L\times SU(2)_R$ and $U(1)_A$, so for exact
partners the dominant short-distance part is common to both channels, and it
only dilutes their splitting.
For free quarks, $t=1$ and $t=N_t-1$ together carry $79\%$ of
$\sum_{t=1}^{N_t-1}C_V(t)$ for $N_t=16$ to $6$, and $84\%$ for $N_t=4$.
For the ensembles of $N_f=2+1+1$ lattice QCD at the physical point
(Sec.~\ref{sec:numerics}), the susceptibility ratio $\kappa_{VT}$ from
Eq.~(\ref{eq:kappa_chi_RAB}), the analog of $\kappa_{CS}(t)$ of
Eq.~(\ref{eq:kappa_CS}) for the emergence of
$SU(2)_{CS}$, is $0.80$--$0.88$ at every temperature from $192$ to $513$~MeV,
against the free-quark value $1$.
Over the same range, the thermal $\kappa_{CS}$ of the correlators at $tT=3/8$
falls from $0.26$ to $0.08$ (Table~\ref{tab:probes_ud}).
The susceptibility ratios are therefore suitable only for probing exact
symmetries of the massless action, such as $U(1)_A$ and $SU(2)_L\times SU(2)_R$.

Within a single channel, the multiplicative renormalization can also be removed
without reference to a partner.
For the scalar and pseudoscalar the factor $m^2$ removes it, since $m\bar qq$ is
RG invariant.
We denote by $\chi_\sigma$, $\chi_{5,\rm disc}$ and $\chi_t$ the flavor-singlet
scalar susceptibility, the quark-disconnected part of the flavor-singlet
pseudoscalar susceptibility (Appendix~\ref{app:flavor}), and the topological
susceptibility, as defined in Ref.~\cite{Chiu:2026upk}, where the index theorem
gives $m^2\chi_{5,\rm disc}=\chi_t$ for chiral fermions.
The RG-invariant combinations $m^2\chi^{\rm reg}_\sigma(T)$ and $m^2\chi^{\rm
reg}_{5,\rm disc}(T)=\chi_t(T)-\chi_t(T_r)$ locate characteristic
temperatures~\cite{Chiu:2026upk}, the peak of the former and the inflection
point of the latter, both independent of $T_r$.
These single-channel quantities do not compare partners.
The renormalized susceptibility (\ref{eq:chiR}) normalizes every channel at the
same reference point $(t_r,T_r)$, so two partners can be compared in $\kAB$ of
Eq.~(\ref{eq:kappa_chi_RAB}).

%======================================================================
\section{Non-chiral lattice fermions}
\label{sec:nonchiral}
%======================================================================

Chirally symmetric fermions realize the multiplet structure of
Sec.~\ref{sec:RAB} directly.
Their chiral partners renormalize identically, $\RAB=1$, and the probes of exact
symmetries are ratios of bare correlators (Sec.~\ref{subsec:exact}).
Such actions are costly, so most finite-temperature simulations use non-chiral
fermions.

Two features recur for every non-chiral action.
First, the channel-dependent renormalization is multiplicative, and the single
ratio $\RAB=Z_A^2/Z_B^2$ of Eq.~(\ref{eq:kappa_RAB}) absorbs it once it is
determined nonperturbatively at a common scale.
Second, the lattice breaking of chiral symmetry is a small correction to the
sizable splittings measured by $\kappa_{CS}$ and $\kappa_{fade}$.
It is of the same size as the signal of the non-emergent (exact) probes,
$\kappa_{VA}$ for $SU(2)_L\times SU(2)_R$ and $\kappa_{PS}$ and
$\kappa_{T_kX_k}$ for $U(1)_A$, which measure whether partners that should be
\emph{degenerate} are split.
We discuss the common actions in turn.

\subsection{Wilson fermions}
\label{subsec:wilson}

The Wilson term~\cite{Wilson:1975id} removes the doublers through a
dimension-five operator, and this operator breaks chiral symmetry explicitly at
$O(a)$.
The breaking is present already at zero bare mass, and it has two consequences,
an additive mass and a splitting of the chiral partners.

The quark mass renormalizes additively, $m_q=Z_m(m_0-m_{cr})$, with a critical
mass that scales as $m_{cr}\sim1/a$.
The chiral limit is the tuned point $m_0=m_{cr}$, where the quark mass defined
by the axial Ward identity (PCAC) vanishes.

The renormalization constants of the chiral partners also split, with $Z_S\ne
Z_P$ and $Z_V\ne Z_A$.
The mismatch is multiplicative, so $\RAB$ absorbs it.
It is a finite function of the bare coupling that differs from unity by
$O(g_0^2)$, and it therefore vanishes only logarithmically as $a\to0$, not as a
power of $a$.
The conserved vector current fixes $Z_V$, the axial Ward identity fixes $Z_A$,
and the rest follow from a nonperturbative scheme.
The emergent ratio $R_{VT}$ runs with the scale and is quoted at the scale of
the reference.
The chiral-partner ratios $R_{VA}$ and $R_{SP}$ do not run, since the Wilson
term is an irrelevant operator and the partners keep a common anomalous
dimension.
These finite ratios must be applied at every lattice spacing.
Finally, $Z_{T_k}=Z_{X_k}$ holds exactly (Sec.~\ref{subsec:exact}), so
$R_{T_kX_k}=1$.

For $\kappa_{CS}$ and $\kappa_{fade}$, the Wilson term is a small perturbation.
It splits the $SU(2)_{CS}$ multiplet by an $O(a)$ artifact on top of the sizable
chiral-spin separation. Full $O(a)$ improvement, the clover
term~\cite{Sheikholeslami:1985ij} together with the improvement coefficients of
the operators, reduces it to $O(a^2)$, and the continuum limit removes it.

For the non-emergent probes, the same $O(a)$ breaking is the whole difficulty.
The probes $\kappa_{VA}$, $\kappa_{PS}$ and $\kappa_{T_kX_k}$ see a spurious
splitting of their would-be degenerate partners, which is present already at
zero mass and is of the same order as the physical breaking they are meant to
measure.
For $\kappa_{T_kX_k}$, full $O(a)$ improvement requires only the coefficient
$c_T$ of the $T_k$ channel, since the $X_k$ channel is automatically $O(a)$
improved at zero momentum~\cite{Chimirri:2024tns}.
Without $c_T$, the $O(a)$ term splits $T_k$ from $X_k$ even though
$Z_{T_k}=Z_{X_k}$, and the ratio $R_{T_kX_k}=1$, being multiplicative, cannot
absorb it.
In general, $\RAB$ removes only the multiplicative renormalization.
The breaking by higher-dimension operators persists at every finite lattice
spacing, at $O(a)$ without improvement and at $O(a^2)$ with it, and only the
continuum limit removes it.

The probe $\kappa_{PS}$ has an extra complication.
In the flavor-singlet and diagonal channels, the scalar density mixes with the
identity through a power-divergent coefficient, which parity forbids for the
pseudoscalar.
This mixing shifts only $\vev{O_S}$, which the vacuum subtraction of
Eq.~(\ref{eq:corr_def}) removes, and it is absent for the off-diagonal
nonsinglets.
The power-divergent contact terms at zero separation differ between $S$ and $P$
at finite $a$, so the $\{S,P\}$ doublet is asymmetric there.
They do not enter the correlators at nonzero separation, and the temperature
subtraction of Sec.~\ref{sec:susceptibility} removes them from the
susceptibilities.
The explicit $O(a^n)$ breaking in $\kappa_{PS}$ and $\kappa_{T_kX_k}$ can mask
the anomalous breaking of $U(1)_A$, whose effective restoration is the question
under study, so a controlled continuum limit is essential.

\subsection{Twisted-mass Wilson fermions}
\label{subsec:tm}

Twisted-mass Wilson fermions~\cite{Frezzotti:2000nk} retain the Wilson term, so
the additive mass persists.
One tunes the untwisted bare mass to $m_{cr}\sim1/a$, which is now the condition
for maximal twist.
The gain is twofold.
The physical mass is carried by the twisted mass $\mu_q$, which renormalizes
only multiplicatively.
At maximal twist the $O(a)$ terms of the Symanzik expansion are odd under a
discrete chiral rotation that leaves the continuum action invariant, so physical
observables are automatically $O(a)$ improved~\cite{Frezzotti:2003ni}.
The cost is an $O(a^2)$ breaking of parity and isospin in physical observables.
A second cost is that the twist rotates the chiral partners into one another, so
every channel must be expressed in the twisted basis.

For $\kappa_{CS}$ and $\kappa_{fade}$, the automatic improvement pushes the
multiplet artifact to $O(a^2)$, and no tuning of the clover coefficient is
needed.
The non-emergent probes gain nothing here over improved Wilson fermions.
At maximal twist the charged scalar and pseudoscalar are not rotated, while the
charged vector and axial vector are exchanged, so $R_{SP}$ and $R_{VA}$ remain
nontrivial, and $R_{T_kX_k}=1$ as before.
The neutral components are rotated into flavor-singlet operators of the twisted
basis, with disconnected contributions and an $O(a^2)$ isospin splitting from
the charged ones, so the probes are best built from the charged components.
Their artifacts are then of $O(a^2)$, the same order as for nonperturbatively
$O(a)$-improved Wilson fermions.
Twisted-mass fermions need no improvement coefficients, such as $c_T$ for
$\kappa_{T_kX_k}$~\cite{Chimirri:2024tns}, but the artifact is still of the
order of the signal, so the continuum limit remains essential.

\subsection{Staggered fermions}
\label{subsec:staggered}

Staggered fermions~\cite{Kogut:1974ag,Susskind:1976jm} keep an exact
$U(1)_\varepsilon$ remnant of chiral symmetry at nonzero $a$, which the
staggered phase $\varepsilon(x)=(-1)^{\sum_\mu x_\mu}$ generates.
It forbids an additive mass, so the quark mass renormalizes only
multiplicatively and the chiral limit sits at $m_0=0$ with no tuning.
Each bilinear carries a spin and a taste label,
$\bar\psi(\gamma_S\otimes\xi_F)\psi$, where $\psi$ is the staggered field, and
the four tastes are degenerate only in the continuum.
One assigns a definite taste to each channel.
The $Z$'s depend on the spin-taste structure, so local and point-split operators
renormalize differently, and the ratio $\RAB$ absorbs the mismatch as before.
The residual artifact is taste breaking at $O(a^2)$, which splits the taste
copies and mixes the multiplets.

For $\kappa_{CS}$ and $\kappa_{fade}$, this is again a small correction to a
sizable signal.
The cancellation of the quark-mass terms between chiral partners, which makes
$M_0-M_1$ and $M_0-M_2$ mass free, holds up to $O(a^2)$ taste breaking, and the
free-fermion reference $\kappa_{fade}=2/3$ is recovered in the continuum.

The exact $U(1)_\varepsilon$ is generated by the taste-nonsinglet
$\gamma_5\otimes\xi_5$ and is not anomalous, so it does not protect the $U(1)_A$
pairs.
For staggered fermions the non-emergent probes $\kappa_{VA}$, $\kappa_{PS}$ and
$\kappa_{T_kX_k}$ realize the chiral and $U(1)_A$ structure only up to taste
breaking, and their spurious $O(a^2)$ splitting is removed only by the continuum
limit.

\subsection{Summary of the non-chiral actions}
\label{subsec:nonchiral_summary}

The three kinds of non-chiral actions share a pattern.
Each introduces a multiplicative channel-dependent renormalization, which $\RAB$
absorbs.
Each also breaks chiral symmetry on the lattice, with artifacts of $O(a)$ for
Wilson fermions and of $O(a^2)$ in the observables of twisted-mass fermions at
maximal twist and of staggered fermions.
Improvement removes the $O(a)$ part, and only the continuum limit removes the
breaking entirely.
For the $SU(2)_{CS}$ probes $\kappa_{CS}$ and $\kappa_{fade}$ the artifact is
small against the signal, and their free-fermion reference $\kappa_{fade}=2/3$
is the same for all three non-chiral actions in the continuum.
The non-emergent probes $\kappa_{PS}$, $\kappa_{VA}$ and $\kappa_{T_kX_k}$
measure a small signal of the same order as the artifact, so they demand a
controlled continuum limit.
For chirally symmetric fermions the artifacts are also of $O(a^2)$, but they
cannot split partners of the exact lattice symmetry.
They only rescale the physical splitting, so the non-emergent probes vanish at
every lattice spacing wherever the symmetry is restored, up to the quark mass
and the residual mass.
The non-emergent probes, in particular those of the anomalous $U(1)_A$, are
therefore cleanest with chirally symmetric fermions.
The renormalized chiral-spin probes of Eqs.~(\ref{eq:kappa_CS}) and
(\ref{eq:kappa_fade}), in contrast, are robust against the lattice breaking of
chiral symmetry, which makes them usable for chiral and non-chiral fermions
alike.

%======================================================================
\section{Numerical results at $a=0.064$ fm}
\label{sec:numerics}
%======================================================================

\subsection{Ensembles and conventions}
\label{subsec:num_setup}

We apply the probes to the $N_f=2+1+1$ ensembles of
Refs.~\cite{Chiu:2023hnm,Chiu:2024jyz}, which were generated with optimal
domain-wall quarks at the physical point.
The gauge coupling is $\beta=6.20$, and the lattice spacing is $a=0.0641(4)$~fm.
The lattices are $32^3\times N_t$ with $N_t=16,12,10,8,6,4$, and the
temperatures are $T=192$, $257$, $308$, $385$, $513$ and $770$~MeV
(Table~\ref{tab:ensembles}).
The $N_t=2$ ensemble, $T=1539$~MeV, lies above the reference temperature. It
enters the probes of the exact symmetries and the bare probes, and it shows the
behavior of the renormalized probes above $T_r$.
The bare quark masses are $m_{ud}a=0.00125$, $m_sa=0.040$ and $m_ca=0.550$.
The sea and the valence quarks are optimal domain-wall quarks~\cite{Chiu:2002ir}
with $L_s=16$, whose weights in the fifth dimension are set by the Zolotarev
optimal rational approximation of the sign function of the Hermitian
Wilson--Dirac operator $H_w=\gamma_5D_w$ for eigenvalues
$\lambda_{min}\le|\lambda|\le\lambda_{max}$.
The sea quarks use $\lambda_{min}/\lambda_{max}=0.05/6.2$, and the valence
quarks use $0.01/6.2$, which reduces their residual chiral breaking.

The correlators are those of Table~\ref{tab:bilinear}, averaged forward and
backward, and they are computed in six flavor sectors, $\bar ud$, $\bar us$,
$\bar uc$, $\bar ss$, $\bar sc$ and $\bar cc$.
Each sector is the quark-line connected correlator of the bilinear $\bar
q_1\Gamma q_2$, with the factor $1/2$ of Eq.~(\ref{eq:offdiag}).
For $q_1\neq q_2$ this is the correlator of the off-diagonal flavor nonsinglet.
The $\bar ss$ and $\bar cc$ sectors are the connected parts that enter the
diagonal nonsinglets $C^8$ and $C^{15}$ and the singlet of
Appendix~\ref{app:flavor}, and their disconnected parts are not computed here.

All correlators, thermal and screening, are averaged over two boundary
conditions, for QCD and for the free quarks below.
The two conditions are periodic and antiperiodic for the quark field in the $z$
direction, which is the prescription of Ref.~\cite{Chiu:2023hnm}.
For the screening correlators the average cancels the interference between the
direct and the wrapped quark propagator, up to the numerical precision of the
quark propagators, while for the thermal correlators it is a finite-volume
correction of order $e^{-\pi TL}$.
In addition, the same code was run on lattice free quarks at every $N_t$ with
the same six mass combinations, and on free overlap quarks at $N_t=16$, $8$, $6$
and $4$.
This gives the free-fermion values of the probes on the lattice
(Appendix~\ref{app:freelat}).

\begin{table}[tbp]
\caption{Ensembles at $\beta=6.20$ on $32^3\times N_t$ lattices. The last two
columns give the numbers of configurations used for the thermal and the
screening correlators.}
\label{tab:ensembles}
\begin{ruledtabular}
\begin{tabular}{cccc}
$N_t$ & $T$ [MeV] & configs ($t$) & configs ($z$)\\
\hline
16 & 192 & 544 & 583\\
12 & 257 & 514 & 514\\
10 & 308 & 476 & 481\\
 8 & 385 & 468 & 468\\
 6 & 513 & 431 & 431\\
 4 & 770 & 991 & 991\\
 2 & 1539 & 770 & 770\\
\end{tabular}
\end{ruledtabular}
\end{table}

All configurations are used.
On every configuration $C_P>C_S$ at every $t$, and on a few of them the excess
is two to four orders of magnitude larger than on the rest, with three such
configurations at $N_t=16$, one at $12$, and two each at $10$, $8$ and $6$.
On the same configurations $V-A$ and $T-X$ stay two to three orders of magnitude
below $P-S$.
This is the signature of near-zero modes of the Dirac operator, which enter $S$
and $P$ with opposite signs and leave $V$ and $A$ nearly untouched
(Sec.~\ref{subsec:outlook} and Ref.~\cite{Chiu:2026upk}).
It is not a fault in the data, and removing these configurations would bias the
$U(1)_A$ tests toward restoration.

The conventions are those of Sec.~\ref{subsec:common}.
For the exact partners $(V,A)$, $(T_k,X_k)$ and $(S,P)$ we set $\RAB=1$ in every
sector, as Sec.~\ref{subsec:exact} prescribes.
The two ratios needed for the emergent pairs are fixed at the thermal midpoint
of the $N_t=4$ ensemble, where $T_r=770$~MeV and $t_r=2$.
We use the $\bar ud$ sector:
\begin{equation}
R_{VT}=\frac{C_V(t_r,T_r)}{C_T(t_r,T_r)},\qquad
R_{VS}=\frac{C_V(t_r,T_r)}{C_S(t_r,T_r)}.
\label{eq:R_sym}
\end{equation}
Since $R_{VA}=R_{T_kX_k}=R_{SP}=1$, also $R_{VX}=R_{AX}=R_{VT}$ and
$R_{VP}=R_{VS}$.
In the $\bar ud$ sector the exact partners are degenerate at the reference, so
forming the ratios with $A$, $X$ or $P$ in place of $V$, $T$ or $S$ changes them
by at most $3\times10^{-5}$, far below their errors.

We check the admissibility of the reference with the midpoint ratios of the
exact partners in each sector.
In the $\bar ud$, $\bar us$, $\bar uc$ and $\bar ss$ sectors, $C_V/C_A$,
$C_T/C_X$ and $C_S/C_P$ at $(t_r,T_r)$ are unity within $1.5\times10^{-3}$, at
$N_t=4$ and at the $N_t=6$ reference of the second scheme below, as they must be
for chiral fermions.
In the $\bar sc$ sector they are $1.003$, $1.002$ and $0.996$ at $N_t=4$.
In the $\bar cc$ sector they are $1.032$, $1.023$ and $0.955$ at $N_t=4$, and
$1.10$, $1.09$ and $0.87$ at $N_t=6$.
A reference in the $\bar cc$ sector would therefore be inadmissible at both
temperatures, since its exact partners are not degenerate.
Had the ratios been fixed in the $\bar cc$ sector, $R_{VT}$ would change by
only $0.1\%$ but $R_{VX}$ by $2.4\%$, and the $U(1)_A$ breaking of the reference
would leak into the chiral-spin probe.

The constant $Z_\Gamma$ is a property of the operator, and in a mass-independent
scheme it does not depend on the quark masses.
Fixing the ratios once, in the $\bar ud$ sector near the chiral limit, and using
them in every sector makes the reference scheme mass independent.
The same two ratios are therefore used in all six sectors and for the screening
probes.
The quark masses then enter the renormalized correlators only through cutoff
effects, of order $(m_qa)^2$ for domain-wall quarks.
These are negligible for the $u$, $d$ and $s$ quarks but not for the charm
quark, with $m_ca=0.550$, and the probes of the charm sectors carry them
together with the explicit mass breaking.

We estimate the scheme dependence by repeating the analysis with a second
reference, the $N_t=6$ ensemble, with $T_r=513$~MeV and $t_r=3$.
For comparison we also give the bare probes, with $\RAB=1$ for all pairs.
For the exact partners this is exact.
For the emergent pairs it is no renormalization: the bare probes carry
$Z_V^2/Z_T^2$, which depends logarithmically on the cutoff, so they have no
continuum limit and only show the size of the renormalization.
The screening probes are evaluated at $zT=1$, that is $z=N_t$, and also at
$zT=1/2$ and $2$.
The thermal probes are evaluated at $tT=1/4$ for $N_t\ge8$ and at $tT=3/8$ for
$N_t\ge6$, so that the separation is at least two lattice units.
Where the separation falls between lattice units, we interpolate linearly in
$t$.
At $N_t=10$, $8$ and $6$ these separations are only $t=2$--$3.75$, and at
$N_t=10$ and $6$ they are interpolated, so the thermal probes there are less
reliable.
The full dependence on $z$ and $t$ is shown in Sec.~\ref{subsec:num_cs}.
The statistical uncertainties come from the jackknife method, with the bin size
at which the statistical error saturates, and the error of the reference ratios
is added in quadrature.
The crossing temperatures $T_{CS}$ and $T_{fade}$ of the criteria
(\ref{eq:CS_criterion}) and (\ref{eq:fade_criterion}) follow by linear
interpolation between neighboring ensembles.
Their errors come from a parametric bootstrap.
In each of $4000$ samples, the probe on every ensemble is drawn from a Gaussian
with its jackknife mean and error, independently for each ensemble, and the
crossing is recomputed.
The restoration temperatures of Table~\ref{tab:hier} use the same procedure
with $2000$ samples.
The error of $\RAB$, which is common to all ensembles, is below $0.2\%$ and
negligible here.

\subsection{Tests of the exact symmetries}
\label{subsec:num_subgroup}

In the $\bar ud$ sector, $\kappa_{TX}$, $\kappa_{T_4X_4}$ and $\kappa_{VA}$ are
below $10^{-3}$ at every temperature and separation, except for
$\kappa_{T_4X_4}=0.002(3)$ at $192$~MeV (Table~\ref{tab:sectors} and
Sec.~\ref{subsec:num_cs}).
Some are resolved from zero, such as $\kappa_{VA}=4.2(5)\times10^{-5}$ at
$257$~MeV, but all are negligible.
For the light sector near $T_c$, the $U(1)_A$ tests $\kappa_{TX}$ and
$\kappa_{T_4X_4}$ are the probes most sensitive to the residual chiral breaking
of the domain-wall valence quarks (Sec.~\ref{subsec:outlook}).

In the $\bar ss$ sector, $\kappa_{T_4X_4}$ is $0.14$ at $192$~MeV and $zT=1$,
$0.01$ at $308$~MeV, and below $0.005$ from $385$~MeV on.
In the $\bar cc$ sector it is $0.90$, $0.75$, $0.60$, $0.42$, $0.22$ and $0.07$
at the six temperatures.
For free charm quarks the $U(1)_A$ test fails as well, with
$\kappa_{T_4X_4}=0.56$--$0.81$ at $z=4$--$16$ on the $N_t=16$ lattice, because
the mass term breaks $U(1)_A$ explicitly.
At $zT=1$ the $\bar cc$ sector passes the $U(1)_A$ test at $\epsilon=0.2$ only
above $534$~MeV (Table~\ref{tab:cswin}).
At $\epsilon=0.05$ it passes only above $845$--$881$~MeV, depending on $zT$.
By the order of tests in Sec.~\ref{subsec:cs}, the chiral-spin symmetry cannot
be certified in the $\bar cc$ sector below $534$~MeV, whatever $\kappa_{CS}$
does.
Across the sectors $\kappa_{T_4X_4}$ is ordered by the product $m_1m_2$ of the
two quark masses (Sec.~\ref{subsec:num_mass}).

\subsection{Hierarchical restoration of the chiral symmetries}
\label{subsec:num_hier}

\begin{table*}[tbp]
\caption{Screening probes at $zT=1$ of the six flavor sectors at the six
temperatures, with the $N_t=4$ reference. The probes $\kappa_{T_4X_4}$ and
$\kappa_{VA}$ test $U(1)_A$ and $SU(2)_L\times SU(2)_R$ with $\RAB=1$. The
probes $\kappa_{CS}$ and $\kappa_{fade}$ use the $\RAB$ of the $\bar ud$ sector.
At $770$~MeV they are evaluated at the reference temperature
(Sec.~\ref{subsec:cs}). The $N_t=2$ ensemble ($1539$~MeV) lies above the
reference temperature and is not used for the renormalized probes.}
\label{tab:sectors}
\setlength{\tabcolsep}{3pt}\footnotesize
\begin{adjustbox}{max width=\linewidth}
\begin{tabular}{|lcc|cccc||lcc|cccc|}
\hline
sector & $N_t$ & $T$ & $\kappa_{T_4X_4}$ & $\kappa_{VA}$ & $\kappa_{CS}$ &
$\kappa_{fade}$ & sector & $N_t$ & $T$ & $\kappa_{T_4X_4}$ & $\kappa_{VA}$ &
$\kappa_{CS}$ & $\kappa_{fade}$\\
\hline
$\bar ud$ & 16 & 192 & $2.3(30)\times10^{-3}$ & $0.8(39)\times10^{-4}$ &
0.502(59) & 0.031(4) & $\bar ss$ & 16 & 192 & 0.143(12) & 0.142(11) & 0.403(32)
& 0.035(2)\\
 & 12 & 257 & $1.3(25)\times10^{-5}$ & $4.2(5)\times10^{-5}$ & 0.318(13) &
 0.071(5) &  & 12 & 257 & 0.029(1) & 0.030(1) & 0.309(13) & 0.071(5)\\
 & 10 & 308 & $3.4(34)\times10^{-5}$ & $1.40(10)\times10^{-5}$ & 0.241(6) &
 0.114(3) &  & 10 & 308 & 0.011(1) & 0.011(1) & 0.239(6) & 0.114(3)\\
 & 8 & 385 & $1.5(11)\times10^{-5}$ & $6.0(10)\times10^{-6}$ & 0.184(4) &
 0.153(4) &  & 8 & 385 & $4.90(8)\times10^{-3}$ & $4.38(6)\times10^{-3}$ &
 0.184(4) & 0.153(4)\\
 & 6 & 513 & $4.0(20)\times10^{-6}$ & $2.00(1)\times10^{-6}$ & 0.142(2) &
 0.224(4) &  & 6 & 513 & $1.79(2)\times10^{-3}$ & $1.60(2)\times10^{-3}$ &
 0.142(2) & 0.224(4)\\
 & 4 & 770 & $<10^{-6}$ & $<10^{-6}$ & 0.096(1) & 0.406(3) &  & 4 & 770 &
 $4.60(2)\times10^{-4}$ & $4.42(2)\times10^{-4}$ & 0.096(1) & 0.406(3)\\
\hline
$\bar us$ & 16 & 192 & $7.0(26)\times10^{-3}$ & $9.0(16)\times10^{-3}$ &
0.436(37) & 0.037(3) & $\bar sc$ & 16 & 192 & 0.389(17) & 0.362(16) & 0.200(15)
& 0.052(4)\\
 & 12 & 257 & $1.03(9)\times10^{-3}$ & $1.15(6)\times10^{-3}$ & 0.314(13) &
 0.072(5) &  & 12 & 257 & 0.160(3) & 0.146(3) & 0.189(6) & 0.091(5)\\
 & 10 & 308 & $3.8(3)\times10^{-4}$ & $3.9(1)\times10^{-4}$ & 0.240(6) &
 0.115(3) &  & 10 & 308 & 0.086(1) & 0.077(1) & 0.172(4) & 0.123(3)\\
 & 8 & 385 & $1.8(1)\times10^{-4}$ & $1.56(6)\times10^{-4}$ & 0.184(4) &
 0.153(4) &  & 8 & 385 & 0.046(1) & 0.041(1) & 0.154(4) & 0.156(4)\\
 & 6 & 513 & $6.5(6)\times10^{-5}$ & $5.6(3)\times10^{-5}$ & 0.142(2) & 0.224(4)
 &  & 6 & 513 & 0.020(1) & 0.018(1) & 0.135(2) & 0.225(4)\\
 & 4 & 770 & $1.40(1)\times10^{-5}$ & $1.40(1)\times10^{-5}$ & 0.096(1) &
 0.406(3) &  & 4 & 770 & $5.71(3)\times10^{-3}$ & $5.57(2)\times10^{-3}$ &
 0.097(1) & 0.405(3)\\
\hline
$\bar uc$ & 16 & 192 & 0.021(2) & 0.023(4) & 0.184(14) & 0.073(5) & $\bar cc$ &
16 & 192 & 0.895(7) & 0.837(7) & 0.320(19) & 0.025(1)\\
 & 12 & 257 & $5.9(2)\times10^{-3}$ & $5.6(2)\times10^{-3}$ & 0.183(6) &
 0.102(5) &  & 12 & 257 & 0.749(5) & 0.677(5) & 0.250(8) & 0.054(2)\\
 & 10 & 308 & $3.2(1)\times10^{-3}$ & $2.83(8)\times10^{-3}$ & 0.168(4) &
 0.130(3) &  & 10 & 308 & 0.596(3) & 0.531(3) & 0.205(5) & 0.085(2)\\
 & 8 & 385 & $1.67(6)\times10^{-3}$ & $1.44(5)\times10^{-3}$ & 0.151(4) &
 0.160(4) &  & 8 & 385 & 0.416(3) & 0.365(2) & 0.174(4) & 0.127(3)\\
 & 6 & 513 & $7.1(5)\times10^{-4}$ & $6.1(3)\times10^{-4}$ & 0.134(2) & 0.227(4)
 &  & 6 & 513 & 0.220(2) & 0.195(1) & 0.145(2) & 0.203(3)\\
 & 4 & 770 & $1.79(1)\times10^{-4}$ & $1.74(1)\times10^{-4}$ & 0.097(1) &
 0.406(3) &  & 4 & 770 & 0.071(1) & 0.070(1) & 0.098(1) & 0.379(3)\\
\hline
\end{tabular}
\end{adjustbox}
\end{table*}

\begin{table*}[tbp]
\caption{Restoration temperatures of the exact symmetries for the generators of
$SU(N_f)$, from the screening correlators. The off-diagonal generators are the
sectors $\bar ud$, $\bar us$, $\bar uc$ and $\bar sc$. The columns $t^8$ and
$t^{15}$ are the diagonal nonsinglets, and $t^0_{uds}$ and $t^0_{udsc}$ are the
$SU(3)$ and $SU(4)$ singlets, from their connected parts
[Eqs.~(\ref{eq:su3_diag}) and (\ref{eq:su4_diag})], without the disconnected
diagrams. The temperature $T_c$ is the one above which $\kappa_{VA}\le\epsilon$,
and $T_1$ the one above which $\kappa_{T_4X_4}\le\epsilon$. The singlets enter
only $T_1$. The thresholds are $\epsilon=0.05$ and $0.01$, at $zT=1/2$, $1$ and
$2$. The ensembles have $N_t=16,\dots,2$ ($192$--$1539$~MeV). Since $\RAB=1$ for
these pairs, the bare and the renormalized probes coincide. These temperatures
are therefore renormalization-group invariant. Crossings are obtained by linear
interpolation of $\log\kappa$ in $T$ between neighboring ensembles. The first
error is the bootstrap error. The second is the interpolation uncertainty, the
difference between the crossings from linear interpolation of $\log\kappa$ and
of $\kappa$~\cite{Chiu:2025sym}. The entry $<T$ means that the criterion holds
already on the lowest ensemble of the scan ($T=192$~MeV, or $385$~MeV at
$zT=2$). A dash means that it is not met below $1539$~MeV.}
\label{tab:hier}
\setlength{\tabcolsep}{3pt}\footnotesize
\begin{adjustbox}{max width=\linewidth}
\begin{tabular}{|cc|cccccc|cccccccc|}
\hline
 & & \multicolumn{6}{c|}{$T_c$ [MeV] ($\kappa_{VA}\le\epsilon$)} &
 \multicolumn{8}{c|}{$T_1$ [MeV] ($\kappa_{T_4X_4}\le\epsilon$)}\\
$\epsilon$ & $zT$ & $\bar ud$ & $\bar us$ & $\bar uc$ & $\bar sc$ & $t^8$ &
$t^{15}$ & $\bar ud$ & $\bar us$ & $\bar uc$ & $\bar sc$ & $t^8$ & $t^{15}$ &
$t^0_{uds}$ & $t^0_{udsc}$\\
\hline
0.05 & $1/2$ & $<192$ & $<192$ & $<192$ & 306(1)(1) & 193(2)(1) & 677(1)(24) &
$<192$ & $<192$ & $<192$ & 337(1)(5) & 202(3)(8) & 721(1)(17) & $<192$ &
$<192$\\
0.05 & $1$ & $<192$ & $<192$ & $<192$ & 360(1)(5) & 214(2)(11) & 733(1)(9) &
$<192$ & $<192$ & $<192$ & 375(1)(3) & 217(3)(12) & 742(1)(9) & $<192$ &
$<192$\\
0.05 & $2$ & $<385$ & $<385$ & $<385$ & 395(2)(5) & $<385$ & $<385$ &
$<385$ & $<385$ & $<385$ & 402(2)(8) & $<385$ & $<385$ & $<385$ & $<385$\\
\hline
0.01 & $1/2$ & $<192$ & $<192$ & 201(4)(6) & 559(1)(23) & 265(1)(3) &
1472(1)(35) &
$<192$ & $<192$ & 209(5)(8) & 605(1)(36) & 273(1)(5) & 1472(2)(37) & 240(1)(8) &
898(2)(90)\\
0.01 & $1$ & $<192$ & $<192$ & 230(4)(10) & 640(1)(36) & 290(1)(5) & 1411(1)(76)
&
$<192$ & $<192$ & 230(3)(9) & 655(1)(38) & 292(1)(5) & 1283(2)(157) & 253(2)(2)
& 842(1)(84)\\
0.01 & $2$ & $<385$ & $<385$ & $<385$ & 676(1)(33) & $<385$ & -- &
$<385$ & $<385$ & $<385$ & 660(1)(40) & $<385$ & 1268(4)(145) & $<385$ &
$<385$\\
\hline
\end{tabular}
\end{adjustbox}
\end{table*}

The $U(1)_A$ and $SU(2)_L\times SU(2)_R$ tests of the six sectors order
themselves by the quark masses, which Ref.~\cite{Chiu:2024jyz} interpreted as a
hierarchical restoration of the chiral symmetries.
First, $SU(2)_L\times SU(2)_R\times U(1)_A$ of the $u,d$ quarks is restored.
Next, $SU(3)_L\times SU(3)_R\times U(1)_A$ of $u,d,s$ is restored, when the
generators with a strange quark, $\bar us$ and $t^8$, pass the tests.
Last, $SU(4)_L\times SU(4)_R\times U(1)_A$ of $u,d,s,c$ is restored, when the
generators with a charm quark, $\bar uc$, $\bar sc$ and $t^{15}$, pass.
With a $b$ quark the sequence continues to $SU(5)_L\times SU(5)_R\times
U(1)_A$~\cite{Chiu:2025sym}.

The restoration temperature of a sector was defined there as the lowest
temperature above which a test passes.
The $SU(2)_L\times SU(2)_R$ test is $\kappa_{VA}\le\epsilon$, with temperature
$T_c^{\bar qQ}$, and the $U(1)_A$ test is $\kappa_{T_4X_4}\le\epsilon$, with
temperature $T_1^{\bar qQ}$.
The thresholds are $\epsilon=0.05$ and $0.01$, at fixed $zT$.
The temperature $T_\chi^{\bar qQ}=\max(T_c^{\bar qQ},T_1^{\bar qQ})$ is the one
above which the sector has the full $SU(2)_L\times SU(2)_R\times U(1)_A$
symmetry.
The restoration temperatures of a flavor content, written $T_c^{ud}$,
$T_c^{uds}$ and $T_c^{udsc}$ in Ref.~\cite{Chiu:2026upk}, are the largest over
the generators of $SU(N_f)$ (Appendix~\ref{app:flavor}).
The off-diagonal generators are the sectors $\bar q_1q_2$ with $q_1\neq q_2$,
and the diagonal nonsinglets are $t^3$, $t^8$ and $t^{15}$.
With the flavors in mass order, this maximum equals, for the connected parts,
the maximum over all elements of the Lie algebra of $SU(N_f)$, so it does not
depend on the choice of basis (Appendix~\ref{app:algmax}).
For $T_1$ the singlet $t^0$ enters as well, since $U(1)_A$ relates its
$(T_4,X_4)$ pair, whereas no chiral transformation relates the singlet $V$ and
$A$.
The $\bar ss$ and $\bar cc$ sectors are not generator correlators, and they
enter only through $t^8$, $t^{15}$ and $t^0$.
They remain sectors of the chiral-spin analysis, as off-diagonal correlators
with a degenerate valence partner (Appendix~\ref{app:flavor}).
We evaluate $t^8$, $t^{15}$ and the singlets from their connected parts,
Eqs.~(\ref{eq:su3_diag}) and (\ref{eq:su4_diag}), without the disconnected
diagrams.
For $t^8$ and $t^{15}$ these cancel only for degenerate quark masses, so the
temperatures below are those of the connected parts (Appendix~\ref{app:flavor}).
In the isospin limit $t^3$ and the connected part of the $SU(2)$ singlet equal
the $\bar ud$ sector, so $T_c^{ud}=T_c^{\bar ud}$.
Without a superscript, $T_c\simeq155$~MeV is the crossover temperature.

Table~\ref{tab:hier} gives these temperatures.
Since $\RAB=1$ for exact partners, $\kappa_{VA}$ and $\kappa_{T_4X_4}$ need no
renormalization, so all ensembles enter, including $N_t=4$ and $2$, and the scan
reaches $1539$~MeV.
The crossings follow by linear interpolation of $\log\kappa$ in $T$, and the
second error is the difference from linear interpolation of $\kappa$ itself,
which estimates the interpolation uncertainty as in Ref.~\cite{Chiu:2025sym}.
This difference is small where the crossing falls between two ensembles close in
$\kappa$, and it is large for $t^{15}$ at $\epsilon=0.01$.
There the crossing falls between the $N_t=4$ and $N_t=2$ ensembles, which are
$770$~MeV apart, and their $\kappa$ values differ by a factor $6$
($\kappa_{VA}$) to $10$ ($\kappa_{T_4X_4}$).

At $\epsilon=0.05$ and $zT=1$, $T_c$ and $T_1$ are $214(2)(11)$ and
$217(3)(12)$~MeV for $t^8$, $360(1)(5)$ and $375(1)(3)$~MeV for $\bar sc$, and
$733(1)(9)$ and $742(1)(9)$~MeV for $t^{15}$.
The sectors $\bar ud$, $\bar us$ and $\bar uc$ and the two singlets pass both
tests already at $192$~MeV.
At $\epsilon=0.01$ only $\bar ud$ and $\bar us$ still pass at $192$~MeV.
Then $t^8$ moves to $290(1)(5)$ and $292(1)(5)$~MeV, the $SU(3)$ singlet to
$253(2)(2)$~MeV, and $t^{15}$ to $1411(1)(76)$ and $1283(2)(157)$~MeV, with the
$SU(4)$ singlet at $842(1)(84)$~MeV.
Hence $T_c^{uds}$ and $T_1^{uds}$ are set by $t^8$, and $T_c^{udsc}$ and
$T_1^{udsc}$ by $t^{15}$.
In $t^8$ and $t^{15}$ the splitting comes from the $\bar ss$ and $\bar cc$
parts, where the explicit breaking by $m_s$ and $m_c$ and the $U(1)_A$ breaking
both enter.
At $zT=1$ and $\epsilon=0.05$, $T_c^{ud}$ and $T_1^{ud}$ lie below $192$~MeV,
$T_c^{uds}=214(2)(11)$ and $T_1^{uds}=217(3)(12)$~MeV, and
$T_c^{udsc}=733(1)(9)$ and $T_1^{udsc}=742(1)(9)$~MeV.
At $\epsilon=0.01$, $T_c^{ud}$ and $T_1^{ud}$ still lie below $192$~MeV,
$T_c^{uds}=290(1)(5)$ and $T_1^{uds}=292(1)(5)$~MeV, and
$T_c^{udsc}=1411(1)(76)$ and $T_1^{udsc}=1283(2)(157)$~MeV.
They lie below the corresponding values of the $\bar ss$ and $\bar cc$ sectors,
$235$ and $903$~MeV for $T_c$ at $\epsilon=0.05$ and $zT=1$, since in $t^8$ and
$t^{15}$ the heavier diagonal correlators enter together with the lighter ones.
The $\kappa$ of $t^{15}$ and of the $SU(4)$ singlet is not monotonic in $T$.
At low temperature the light connected correlators dominate at $zT=1$, since the
$\bar cc$ correlator falls much faster in $z$, and $\kappa$ rises again as the
weight of $\bar cc$ grows with $T$.
The restoration temperature is the lowest temperature above which the test holds
at all higher temperatures.
For $\bar ud$, both quarks belong to the restored $SU(2)_L\times SU(2)_R$, and
the $U(1)_A$ test is independent of the $SU(2)_L\times SU(2)_R$ test.
For a heavy-light sector, such as $\bar us$ or $\bar uc$, it is not.
The splittings of its exact partners are ordered by the product $m_1m_2$ of the
two quark masses (Sec.~\ref{subsec:num_subgroup}), so they become small once the
lighter quark is effectively chiral.
For $t^8$ and $t^{15}$ the test is not independent either, since with three or
more massless flavors $SU(N_f)_L\times SU(N_f)_R$ alone equates the nonsinglet
$T_4$ and $X_4$ (Sec.~\ref{subsec:exact}).
Hence $T_1^{uds}$ and $T_1^{udsc}$ are the temperatures at which the $U(1)_A$
partners become degenerate, and their $U(1)_A$ content enters only through the
nonzero quark masses~\cite{Chiu:2026upk}.
The $\epsilon=0.05$ values of the sectors reproduce those of
Ref.~\cite{Chiu:2024jyz} within the interpolation uncertainty.

What the present paper adds is that these temperatures are invariant under
renormalization, since $\RAB=1$ for exact partners.
The hierarchy of Ref.~\cite{Chiu:2024jyz} therefore stands as it was obtained
there, from the bare lattice correlators, and no renormalization can move one
sector past another.

The physical content is the one drawn in Refs.~\cite{Chiu:2024jyz,Chiu:2025sym}.
A $\bar qQ$ meson dissolves when $\bar q$ and $Q$ are deconfined, which is when
its screening mass exceeds that of free quarks of the same masses.
It is expected at $T_d^{\bar qQ}\gtrsim T_{\chi}^{\bar qQ}$, after the chiral
symmetry of the sector is restored.
The sector of the heavier quark pair restores later, so the mesons are expected
to dissolve in the same order,
\begin{equation}
T_d^{\bar ud}<T_d^{\bar us}<T_d^{\bar uc}<T_d^{\bar ss}<T_d^{\bar sc}<T_d^{\bar
cc}.
\label{eq:Td_hier}
\end{equation}
The suppression of mesons in the quark-gluon plasma should follow the same
hierarchy.
The suppression of the $J/\psi$~\cite{Matsui:1986dk} is the heaviest case.
The present analysis makes \emph{this ordering a scheme-independent statement
about the exact symmetries}, and Sec.~\ref{subsec:num_mass} gives its origin,
which is the explicit breaking by the mass term that grows with $m_1m_2$.

\subsection{Chiral-spin probes}
\label{subsec:num_cs}

\begin{figure}[tbp]
\includegraphics[width=\linewidth]{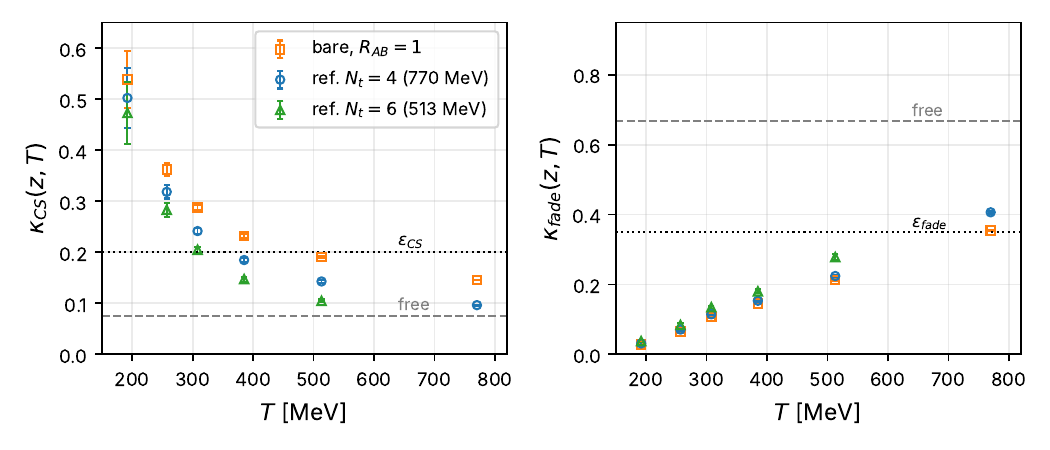}
\caption{Screening probes of the $\bar ud$ sector at $zT=1$ versus temperature,
in the two reference schemes and for the bare probes ($\RAB=1$).
Dashed lines are the continuum free-fermion values.
The lattice free-fermion values of $\kappa_{CS}$ at the same $N_t$
(Appendix~\ref{app:freelat}) lie below the dashed line by $48\%$ and $10\%$ at
$N_t=16$ and $12$.
This is a transverse-volume effect.
They lie above it by $3$, $11$, $21$ and $33\%$ at $N_t=10$, $8$, $6$ and $4$.
This is a cutoff effect.
Dotted lines are the thresholds $\epsilon_{CS}=0.2$ and $\epsilon_{fade}=0.35$.}
\label{fig:probes_T}
\end{figure}

\begin{figure}[tbp]
\includegraphics[width=\linewidth]{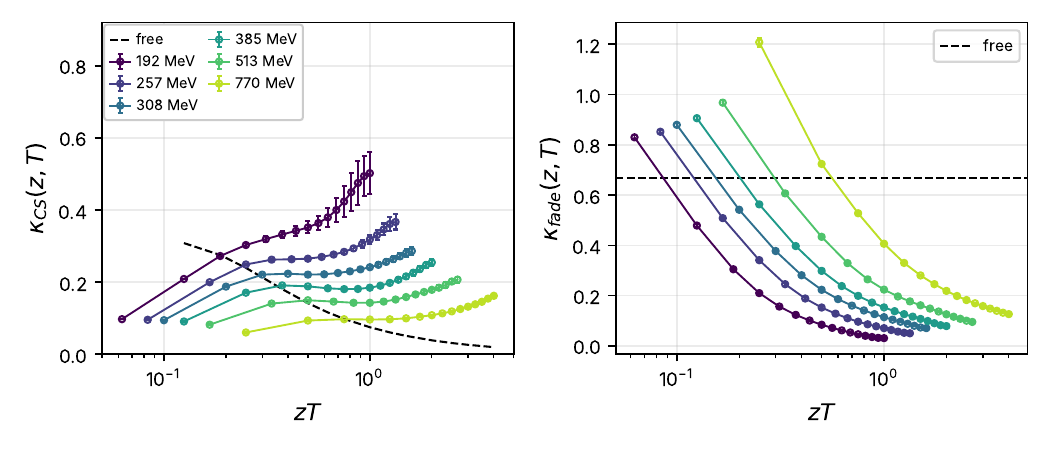}
\caption{Dependence of the screening probes on the separation, $\bar ud$ sector,
$N_t=4$ reference.
The dashed line in the left panel is the free-fermion $\kappa_{CS}(zT)$ of
Appendix~\ref{app:free}.}
\label{fig:probes_z}
\end{figure}

\begin{figure}[tbp]
\includegraphics[width=\linewidth]{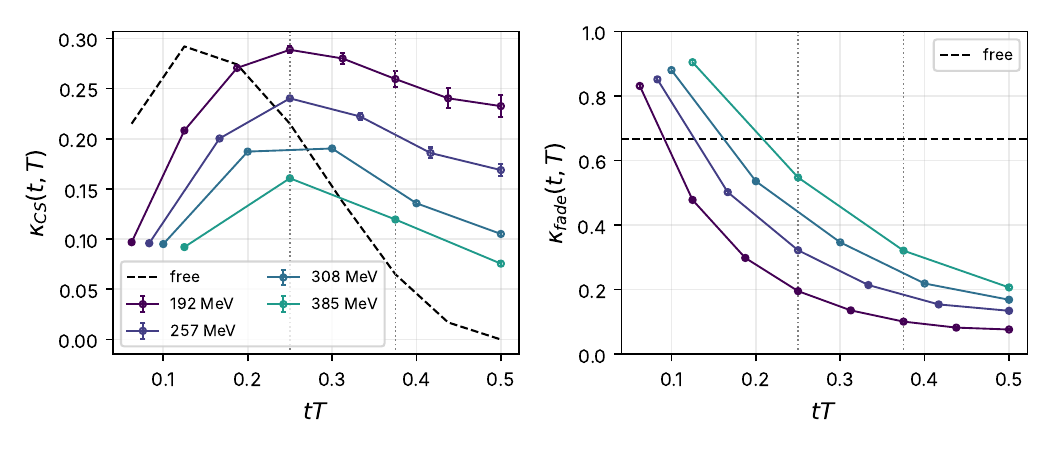}
\caption{Thermal probes of the $\bar ud$ sector versus $tT$ for
$N_t=16,12,10,8$, with the $N_t=4$ reference.
The dashed line in the left panel is $\kappa_{CS}$ of the lattice free quarks at
$N_t=16$.
In the right panel it is the free value $\kappa_{fade}=2/3$.
The dotted lines mark $tT=1/4$ and $3/8$, where the probes are evaluated.}
\label{fig:probes_t}
\end{figure}

\begin{table*}[tbp]
\caption{Renormalized probes of the $\bar ud$ sector at $\beta=6.20$ on
$32^3\times N_t$, with $\RAB$ from the thermal midpoint of the $N_t=4$ ensemble.
The thermal probes are at $tT=1/4$ and $tT=3/8$. The screening probes are at
$zT=1$ ($z=N_t$). Where the separation falls between lattice units ($tT=1/4$ at
$N_t=10$, and $tT=3/8$ at $N_t=12$, $10$ and $6$), the probe is interpolated
linearly in $t$. At $N_t=10$, $8$ and $6$ these separations are $t=2$--$3.75$
lattice units. A dash means that the thermal probe is not evaluated at that
$N_t$. The screening probes at $770$~MeV are evaluated at the reference
temperature (Sec.~\ref{subsec:cs}). The $N_t=2$ ensemble ($1539$~MeV) lies above
the reference temperature and is not used for the renormalized probes. An entry
$<10^{-4}$ means that the probe is below $10^{-4}$. The errors are statistical.
The free-fermion values are $\kappa_{CS}(tT=1/4)=0.217$,
$\kappa_{CS}(tT=3/8)=0.068$, $\kappa_{CS}(zT=1)=0.075$ and $\kappa_{fade}=2/3$.}
\label{tab:probes_ud}
\setlength{\tabcolsep}{4pt}
\begin{adjustbox}{max width=\linewidth}
\begin{tabular}{cc|ccc|ccc|ccc}
\hline\hline
 & & \multicolumn{3}{c|}{thermal, $tT=1/4$} & \multicolumn{3}{c|}{thermal,
 $tT=3/8$} & \multicolumn{3}{c}{screening, $zT=1$}\\
$N_t$ & $T$ [MeV] & $\kappa_{TX}$ & $\kappa_{CS}$ & $\kappa_{fade}$ &
$\kappa_{TX}$ & $\kappa_{CS}$ & $\kappa_{fade}$ & $\kappa_{T_4X_4}$ &
$\kappa_{CS}$ & $\kappa_{fade}$\\
\hline
16 & 192 & 0.0001(1) & 0.289(4) & 0.195(4) & 0.0002(1) & 0.259(8) & 0.101(3) &
0.0023(30) & 0.502(59) & 0.031(4)\\
12 & 257 & $<10^{-4}$ & 0.240(2) & 0.322(5) & $<10^{-4}$ & 0.204(3) & 0.184(3) &
$<10^{-4}$ & 0.318(13) & 0.071(5)\\
10 & 308 & $<10^{-4}$ & 0.189(1) & 0.441(2) & $<10^{-4}$ & 0.149(2) & 0.251(2) &
$<10^{-4}$ & 0.241(6) & 0.114(3)\\
8 & 385 & $<10^{-4}$ & 0.161(1) & 0.548(4) & $<10^{-4}$ & 0.120(2) & 0.321(4) &
$<10^{-4}$ & 0.184(4) & 0.153(4)\\
6 & 513 & -- & -- & -- & $<10^{-4}$ & 0.082(1) & 0.470(4) & $<10^{-4}$ &
0.142(2) & 0.224(4)\\
4 & 770 & -- & -- & -- & -- & -- & -- & $<10^{-4}$ & 0.096(1) & 0.406(3)\\
\hline\hline
\end{tabular}
\end{adjustbox}
\end{table*}

\begin{table}[tbp]
\caption{Scheme dependence of the screening probes of the $\bar ud$ sector at
$zT=1$. The two reference schemes take $\RAB$ from the thermal midpoint of the
$N_t=4$ ($T_r=770$~MeV) and $N_t=6$ ($T_r=513$~MeV) ensembles, and the bare
probes have $\RAB=1$. The renormalized probes are given for $T\le T_r$.}
\label{tab:scheme}
\begin{ruledtabular}
\begin{tabular}{cc|ccc|ccc}
 & & \multicolumn{3}{c|}{$\kappa_{CS}(z,T)$} &
 \multicolumn{3}{c}{$\kappa_{fade}(z,T)$}\\
$N_t$ & $T$ [MeV] & bare & $N_t^r=4$ & $N_t^r=6$ & bare & $N_t^r=4$ &
$N_t^r=6$\\
\hline
16 & 192 & 0.538(56) & 0.502(59) & 0.473(61) & 0.027(3) & 0.031(4) & 0.037(4)\\
12 & 257 & 0.362(13) & 0.318(13) & 0.283(13) & 0.064(4) & 0.071(5) & 0.084(6)\\
10 & 308 & 0.288(6) & 0.241(6) & 0.204(6) & 0.106(2) & 0.114(3) & 0.135(4)\\
8 & 385 & 0.232(4) & 0.184(4) & 0.147(5) & 0.145(3) & 0.153(4) & 0.180(5)\\
6 & 513 & 0.191(2) & 0.142(2) & 0.104(3) & 0.212(3) & 0.224(4) & 0.278(7)\\
4 & 770 & 0.145(1) & 0.096(1) & -- & 0.354(1) & 0.406(3) & --\\
\end{tabular}
\end{ruledtabular}
\end{table}

\begin{table}[tbp]
\caption{Crossing temperatures of the $\bar ud$ sector. The temperature $T_{CS}$
is where $\kappa_{CS}$ falls through $\epsilon_{CS}=0.2$. The temperature
$T_{fade}$ is where $\kappa_{fade}$ rises through $\epsilon_{fade}=0.35$. Both
follow from linear interpolation between neighboring ensembles, in the two
reference schemes and for the bare probes ($\RAB=1$), at several separations.
The errors are bootstrap errors from the statistical errors of the probes. The
renormalized probes are evaluated for $T\le T_r$ only. An entry $>T$ means that
the crossing lies above the highest temperature of the scan.}
\label{tab:crossings}
\begin{ruledtabular}
\begin{tabular}{lc|ccc|ccc}
 & & \multicolumn{3}{c|}{$T_{CS}$ [MeV]} & \multicolumn{3}{c}{$T_{fade}$
 [MeV]}\\
 & sep. & bare & $N_t^r=4$ & $N_t^r=6$ & bare & $N_t^r=4$ & $N_t^r=6$\\
\hline
screening & $zT=1/2$ & 506(3) & 357(4) & 289(3) & 493(3) & 434(3) & 351(3)\\
 & $zT=1$ & 485(6) & 364(5) & 314(6) & 762(2) & 691(4) & $>513$\\
thermal & $tT=1/4$ & $>385$ & 297(1) & 260(2) & 298(1) & 269(2) & 227(2)\\
 & $tT=3/8$ & 306(2) & 261(3) & 219(6) & 457(1) & 410(3) & 321(2)\\
\end{tabular}
\end{ruledtabular}
\end{table}

The chiral-spin probes depend on the reference ratios of Eq.~(\ref{eq:R_sym}),
which are $R_{VT}=1.1047(14)$ and $R_{VS}=0.8459(12)$ at $N_t=4$, and
$R_{VT}=1.193(4)$ and $R_{VS}=0.674(3)$ at $N_t=6$.

On the lattice free quarks, the probes take their free-fermion values of
Appendix~\ref{app:freelat}.
We find $\kappa_{fade}=0.6667$ at every $z$ and at every $t$ except the
midpoint, in all six mass combinations from $\bar ud$ to $\bar cc$.
At the midpoint, $\kappa_{fade}$ is the ratio $0/0$.
The result confirms the mass independence of $\kappa_{fade}$ derived in
Appendix~\ref{app:free}, now at finite lattice spacing.
The thermal $\kappa_{CS}(t)$ vanishes at the midpoint at every $N_t$.
In the $\bar ud$ sector at $N_t=16$, $\kappa_{CS}(z)=0.239$, $0.127$, $0.069$,
$0.039$ at $z=4,8,12,16$, while the continuum formulas in infinite volume give
$0.239$, $0.142$, $0.098$, $0.075$.
With the discrete transverse momenta $k_x,k_y=2\pi n/32$ of the $32^3$ box, they
give $0.233$, $0.123$, $0.068$, $0.037$ (Appendix~\ref{app:freelat}).
The free overlap quarks give the same six correlators as the free domain-wall
quarks, with agreement of $3\times10^{-6}$ in the $\bar ud$ sector and
$6\times10^{-4}$ in $\bar cc$, so the two actions share one set of free-fermion
values.
For free charm quarks on the same lattice, $\kappa_{CS}(z)$ is $0.24$--$0.32$,
near the upper bound $1/3$ of Appendix~\ref{app:free}.

Tables~\ref{tab:probes_ud}, \ref{tab:scheme} and \ref{tab:crossings} and
Fig.~\ref{fig:probes_T} give the probes of the $\bar ud$ sector, and we begin
with the screening correlators at $zT=1$.
The probe $\kappa_{CS}=\kappa_{AX_4}$ falls from $0.50(6)$ at $192$~MeV to
$0.096(1)$ at $770$~MeV. It crosses $\epsilon_{CS}=0.2$ at $T_{CS}=364(5)$~MeV.
With the $N_t=6$ reference it is $314(6)$~MeV, and the bare probe crosses at
$485(6)$~MeV (Table~\ref{tab:crossings}).
In the $N_t=4$ scheme, $\kappa_{CS}$ stays above the continuum free-fermion
value $0.075$ at every temperature.
The lattice free quarks at the same $N_t$ give a comparison
(Appendix~\ref{app:freelat}), with $\kappa_{CS}(zT=1)=0.067$--$0.090$ for
$N_t=12$ to $6$ and $0.100$ at $N_t=4$.
The measured probe is larger by factors $4.7$ to $1.6$ from $257$ to $513$~MeV,
and it reaches the lattice free-fermion value at $770$~MeV, with $0.096(1)$
against $0.100$.
At fixed temperature, $\kappa_{CS}$ grows with $z$ (Fig.~\ref{fig:probes_z})
while the free value falls with $z$.
The members of the triplet therefore have different screening masses, and the
mesons $A$ and $V$ are lighter than $T_4$ and $X_4$ by about $2\%$ at $385$~MeV.
Their amplitudes also differ, and the ratio $C_A/C_{X_4}$ at $zT=1$ is $3.3$,
$2.1$, $1.8$, $1.6$, $1.5$ and $1.3$ from $192$ to $770$~MeV.

The same correlators show how much a comparison of screening masses would miss.
We take the screening masses $m_A$ and $m_{X_4}$ from the plateaus of the
effective masses between $z=\min(N_t,12)$ and $15$, and define the mass analog
of the probe, $\kappa^m_{CS}=|m_A-m_{X_4}|/(m_A+m_{X_4})$, which needs no
renormalization.
It is $0.10(3)$, $0.038(10)$, $0.014(2)$, $0.012(2)$, $0.006(1)$ and $0.0036(2)$
from $192$ to $770$~MeV.
The renormalized probe $\kappa_{CS}=\kappa_{AX_4}$ at $zT=1$, built from the
same two correlators, is $0.502(59)$, $0.318(13)$, $0.241(6)$, $0.184(4)$,
$0.142(2)$ and $0.096(1)$, larger than $\kappa^m_{CS}$ by a factor that grows
from $5$ to $27$.
For lattice free quarks at the same $N_t$, $\kappa^m_{CS}$ is $0.028$, $0.015$,
$0.010$, $0.007$, $0.006$ and $0.005$, and $\kappa_{CS}(zT=1)$ is $0.039$,
$0.067$, $0.077$, $0.083$, $0.090$ and $0.100$.
From $308$~MeV on, $\kappa^m_{CS}$ is within a factor $1.7$ of its free value,
whereas $\kappa_{CS}$ is $3.1$, $2.2$ and $1.6$ times its free value at $308$,
$385$ and $513$~MeV.
The screening masses of $A$ and $X_4$ are thus close to their free values, while
the correlators still carry the splitting of the triplet in their amplitudes.

The mass analog of the fade probe is
$\kappa^m_{fade}=|m_A-m_{X_4}|/|m_{M_0}-m_{M_2}|$, where $m_{M_0}$ and $m_{M_2}$
are the averages of the screening masses of the channels of $M_0$ and $M_2$, and
it falls from $0.33(10)$ to $0.14(1)$ over the same range.
Meanwhile the renormalized $\kappa_{fade}$ rises from $0.031(4)$ to $0.406(3)$.
The reason is that the denominator of the mass analog is driven by the $S,P$
screening mass closing on $2\pi T$, not by the multiplet structure.
In the $\bar cc$ sector, the mass analog of the $U(1)_A$ test is
$|m_{T_4}-m_{X_4}|/(m_{T_4}+m_{X_4})$, which is $0.018(1)$ at $308$~MeV and
$0.0004$ at $770$~MeV, where $\kappa_{T_4X_4}=0.596(3)$ and $0.071$.
By its screening masses, charm would pass the $U(1)_A$ test at $\epsilon=0.05$
above $300$~MeV, but $\kappa_{T_4X_4}$ from its correlators is $30$ to $180$
times larger than the mass analog and passes that test only above
$845$--$881$~MeV.
Degenerate screening masses are implied by the symmetry, and are therefore
necessary, but the free theory already provides them in every channel.
The symmetry information resides in the amplitudes and the power-law prefactors,
and the renormalized correlators retain them (Sec.~\ref{subsec:cs},
Appendix~\ref{app:free}).

The fade probe $\kappa_{fade}(z)$ rises from $0.031(4)$ at $192$~MeV to
$0.406(3)$ at $770$~MeV, and it stays far below the free value $2/3$ up to
$513$~MeV. It crosses $\epsilon_{fade}=0.35$ at $T_{fade}=691(4)$~MeV, with a
systematic uncertainty of $+20$~MeV from the evaluation at the reference
temperature (see below), while the bare probe crosses at $762$~MeV. In the
$N_t=6$ scheme, $\kappa_{fade}$ is still $0.28$ at its reference temperature
$513$~MeV, the highest temperature at which the probes of that scheme are
evaluated, so its crossing is not resolved.
The chirally completed multiplet $M_2$ thus stays well separated from $M_0$
throughout the range, and the $S,P$ screening masses remain far below those of
$V,A,T_4,X_4$.

With the thresholds of Sec.~\ref{subsec:cs}, the chiral-spin window of the
screening correlators at $zT=1$ is $364$--$691$~MeV. With the $N_t=6$ reference
the lower end moves down by $50$~MeV, to $314$~MeV, and the upper end is not
resolved, since the probes of that reference are evaluated only up to
$513$~MeV. The bare probes
give $485$--$762$~MeV.
The window moves with $T_r$, as expected for an emergent symmetry
(Sec.~\ref{subsec:emergent}).
At $zT=1/2$ the lower end moves little, with $T_{CS}=357(4)$~MeV, while the
upper end comes down to $T_{fade}=434(3)$~MeV. At shorter distance the $S,P$
correlators are closer to the triplet, so the fade sets in earlier.
The window depends on the separation at which the probes are evaluated, as a
window of an
emergent symmetry must, because the separation is part of its statement.

The thermal probes (Fig.~\ref{fig:probes_t}) are evaluated at $tT=1/4$ and
$3/8$.
At $tT\ge3/8$ the interacting $\kappa_{CS}(t)$ lies above the free-fermion curve
at every temperature, since the free value falls to zero at the midpoint while
the interacting one remains at $0.08$--$0.23$ there.
The fade probe decreases with $t$ at fixed $T$, from values above $2/3$ at $t=1$
to $0.08$--$0.21$ at the midpoint.
At $tT=1/4$ the two criteria do not define a window, because $\kappa_{fade}$
rises through $\epsilon_{fade}$ at $269(2)$~MeV, before $\kappa_{CS}$ falls
through $\epsilon_{CS}$ at $297(1)$~MeV. The short separation still weights the
high-frequency part of the spectral function, where the $S,P$ and the triplet
channels approach each other.
At $tT=3/8$ a thermal window opens, running from $T_{CS}=261(3)$~MeV to
$T_{fade}=410(3)$~MeV. With the $N_t=6$ reference the window is
$219$--$321$~MeV, and the bare probes give $306$--$457$~MeV, so this window lies
lower than the screening window and partly overlaps it.

The thermal and screening probes thus do not give the same picture.
At fixed $T$ the thermal correlator probes the spectral function at separations
$t\le1/2T$, while the screening correlator probes the screening masses at
$z\gtrsim1/T$.
The thermal numbers at $N_t\le10$ rest on separations of at most four lattice
units and on interpolation between them, so finer lattices are needed before the
thermal window is quoted with the same confidence as the screening one.

Above the reference temperature the fade probe loses its meaning.
At the reference point the renormalized $M_0$ and $M_1$ coincide by
construction, and as $T$ increases past $T_r$ the renormalized difference
$M_0-M_1$ passes through zero and changes sign.
For the screening correlators the same holds for $M_0-M_2$.
The denominator of $\kappa_{fade}$ then vanishes somewhere above $T_r$, so the
probe has a pole there.
In the $N_t=4$ scheme, $\kappa_{fade}(zT=1)$ is $0.41$ at $770$~MeV and $6(1)$
at $1539$~MeV. In the $N_t=6$ scheme it is $0.87(4)$ already at $770$~MeV and
$0.13(1)$ at $1539$~MeV, which is on the far side of the pole.
This is why $T_r$ is chosen as the highest admissible temperature of the set,
and the renormalized probes are evaluated for $T<T_r$, with the screening probes
at $zT\ge1$ also at $T=T_r$ (Sec.~\ref{subsec:cs}).
On the reference ensemble the normalization at $t_r$ makes the renormalized
$M_0-M_2$ small at $z\lesssim t_r$, where $\kappa_{fade}$ exceeds $2/3$, and the
effect decreases with $z$.
At $zT=1$ we estimate it from the two references at $513$~MeV, which is the
reference temperature of the $N_t=6$ scheme.
The ratio of $\kappa_{fade}(zT=1)$ in the $N_t=6$ and $N_t=4$ schemes is $1.18$
from $257$ to $385$~MeV ($1.22$ at $192$~MeV, with large errors) and $1.24$ at
$513$~MeV, so the evaluation at $T_r$ raises $\kappa_{fade}(zT=1)$ by about
$5\%$,
while $\kappa_{CS}$ shows no such step.
For the $N_t=4$ scheme this lowers $T_{fade}(zT=1)$ by about $20$~MeV, which we
quote as a systematic uncertainty of the upper end of the window.
For the bare probes there is no pole, and at $1539$~MeV the bare probes reach
$\kappa_{CS}(zT=1)=0.073$ and $\kappa_{fade}=0.59$, which are close to the
continuum free values $0.075$ and $0.667$.
This is expected when the interactions become weak and $Z_\Gamma\to1$ at tree
level.
The lattice free-fermion value at $N_t=2$ is $0.086$
(Appendix~\ref{app:freelat}), and the bare QCD value is $15\%$ below it, but the
cutoff effects on the free correlators are themselves large there.

\subsection{From light to heavy sectors}
\label{subsec:num_mass}

\begin{figure*}[tbp]
\includegraphics[width=\linewidth]{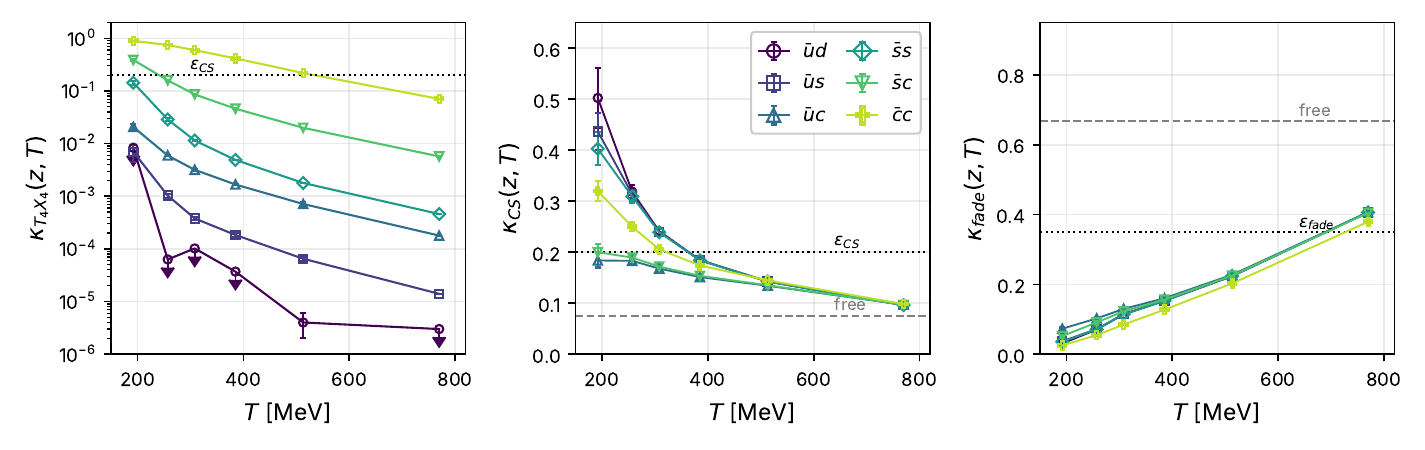}
\caption{Screening probes at $zT=1$ of the six flavor sectors versus
temperature, with the $N_t=4$ reference (Table~\ref{tab:sectors}).
The panels show the $U(1)_A$ test $\kappa_{T_4X_4}$ (log scale), the intrinsic
splitting $\kappa_{CS}$ and the fade probe $\kappa_{fade}$.
Values of $\kappa_{T_4X_4}$ consistent with zero within two standard deviations
are shown as upper limits (symbols with arrows).
They are placed at the value plus two standard deviations, or at
$3\times10^{-6}$ when that is smaller.
Dotted lines are the thresholds.
Dashed lines are the free-fermion values of the light sector.}
\label{fig:sectors}
\end{figure*}

Figure~\ref{fig:sectors} shows how the three probes change from the light to the
heavy sectors, and Table~\ref{tab:sectors} gives the numbers.
The six sectors order themselves by the product $m_1m_2$ of the two quark
masses.
In lattice units it is $1.6\times10^{-6}$, $5\times10^{-5}$, $6.9\times10^{-4}$,
$1.6\times10^{-3}$, $2.2\times10^{-2}$ and $0.30$ for $\bar ud$, $\bar us$,
$\bar uc$, $\bar ss$, $\bar sc$ and $\bar cc$.

The $U(1)_A$ test follows this product.
At $385$~MeV, $\kappa_{T_4X_4}$ is below $10^{-4}$, $2\times10^{-4}$,
$1.7\times10^{-3}$, $4.9\times10^{-3}$, $0.046$ and $0.42$ in the six sectors,
and the $\bar ud$ value is consistent with zero.
The other five are $2$ to $4$ times $m_1m_2a^2$ from $\bar us$ to $\bar sc$,
over nearly three orders of magnitude, and the relation bends below
proportionality only for $\bar cc$, at $1.4$ times, as it approaches its bound
of unity.
This is the explicit breaking by the mass term.
In the free traces of Appendix~\ref{app:free}, the partners $T_4$ and $X_4$
differ only in the sign of the $m_1m_2$ term, and the same holds for $V$ and
$A$.
The $SU(2)_L\times SU(2)_R$ test $\kappa_{VA}$ indeed tracks $\kappa_{T_4X_4}$
sector by sector (Table~\ref{tab:sectors}).
In every sector the $U(1)_A$ probe $\kappa_{T_4X_4}$ at $zT=1$ decreases with
temperature between $192$ and $770$~MeV, by a factor $13$ for $\bar cc$ and a
factor $68$ to $500$ for $\bar us$, $\bar uc$, $\bar ss$ and $\bar sc$.

The intrinsic splitting shows a different pattern.
The three light sectors $\bar ud$, $\bar us$ and $\bar ss$ agree within errors
at every temperature, with $\kappa_{CS}$ falling from $0.40$--$0.50$ at
$192$~MeV to $0.096$ at $770$~MeV, although $m_1m_2$ differs by three orders of
magnitude between them.
The two sectors with one charm quark vary more slowly, with $\kappa_{CS}$ going
from $0.18$--$0.20$ at $192$~MeV to $0.097$ at $770$~MeV. It lies below the
light sectors up to $513$~MeV and joins them at $770$~MeV.
The $\bar cc$ sector is $0.32(2)$ at $192$~MeV, $15$--$36\%$ below the light
sectors up to $308$~MeV, and within $6\%$ of them from $385$~MeV on.
At $770$~MeV all six sectors coincide, at $0.096$--$0.098$.

The slow variation of the heavy-light $\kappa_{CS}$ is not a free-fermion
effect.
On the lattice free quarks at $N_t=16$, $\kappa_{CS}(z)$ of $\bar uc$ is $0.12$
and $0.08$ at $zT=1/2$ and $3/4$, against $0.13$ and $0.07$ for $\bar ud$.
The heavy-light triplet $(A,T_4,X_4)$ is simply more degenerate than the light
one at low temperature.
The thermal probes show the same ordering at $192$~MeV, more weakly:
$\kappa_{CS}(tT=3/8)$ of the triplet $(V,T,X)$ is $0.17$ for $\bar uc$ against
$0.26$ for $\bar ud$, while the free values are $0.08$ and $0.07$.
But $\kappa_{fade}$ of $\bar uc$ rises faster, $0.19$ against $0.10$, so the
thermal heavy-light window is not wider than the light one.
The $\kappa_{T_4X_4}$ of $\bar uc$ passes the $U(1)_A$ test by a wide margin, so
the ordered criteria would certify the chiral-spin symmetry in the $\bar uc$
sector already at $192$~MeV. What this degeneracy means for a heavy-light system
is a question for a dedicated study, since the heavy quark line is far from
chiral there.
The mass-dependent free benchmark of Appendix~\ref{app:free} is the starting
point.

The fade probe is the most mass independent of the three.
All six sectors lie within $0.05$ of each other at every temperature, with
values from $0.02$--$0.07$ at $192$~MeV to $0.38$--$0.41$ at $770$~MeV. Only
$\bar cc$ is systematically low, by $17$--$25\%$ below $400$~MeV. The distance
of the chiral-spin triplet from the $U(1)_A$ multiplet, in units of the
splitting, is thus nearly the same from light to heavy quarks.

In summary, the quark masses rise from $\bar ud$ to $\bar cc$, and the explicit
$U(1)_A$ breaking grows by more than four orders of magnitude.
Over the same change, $\kappa_{fade}$ varies by at most a factor $2.9$ at the
lowest temperature and by at most $26\%$ for $T\ge385$~MeV. The probe
$\kappa_{CS}$ varies by a factor $2.7$ at the lowest temperature and by less
than $10\%$ from $513$~MeV on. The mass independence of $\kappa_{fade}$, exact
for free quarks (Appendix~\ref{app:free}), thus holds approximately in
QCD, $\kappa_{CS}$ becomes nearly mass independent for
$T\gtrsim400$~MeV, and this allows the same $\epsilon_{CS}$ and
$\epsilon_{fade}$ from light to heavy quarks.
The $U(1)_A$ test decides, sector by sector, whether the chiral-spin probes
test $SU(2)_{CS}$ at all.

\subsection{Chiral-spin windows of the six sectors}
\label{subsec:num_cswin}

\begin{table*}[tbp]
\caption{Chiral-spin crossing temperatures of the six flavor sectors from the
screening correlators at $zT=1/2$, $1$ and $2$. The temperature $T_1^{\bar qQ}$
is the one above which the $U(1)_A$ test passes,
$\kappa_{T_4X_4}\le\epsilon_{CS}=0.2$. It is the same for the bare and the
renormalized probes, and its errors are as in Table~\ref{tab:hier}. The
temperatures $T_{CS}$ and $T_{fade}$ are the crossings of $\kappa_{CS}$ through
$\epsilon_{CS}$ and of $\kappa_{fade}$ through $\epsilon_{fade}=0.35$. They are
given for the bare probes ($\RAB=1$, ensembles up to $1539$~MeV) and for the
renormalized probes with the $N_t=4$ reference (evaluated for
$T\le770$~MeV). By the
ordered criteria of Sec.~\ref{subsec:cs}, the chiral-spin window of a sector is
$\max(T_1,T_{CS})<T<T_{fade}$. The entry $<T$ means that the criterion holds
already on the lowest ensemble of the scan ($192$~MeV, or $385$~MeV at $zT=2$).
The entry $>T$ means that the crossing lies above the scan.}
\label{tab:cswin}
\begin{ruledtabular}
\begin{tabular}{cl|cccccc}
$zT$ & & $\bar ud$ & $\bar us$ & $\bar uc$ & $\bar ss$ & $\bar sc$ & $\bar cc$\\
\hline
$1/2$ & $T_1^{\bar qQ}$ & $<192$ & $<192$ & $<192$ & $<192$ & 211(2)(6) &
476(1)(8)\\
 & $T_{CS}$, bare & 506(3) & 505(3) & 469(4) & 506(3) & 479(3) & 527(3)\\
 & $T_{fade}$, bare & 493(2) & 493(3) & 471(2) & 493(3) & 475(2) & 493(2)\\
 & $T_{CS}$, $N_t^r=4$ & 357(4) & 356(4) & $<192$ & 357(4) & 263(8) & 399(4)\\
 & $T_{fade}$, $N_t^r=4$ & 434(3) & 433(3) & 407(3) & 433(3) & 409(3) & 410(3)\\
\hline
$1$ & $T_1^{\bar qQ}$ & $<192$ & $<192$ & $<192$ & $<192$ & 241(1)(5) &
534(2)(13)\\
 & $T_{CS}$, bare & 485(6) & 484(6) & 382(17) & 484(6) & 398(17) & 482(7)\\
 & $T_{fade}$, bare & 762(2) & 762(2) & 762(2) & 762(2) & 765(2) & 836(3)\\
 & $T_{CS}$, $N_t^r=4$ & 364(5) & 363(5) & $<192$ & 363(5) & $<192$ & 321(9)\\
 & $T_{fade}$, $N_t^r=4$ & 691(4) & 691(4) & 689(4) & 691(4) & 692(4) & 728(4)\\
\hline
$2$ & $T_1^{\bar qQ}$ & $<385$ & $<385$ & $<385$ & $<385$ & $<385$ &
549(2)(22)\\
 & $T_{CS}$, bare & 610(15) & 610(14) & 558(18) & 610(15) & 558(18) & 509(23)\\
 & $T_{fade}$, bare & 1173(3) & 1174(3) & 1183(3) & 1174(3) & 1190(3) &
 1287(3)\\
 & $T_{CS}$, $N_t^r=4$ & 477(9) & 476(9) & $<385$ & 476(9) & $<385$ & $<385$\\
 & $T_{fade}$, $N_t^r=4$ & $>770$ & $>770$ & $>770$ & $>770$ & $>770$ & $>770$\\
\end{tabular}
\end{ruledtabular}
\end{table*}

For the emergent symmetry the renormalization does matter.
Table~\ref{tab:cswin} sets the crossing temperatures of the bare probes next to
those of the renormalized probes with the $N_t=4$ reference, sector by sector.

The bare crossings reproduce the windows of Ref.~\cite{Chiu:2024jyz} where the
thresholds coincide.
At $zT=1$ the bare $T_{CS}$ is $485(6)$, $484(6)$ and $484(6)$~MeV in the light
sectors, and $382(17)$ and $398(17)$~MeV in $\bar uc$ and $\bar sc$, against
$480$, $485$, $475$, $370$ and $385$~MeV there.
At $zT=2$ it is $610$~MeV in the light sectors in both analyses.

The renormalization moves the light-sector window down at both ends, from
$485$--$762$ to $364$--$691$~MeV at $zT=1$.
Two reasons explain this.
The reference ratio $R_{VT}=1.10$ removes from $\kappa_{CS}$ the part of the
$V$--$T$ splitting that is the normalization of the two operators, and the ratio
$R_{VS}=0.85$ moves $M_0$ relative to $M_2$ in $\kappa_{fade}$.
The upper end is then nearly the same in all six sectors, at $689$--$692$~MeV
for the five sectors with at most one charm quark and $728(4)$~MeV for $\bar
cc$.

The lower end is where the sectors differ, and in two different ways.
In the light sectors $\kappa_{CS}$ sets it, at $364$~MeV. In the sectors with
one charm quark, $\kappa_{CS}$ is at or below $\epsilon_{CS}$ at nearly every
temperature (Sec.~\ref{subsec:num_mass}), so the $U(1)_A$ test sets the lower
end.
It is $241(1)$~MeV for $\bar sc$, against the bare $T_{CS}=398(17)$~MeV, and for
$\bar uc$ it lies below $192$~MeV, where $\kappa_{CS}=0.184(14)$ sits at the
threshold and the lower end is not resolved.
In the $\bar cc$ sector the lower end is the $U(1)_A$ certification, $T_1^{\bar
cc}=534(2)$~MeV, which is well above the $\kappa_{CS}$ crossing at $321(9)$~MeV
and the same for the bare and the renormalized probes.
The renormalized $\bar cc$ window is $534$--$728$~MeV, against $534$--$836$~MeV
bare, with the same systematic uncertainty of about $+20$~MeV on the upper end
as in the light sector.

At $zT=1/2$ the windows end early, with $T_{fade}$ at $407$--$434$~MeV. At
$zT=2$ their upper end lies above the reference temperature, because
$\kappa_{fade}$ does not reach $\epsilon_{fade}$ below $770$~MeV in any sector,
while the bare probes give $T_{fade}=1173$--$1287$~MeV.

The physical interpretation follows
Refs.~\cite{Glozman:2015qva,Chiu:2023hnm,Chiu:2024jyz}.
In this interpretation, the emergence of $SU(2)_{CS}$ (Sec.~\ref{subsec:cs}) in
the meson correlators means that, in the window, the color-singlet $\bar qQ$
pairs are bound predominantly by the chromoelectric interaction.
They form hadron-like objects and do not propagate as quasi-free quarks.
Above $T_{fade}$ the $U(1)_A$ multiplet merges into the chiral-spin multiplet
and the correlators approach the pattern of free quarks, which is the
dissolution of these objects.

The renormalized probes place the onset of this regime in the light sector at
$364(5)$~MeV, which is about $2.3\,T_c$ with $T_c\simeq155$~MeV and lower than
the bare onset by $120$~MeV. The end of the regime is at $691(4)$~MeV, about
$4.5\,T_c$, for every sector but $\bar cc$.
For the heavy quarks, the picture of Ref.~\cite{Chiu:2024jyz} survives the
renormalization and sharpens.
The $U(1)_A$ breaking in the charm sector is the explicit one of the mass term.
The $\bar cc$ window $534$--$728$~MeV is therefore the range in which the
charm-anticharm pair in the vector channels is at once chirally symmetric to
$20\%$ and degenerate in its $SU(2)_{CS}$ multiplets to within $\epsilon_{CS}$,
and it is the cleanest target for the phenomenology of the chiral-spin regime in
heavy-ion collisions suggested in Ref.~\cite{Chiu:2024jyz}.

Two further points of Refs.~\cite{Chiu:2023hnm,Chiu:2025sym} follow from the
same tables.
First, the emergent symmetry is approximate, and the intrinsic splitting is its
measure.
Ref.~\cite{Chiu:2023hnm} noted that $SU(2)_{CS}$ is not exact, so the
chromomagnetic interaction cannot be neglected in the binding of these objects,
and it asked for the relation between the degree of chromoelectric dominance and
the threshold $\epsilon_{CS}$.
In the renormalized probes, that relation is $\kappa_{CS}$ itself, compared
with its free-fermion value.
Inside the window at $zT=1$ the splitting is $0.10$--$0.18$, against $0.075$ for
free continuum quarks and $0.08$--$0.10$ for lattice free quarks at the same
$N_t$ (Appendix~\ref{app:free}).
The triplet is therefore never more degenerate than for free quarks at this
separation.
The emergence is the compactness of the triplet relative to its distance from
$M_0$, which is $\kappa_{fade}\le0.35$ against $2/3$ for free quarks, and it is
not a degeneracy beyond the free one.

Second, Ref.~\cite{Chiu:2025sym} compared the precision of the chiral and the
chiral-spin symmetry through $\kappa_{VA}/\kappa_{CS}$, and it found this ratio
below $10^{-2}$ in the light sectors, below $0.5$ in the heavy-light sectors and
of order one for $\bar cc$.
In the present language, this is the distinction between the exact and the
emergent symmetry.
The probe $\kappa_{VA}$ is at the level of the statistics in the light sectors,
where $\kappa_{CS}$ is $0.1$--$0.5$, while in $\bar cc$ the explicit breaking by
the mass term makes the two comparable, at $0.07$ and $0.10$ at $770$~MeV.

The most interesting feature of the windows is the one found in
Ref.~\cite{Chiu:2024jyz} with the bare probes.
There the windows of the vector mesons with a charm quark remain wider than
those of the light mesons as the precision of the symmetry is raised, and
Ref.~\cite{Chiu:2025sym} sharpened this with a $b$ quark.
The widest chiral-spin windows belong to the sectors that pair the heaviest
quark with a light one, which are $\bar uc$ and $\bar sc$ here and $\bar ub$ and
$\bar sb$ with a $b$ quark.
Table~\ref{tab:cswin_eps} shows what survives the renormalization.

\begin{table}[tbp]
\caption{Chiral-spin windows $\max(T_1^{\bar qQ},T_{CS})<T<T_{fade}$ [MeV] of
the six sectors from the screening correlators at $zT=1$, with the $N_t=4$
reference. The threshold $\epsilon_{CS}$ of the two ordered criteria is lowered
at fixed $\epsilon_{fade}=0.35$. The entry $<192$ means that the lower end lies
at or below the lowest ensemble. A dash means that no temperature below $T_r$
satisfies both criteria.}
\label{tab:cswin_eps}
\begin{ruledtabular}
\begin{tabular}{c|cccccc}
$\epsilon_{CS}$ & $\bar ud$ & $\bar us$ & $\bar uc$ & $\bar ss$ & $\bar sc$ &
$\bar cc$\\
\hline
0.20 & 364--691 & 363--691 & $<192$--689 & 363--691 & 241--692 & 534--728\\
0.15 & 490--691 & 490--691 & 393--689 & 490--691 & 410--692 & 599--728\\
0.10 & -- & -- & -- & -- & -- & --\\
\end{tabular}
\end{ruledtabular}
\end{table}

\emph{At $\epsilon_{CS}=0.2$ the $\bar uc$ and $\bar sc$ windows are the
widest.}
Their lower ends are set by the $U(1)_A$ test.
In a heavy-light sector the splitting $T_4-X_4$ becomes small once the lighter
quark is effectively chiral, which keeps the lower ends low
(Sec.~\ref{subsec:num_hier}).
Their upper ends coincide with those of the light sectors.
At $\epsilon_{CS}=0.15$ the light windows shrink to $490$--$691$~MeV and the
$\bar cc$ window to $599$--$728$~MeV, while the $\bar uc$ and $\bar sc$ windows
keep $393$--$689$ and $410$--$692$~MeV. They remain about three hundred MeV
wide.
At $\epsilon_{CS}=0.1$ every window is empty, because $T_{CS}$ then lies above
$T_{fade}$.
The value of $\kappa_{CS}$ at $T_{fade}$ is $0.106$--$0.110$ in all sectors, so
every window is empty for $\epsilon_{CS}\lesssim0.11$.
At this separation the free-fermion value of $\kappa_{CS}$ is $0.075$ in the
continuum and $0.07$--$0.10$ on the lattice at $N_t\le12$, so a threshold of
$0.05$ is out of reach with any reference temperature.
The heavy-light windows therefore do not survive to smaller $\epsilon_{CS}$
than the light one, but they are wider at $\epsilon_{CS}=0.2$ and $0.15$.

The mechanism is the one of Sec.~\ref{subsec:num_mass}.
The heavy-light intrinsic splitting varies slowly, from $0.19$ to $0.10$ over
the whole range, so it passes a tightened threshold earlier than the
light-sector splitting, which falls from $0.5$.
It does so without the $U(1)_A$ obstruction of $\bar cc$.
In the terms of Ref.~\cite{Chiu:2024jyz}, the vector mesons with one charm quark
are the most attractive channels in which to look for the chiral-spin regime,
since their windows are the widest and begin lowest.
This is now a statement about renormalized correlators.

Two qualifications go with it.
First, the lower end of the $\bar uc$ window is not resolved, since
$\kappa_{CS}=0.184(14)$ already sits at the threshold at $192$~MeV.
Second, a pair with one heavy, non-chiral quark line has a chiral-spin
degeneracy that does not have the same meaning as in the light sector, where
both lines are chirally symmetric.
The $\bar cc$ sector remains the one whose window is best controlled, because
its lower end is the scheme-independent $U(1)_A$ certification, and it is also
experimentally the most accessible, through the $J/\psi$.
Two general caveats apply: the window depends on the separation at which the
probes are evaluated, and all numbers are at a single lattice spacing.

%======================================================================
\section{Summary and outlook}
\label{sec:summary}
%======================================================================

\subsection{The construction}

Our construction of the renormalized symmetry probes $\kappa_{AB}(t,T)$ and
$\kappa_{AB}(z,T)$ is economical.
For each channel, one measures the folded correlators $C_\Gamma(t,T)$ and
$C_\Gamma(z,T)$ and records the reference value $C_\Gamma(t_r,T_r)$ in the
$\bar ud$ light sector.
One then forms $\kAB$ from Eq.~(\ref{eq:kappa_RAB}) with the single ratio
$\RAB$, for both $\kappa_{AB}(t,T)$ and $\kappa_{AB}(z,T)$, and for every flavor
sector.

For partners of an exact symmetry of the massless action, $\RAB=1$ by symmetry,
so $\kAB$ is RG invariant and scheme independent in mass-independent schemes.
Among them the tensor pair $(T_k,X_k)$ keeps $R_{T_kX_k}=1$ exactly also for
Wilson-type fermions on isotropic lattices, because $T_k$ and $X_k$ are
components of one tensor, so $\kappa_{T_kX_k}$ is a $U(1)_A$ probe that needs no
renormalization on Wilson-type actions (Sec.~\ref{subsec:exact}).
For these pairs the measured reference ratio serves only to check that the
reference temperature is admissible.
For partners of an emergent symmetry, $\RAB$ carries their relative logarithmic
cutoff dependence, and it is fixed once at the reference point.
The residual scheme dependence of $\kAB$ is the scale dependence of the emergent
degeneracy itself, and we quote it with the results.

The susceptibility ratio of Eq.~(\ref{eq:kappa_chi_RAB}) follows by summation
with the same $\RAB$, but it needs the temperature subtraction
$\chi_\Gamma(T)-\chi_\Gamma(T_r)$, which removes the additive divergences.
The correlators $C(t)$ and $C(z)$ at nonzero separation do not need this
subtraction, so the renormalized $\kAB(zT)$ and $\kAB(tT)$ are the fundamental
probes.
They serve exact and emergent symmetries alike, namely $U(1)_A$,
$SU(N_f)_L\times SU(N_f)_R$ and $SU(2)_{CS}$.

Screening masses are the asymptotic exponents of the correlators, and
susceptibilities are their sums.
Both are weaker degeneracy tests than the correlators, already for the exact
symmetries.
In the $\bar cc$ sector, the mass analog of $\kappa_{T_4X_4}$ passes the
$U(1)_A$ test at $\epsilon=0.05$ above $300$~MeV, while $\kappa_{T_4X_4}$ from
the correlators is $30$ to $180$ times larger and passes it only above
$845$--$881$~MeV.
For the chiral-spin symmetry both screening-mass and susceptibility probes fail:
the screening masses discard the amplitudes, and the susceptibilities are
dominated by the shortest separations, where the kinetic term breaks
$SU(2)_{CS}$ maximally.

For the chiral-spin symmetry the tests are ordered.
The $U(1)_A$ test $\kappa_{T_4X_4}$ comes first.
Then the intrinsic splitting $\kappa_{CS}$ (\ref{eq:kappa_CS}) decides the
emergence, and the fade probe $\kappa_{fade}$ (\ref{eq:kappa_fade}), which
measures the splitting against the distance of $M_1/M_2$ from $M_0$, decides the
fading.

For free quarks of any mass, we derive $\kappa_{fade}=2/3$
(Appendix~\ref{app:free}).
The renormalized screening probe at $zT=1$ rises to $0.41$ at $T_r=770$~MeV,
and the bare one reaches $0.59$ at $1539$~MeV, both below this value, and the
threshold $\epsilon_{fade}=0.35$ used in this paper is about half of it.
In the noninteracting theory the numerator of $\kappa_{CS}$ has no explicit
$m_1m_2$ term, which motivates one threshold, $\epsilon_{CS}=0.2$, for light and
heavy quarks in this paper.
Free quarks lack the chiral-spin symmetry because their $\kappa_{fade}$ is too
large, not because their $\kappa_{CS}$ is too large.
On lattices with gauge links equal to the identity, optimal domain-wall quarks
give $\kappa_{fade}=2/3$ for all six mass combinations, in agreement with
continuum free quarks.

\subsection{Lattice results}

We applied the probes to $N_f=2+1+1$ lattice QCD with optimal domain-wall quarks
at $a=0.064$~fm.
The $U(1)_A$ and $SU(2)_L\times SU(2)_R$ tests of the $\bar ud$ sector are
consistent with zero at every temperature and separation, and below $10^{-3}$
except for $\kappa_{T_4X_4}=0.002(3)$ at $192$~MeV.
Across the six sectors the violation of $U(1)_A$ grows with the product
$m_1m_2$ of the two quark masses (Sec.~\ref{subsec:num_mass}).
The ratios $\RAB$ of the emergent $SU(2)_{CS}$ pairs are fixed in the $\bar ud$
sector and used in every sector, which makes the scheme mass independent.

The restoration temperatures of the exact symmetries, for the generators of
$SU(N_f)$, do not depend on the reference temperature $T_r$ and coincide with
those of the bare probes, since $\RAB=1$ for their probes, and so does the
hierarchy they express,
from $SU(2)_L\times SU(2)_R\times U(1)_A$ to $SU(3)_L\times SU(3)_R\times
U(1)_A$ and $SU(4)_L\times SU(4)_R\times U(1)_A$.
The chiral-spin windows depend on $T_r$: the crossing temperatures of the
emergent probes shift by $12$--$22\%$ from $N_t^r=4$ to $N_t^r=6$
(Table~\ref{tab:crossings}).
Relative to the bare probes, the renormalization lowers both ends of the
light-sector window by $70$--$120$~MeV.
The lower end of the $\bar cc$ window is set by the $U(1)_A$ test and stays at
$534$~MeV.
The lower end of the $\bar sc$ window falls from the bare $T_{CS}=398$~MeV to
$T_1^{\bar sc}=241$~MeV, where the $U(1)_A$ test sets it.
The lower end of the $\bar uc$ window is not resolved ($<192$~MeV).
The chiral-spin probes of the $\bar ud$, $\bar us$ and $\bar ss$ sectors agree
within $3\%$ at $257$~MeV and within $1\%$ from $308$~MeV on, which is the mass
independence of the intrinsic splitting, now in QCD.

The sectors with one charm quark have the widest chiral-spin windows at
$\epsilon_{CS}=0.2$ and $0.15$, for two reasons.
First, the intrinsic splitting of the heavy-light triplet $(A,T_4,X_4)$ is small
and varies slowly, at $0.18$--$0.20$ already at $192$~MeV against $0.40$--$0.50$
in the light sectors.
The criterion $\kappa_{CS}\le0.2$ is therefore met from about the lowest
temperature on, while the light sectors reach it only at $364$~MeV.
This small splitting of the screening probe at $zT=1$ is not a free-fermion
effect, since free quarks give nearly the same $\kappa_{CS}$ for $\bar uc$ and
$\bar ud$ (Sec.~\ref{subsec:num_mass}), and its physical origin remains open.
Second, the lighter quark keeps $m_1m_2$ small enough for the $U(1)_A$ test at
$\epsilon_{CS}=0.2$ to be passed early, whereas the $\bar cc$ sector passes it
only at $534$~MeV.
The wide heavy-light windows were found with the bare probes in
Refs.~\cite{Chiu:2024jyz,Chiu:2025sym}, and they survive the renormalization.
A window is empty when $T_{CS}$ lies above $T_{fade}$, that is, when
$\epsilon_{CS}$ is below the value of $\kappa_{CS}$ at $T_{fade}$.
This value is $0.106$--$0.110$ in all six sectors, so every window is empty for
$\epsilon_{CS}\lesssim0.11$.

With the screening correlators at $zT=1$, the light-sector window is
$364(5)$--$691(4)$~MeV with $N_t^r=4$, $T_r=1/(4a)=770$~MeV
(Table~\ref{tab:crossings}).
Its upper end may lie up to $20$~MeV higher, because it is interpolated to
$\kappa_{fade}$ at the reference temperature, which the normalization distorts
by about $5\%$.
The second reference, $N_t^r=6$ with $T_r=1/(6a)=513$~MeV, lowers the lower end
by $14\%$, to $314(6)$~MeV, the size expected from the $O(\alpha_s(T_r))$
departure of the reference midpoint from channel-blindness.
The intrinsic splitting stays above its continuum free-fermion value $0.075$ and
grows with $z$ beyond $zT\approx1$.
The triplet members differ in screening mass and in amplitude.
Only the correlators retain the amplitudes, and only the renormalized
correlators compare them across channels.
The analog of $\kappa_{CS}$ built with the screening masses is smaller by a
factor $5$ to $27$, and from $308$~MeV on it lies within a factor $2$ of its
free-quark value.
The thermal correlators at $tT=3/8$ give a lower window, $261(3)$--$410(3)$~MeV,
and they give none at $tT=1/4$ (Table~\ref{tab:crossings}).
The two kinds of correlator project the spectral function differently, so they
must agree only in the limit of exact restoration, a check beyond the present
data.

\subsection{Outlook}
\label{subsec:outlook}

Three ongoing studies follow from these results.

The first concerns the valence quarks.
The optimal domain-wall operator at finite $L_s$ is the overlap operator with
the sign function of $H_w$ replaced by its Zolotarev optimal rational
approximation on $[\lambda_{min},\lambda_{max}]$~\cite{Chiu:2002ir}.
The approximation is inexact for the eigenmodes of $H_w$ with
$|\lambda|<\lambda_{min}$.
The residual chiral breaking therefore grows with the density of the near-zero
modes of $H_w$ and as its spectral gap shrinks, and it decreases as
$\lambda_{min}$ is lowered.
It is largest in the light sectors near $T_c$, and it enters $\kappa_{PS}$ most
directly.
With the domain-wall valence quarks, the screening probes $\kappa_{PS}$ and
$\kappa_{T_4X_4}$ give restoration temperatures $T_1^{\bar qQ}$ that agree
within $6\%$ for $\bar ss$ and $\bar sc$, and within the interpolation
uncertainty of $137$~MeV for $\bar cc$.
At $\epsilon=0.05$ and $zT=1$, the two probes give $248$ and $235$~MeV for
$\bar ss$, $369$ and $375$~MeV for $\bar sc$, and $980$ and $881$~MeV for
$\bar cc$, and both lie below $192$~MeV for $\bar ud$, $\bar us$ and $\bar uc$.
At $\epsilon=0.01$, $\kappa_{PS}(zT=1)$, which is the most sensitive to the
near-zero modes, gives $208$ and $214$~MeV for $\bar ud$ and $\bar us$, while
$\kappa_{T_4X_4}$ gives $<192$~MeV for both, since $\kappa_{PS}=0.023(14)$ at
$192$~MeV against $\kappa_{T_4X_4}=0.002(3)$.
On the other hand, the thermal $\kappa_{PS}$ and $\kappa_{TX}$ at $tT=1/4$ both
pass $\epsilon=0.01$ below $192$~MeV in the light sectors.
To quantify the effect of the residual chiral symmetry breaking, we will repeat
the renormalized probes and the analysis with overlap valence quark propagators,
which are computed with the low modes of the overlap operator projected out
exactly and the remainder solved on the deflated space~\cite{Chiu:inprep}.
The comparison will show how much the probes of this paper, in particular
$\kappa_{PS}$ and $\kappa_{T_4X_4}$ of the light sectors, depend on the residual
chiral breaking of the valence quarks.

The second study is the continuum limit at fixed temperature.
The present single lattice spacing fixes $T$ by $N_t$, so each temperature is
realized once.
We will use ensembles at several lattice spacings tuned to the same set of
physical temperatures, and we will hold $T_r$ fixed in physical units.
Each $\kAB(T)$ is then extrapolated to $a\to0$ at fixed $T$~\cite{Chiu:inprep}.
The thermal window at $tT=3/8$ rests on separations of $2.25$--$4.5$ lattice
units ($N_t=12$ to $6$), against $4$--$10$ for the screening window at $zT=1$,
so it needs this extrapolation before it is quoted with the confidence of the
screening window.

The third study is the baryon sector.
Baryon correlators carry an open Dirac index, and their chiral-spin multiplets
differ from the mesonic ones.
Their operators are three-quark operators with their own anomalous dimensions,
but the $\RAB$ bookkeeping applies to them without change.
The multiplet structure and the free-fermion values have to be worked out
afresh, and we will present them in a separate paper.

Beyond our own program, the renormalized probes $\kAB(t,T)$ and $\kAB(z,T)$ can
be applied to the finite-temperature ensembles of other lattice groups, and most
of them use non-chiral lattice fermions.
By Sec.~\ref{sec:nonchiral}, the chiral-spin windows can be determined on
Wilson, twisted-mass and staggered ensembles at a single lattice spacing, with
the continuum limit as a check, while the probes of the exact symmetries require
the
continuum limit.
On the existing staggered and Wilson-type ensembles, $\kAB(t,T)$ and
$\kAB(z,T)$ can thus be evaluated directly.
The heavy-light chiral-spin windows found here can then be tested with the same
thresholds at a single lattice spacing, and the hierarchy of restoration
temperatures after the continuum limit.

%======================================================================
\begin{acknowledgments}
This work was supported by the National Science and Technology Council
(Grants No.~108-2112-M-003-005, No.~109-2112-M-003-006, No.~110-2112-M-003-009),
and Academia Sinica Grid Computing Centre (Grant No.~AS-CFII-112-103).
\end{acknowledgments}

%======================================================================

\appendix

\section{Flavor generators and the diagonal channels}
\label{app:flavor}

This appendix collects the conventions for the $SU(N_f)$ flavor structure of the
operators $O^a_\Gamma=\bar q\Gamma t^aq$ of Eq.~(\ref{eq:Oa_def}).
It gives the flavor-singlet and the diagonal nonsinglet correlators in terms of
the correlators of the quark bilinears $\bar q_1\Gamma q_2$, for $N_f=2$, $3$
and $4$.

\subsection{Generators}

The generators $t^a$ ($a=1,\dots,N_f^2-1$) of $SU(N_f)$ are hermitian and
traceless, and they are normalized as
\begin{equation}
\tr_F(t^at^b)=\tfrac12\,\delta^{ab}.
\label{eq:tnorm}
\end{equation}
They satisfy the commutation and anticommutation relations
\begin{align}
[t^a,t^b] &= i f^{abc}\,t^c,
\nonumber\\
\{t^a,t^b\} &= \sqrt{\tfrac{2}{N_f}}\;\delta^{ab}\,t^0 + d^{abc}\,t^c,
\qquad
\{t^a,t^0\} = \sqrt{\tfrac{2}{N_f}}\;t^a.
\label{eq:anticomm}
\end{align}
Here $f^{abc}$ are the totally antisymmetric structure constants of $SU(N_f)$,
and the $d^{abc}$ are the totally symmetric ones, which vanish identically for
$N_f=2$.
The flavor-singlet generator is
\begin{equation}
t^0=\frac{\Id}{\sqrt{2N_f}}.
\label{eq:t0}
\end{equation}
Then $\tr_F(t^0t^0)=1/2$, the same normalization as Eq.~(\ref{eq:tnorm}), and
$\tr_F(t^0t^a)=0$ for $a\neq0$.
The $N_f(N_f-1)$ off-diagonal generators are obtained by placing the Pauli
matrices $\sigma_1/2$ and $\sigma_2/2$ in the rows and columns $(i,j)$, for
every pair $i<j$.
The $N_f-1$ diagonal generators are
\begin{equation}
t^{n^2-1}=\frac{\mathrm{diag}(\,\underbrace{1,\dots,1}_{n-1},\,-(n-1),\,0,%
\dots,0\,)}
              {\sqrt{2n(n-1)}},
\qquad n=2,\dots,N_f .
\label{eq:tdiag}
\end{equation}
For $N_f=2$, the diagonal generator is $t^3=\mathrm{diag}(1,-1)/2$.
For $N_f=3$, they are $t^3=\mathrm{diag}(1,-1,0)/2$ and
$t^8=\mathrm{diag}(1,1,-2)/(2\sqrt3)$.
For $N_f=4$, they are $t^3=\mathrm{diag}(1,-1,0,0)/2$,
$t^8=\mathrm{diag}(1,1,-2,0)/(2\sqrt3)$ and
$t^{15}=\mathrm{diag}(1,1,1,-3)/(2\sqrt6)$.
The flavors are ordered by mass, $(u,d,s,c)$, so $SU(2)\subset SU(3)\subset
SU(4)$ act on $(u,d)$, $(u,d,s)$ and $(u,d,s,c)$, and the last diagonal
generator isolates the heaviest flavor (Appendix~\ref{app:algmax}).
The generators of a subgroup are those of the larger group with the extra rows
and columns set to zero.

\subsection{Connected and disconnected parts}

Write $t^a_{ij}$ for the matrix elements of $t^a$, so that the operator is
$O^a_\Gamma=\sum_{ij}t^a_{ij}\,\bar q_i\Gamma q_j$.
The correlator of Eq.~(\ref{eq:corr_def}) decomposes by Wick's theorem into a
quark-line connected and a quark-line disconnected part:
\begin{equation}
C^a_\Gamma=\sum_{i,j}\,|t^a_{ij}|^2\;C^{\rm c}_{\bar q_iq_j}
          +\sum_{i,k}\,t^a_{ii}\,t^a_{kk}\;D_{q_iq_k}.
\label{eq:conn_disc}
\end{equation}
Here $C^{\rm c}_{\bar q_iq_j}$ is the connected correlator of the bilinear $\bar
q_i\Gamma q_j$, with a single quark loop through the source and the sink, in
which $q_j$ propagates from the source to the sink and $q_i$ propagates back.
The $D_{q_iq_k}$ is the disconnected correlator of two separate loops, the loop
of $q_i$ at the sink and the loop of $q_k$ at the source, with the product of
the vacuum expectation values subtracted as in Eq.~(\ref{eq:corr_def}).
Both carry the Dirac structure $\Gamma$, which is suppressed.
Both are symmetric functions of the two quark masses, $C^{\rm c}_{\bar
q_iq_j}=C^{\rm c}_{\bar q_jq_i}$ and $D_{q_iq_k}=D_{q_kq_i}$, and they depend on
the flavors only through the masses.

For an off-diagonal generator the diagonal elements vanish, so the disconnected
part is absent.
With $|t^a_{ij}|^2=|t^a_{ji}|^2=1/4$,
\begin{equation}
C^a_\Gamma=\tfrac12\,C^{\rm c}_{\bar q_iq_j}\qquad(i\neq j).
\label{eq:offdiag}
\end{equation}
This is the factor $1/2$ of the flavor normalization quoted in
Sec.~\ref{sec:numerics} and in the tables of Appendix~\ref{app:tables}.
The six flavor sectors of this paper are the connected correlators
$\tfrac12C^{\rm c}_{\bar q_1q_2}$ of the six mass combinations of $u$, $s$ and
$c$.
For $q_1\neq q_2$ they are the correlators of the off-diagonal nonsinglets.
For $q_1=q_2$, the $\bar ss$ and $\bar cc$ sectors, they are not by themselves
the correlators of any flavor operator, because a flavor-diagonal bilinear $\bar
q_iq_i$ enters only through the diagonal generators and the singlet.
Their correlators are the combinations given below.
The connected $\bar ss$ correlator is, however, the correlator of the
off-diagonal operator $\bar s'\Gamma s$ with a valence quark $s'$ degenerate
with
$s$, in the partially quenched sense, and likewise for $\bar cc$.
In this sense $\bar ss$ and $\bar cc$ are nonsinglet sectors, and we keep them
in the chiral-spin analysis.

\subsection{Singlet and diagonal nonsinglets}

For a diagonal generator $t^a=\mathrm{diag}(d_1,\dots,d_{N_f})$, or for the
singlet, Eq.~(\ref{eq:conn_disc}) reads
\begin{equation}
C^a_\Gamma=\sum_i d_i^2\,C^{\rm c}_{\bar q_iq_i}
          +\sum_i d_i^2\,D_{q_iq_i}
          +2\sum_{i<k} d_id_k\,D_{q_iq_k}.
\label{eq:diag_general}
\end{equation}
This holds for any quark masses, and with the generators of Eq.~(\ref{eq:tdiag})
it gives the following.

For $SU(2)$ with flavors $(u,d)$:
\begin{align}
C^3 &= \tfrac14\big(C^{\rm c}_{\bar uu}+C^{\rm c}_{\bar dd}\big)
      +\tfrac14\big(D_{uu}+D_{dd}-2D_{ud}\big),
\nonumber\\
C^0 &= \tfrac14\big(C^{\rm c}_{\bar uu}+C^{\rm c}_{\bar dd}\big)
      +\tfrac14\big(D_{uu}+D_{dd}+2D_{ud}\big).
\label{eq:su2_gen}
\end{align}
For $SU(3)$ with flavors $(u,d,s)$, $C^3$ is as in $SU(2)$, and
\begin{align}
C^8 &= \tfrac1{12}\big(C^{\rm c}_{\bar uu}+C^{\rm c}_{\bar dd}
       +4C^{\rm c}_{\bar ss}\big)
\nonumber\\
    &\quad+\tfrac1{12}\big(D_{uu}+D_{dd}+4D_{ss}+2D_{ud}-4D_{us}-4D_{ds}\big),
\nonumber\\
C^0 &= \tfrac16\big(C^{\rm c}_{\bar uu}+C^{\rm c}_{\bar dd}
       +C^{\rm c}_{\bar ss}\big)
\nonumber\\
    &\quad+\tfrac16\big(D_{uu}+D_{dd}+D_{ss}+2D_{ud}+2D_{us}+2D_{ds}\big).
\label{eq:su3_gen}
\end{align}
For $SU(4)$ with flavors $(u,d,s,c)$, $C^3$ and $C^8$ are as above, and
\begin{align}
C^{15} &= \tfrac1{24}\big(C^{\rm c}_{\bar uu}+C^{\rm c}_{\bar dd}
          +C^{\rm c}_{\bar ss}+9C^{\rm c}_{\bar cc}\big)
\nonumber\\
       &\quad+\tfrac1{24}\big(D_{uu}+D_{dd}+D_{ss}+9D_{cc}
          +2D_{ud}+2D_{us}+2D_{ds}
\nonumber\\
       &\qquad\qquad-6D_{uc}-6D_{dc}-6D_{sc}\big),
\nonumber\\
C^0 &= \tfrac18\big(C^{\rm c}_{\bar uu}+C^{\rm c}_{\bar dd}
       +C^{\rm c}_{\bar ss}+C^{\rm c}_{\bar cc}\big)
\nonumber\\
    &\quad+\tfrac18\Big(\sum_{i}D_{q_iq_i}+2\sum_{i<k}D_{q_iq_k}\Big).
\label{eq:su4_gen}
\end{align}

Two sum rules follow from Eq.~(\ref{eq:diag_general}), and each line above
satisfies them.
First, $\sum_id_i^2=\tr_F(t^at^a)=1/2$ by Eq.~(\ref{eq:tnorm}), so the
coefficients of the connected correlators add up to $1/2$ in every channel,
singlet or nonsinglet.
Second, the coefficients of the disconnected correlators add up to
$\sum_id_i^2+2\sum_{i<k}d_id_k=(\sum_id_i)^2=(\tr_Ft^a)^2$, which vanishes for
the traceless nonsinglet generators and equals $N_f/2$ for $t^0$.
The disconnected correlators therefore enter every nonsinglet with coefficients
of both signs, and they cancel when all $D_{q_iq_k}$ are equal, that is, for
degenerate quark masses.
In the singlet they add up with the total weight $N_f/2$.

In the isospin limit $m_u=m_d$, the connected correlators of $\bar uu$, $\bar
dd$ and $\bar ud$ coincide:
$C^{\rm c}_{\bar uu}=C^{\rm c}_{\bar dd}=C^{\rm c}_{\bar ud}$.
The disconnected ones reduce to $D_{uu}=D_{dd}=D_{ud}\equiv D_{\ell\ell}$,
$D_{us}=D_{ds}\equiv D_{\ell s}$ and $D_{uc}=D_{dc}\equiv D_{\ell c}$.
Then
\begin{align}
C^3 &= \tfrac12\,C^{\rm c}_{\bar ud},
\nonumber\\
C^0_{SU(2)} &= \tfrac12\,C^{\rm c}_{\bar ud}+D_{\ell\ell},
\label{eq:su2_diag}
\end{align}
\begin{align}
C^8 &= \tfrac16\,C^{\rm c}_{\bar ud}+\tfrac13\,C^{\rm c}_{\bar ss}
       +\tfrac13\big(D_{\ell\ell}+D_{ss}-2D_{\ell s}\big),
\nonumber\\
C^0_{SU(3)} &= \tfrac13\,C^{\rm c}_{\bar ud}+\tfrac16\,C^{\rm c}_{\bar ss}
       +\tfrac16\big(4D_{\ell\ell}+4D_{\ell s}+D_{ss}\big),
\label{eq:su3_diag}
\end{align}
\begin{align}
C^{15} &= \tfrac1{12}\,C^{\rm c}_{\bar ud}+\tfrac1{24}\,C^{\rm c}_{\bar ss}
          +\tfrac38\,C^{\rm c}_{\bar cc}
\nonumber\\
       &\quad+\tfrac1{24}\big(4D_{\ell\ell}+4D_{\ell s}+D_{ss}
          +9D_{cc}-12D_{\ell c}-6D_{sc}\big),
\nonumber\\
C^0_{SU(4)} &= \tfrac14\,C^{\rm c}_{\bar ud}+\tfrac18\,C^{\rm c}_{\bar ss}
       +\tfrac18\,C^{\rm c}_{\bar cc}
\nonumber\\
       &\quad+\tfrac18\big(4D_{\ell\ell}+D_{ss}+D_{cc}
          +4D_{\ell s}+4D_{\ell c}+2D_{sc}\big).
\label{eq:su4_diag}
\end{align}
In $C^3$ the disconnected part cancels between $D_{uu}+D_{dd}$ and $-2D_{ud}$,
so the isovector correlator is the connected $\bar ud$ correlator.
The $SU(2)$ singlet differs from it only by $D_{\ell\ell}$.

Away from the isospin limit, the connected and the disconnected correlators
remain symmetric functions of the two quark masses.
The difference $\tfrac14(C^{\rm c}_{\bar uu}+C^{\rm c}_{\bar dd})-\tfrac12C^{\rm
c}_{\bar ud}$ is then of second order in $m_u-m_d$, and so is the residual
disconnected part $\tfrac14(D_{uu}+D_{dd}-2D_{ud})$ of $C^3$.
The same holds for the corrections to Eqs.~(\ref{eq:su3_diag}) and
(\ref{eq:su4_diag}).
In the limit of $N_f$ degenerate flavors, there is a single connected correlator
$C^{\rm c}$ and a single disconnected correlator $D$, and the two sum rules give
$C^a=\tfrac12C^{\rm c}$ for every nonsinglet and $C^0=\tfrac12C^{\rm
c}+\tfrac{N_f}{2}D$ for the singlet.

The $\bar ss$ and $\bar cc$ sectors of Sec.~\ref{sec:numerics} enter the $SU(3)$
and $SU(4)$ nonsinglets $C^8$ and $C^{15}$ and the singlet $C^0$, with the
weights of Eqs.~(\ref{eq:su3_diag}) and (\ref{eq:su4_diag}), together with
disconnected parts that are not computed in this work.
The symmetry relations of Sec.~\ref{sec:RAB} hold for the connected correlators
by themselves, because the symmetry transformations act on the quark fields and
the Wick contractions are taken after the transformation.
The probes of the $\bar ss$ and $\bar cc$ sectors therefore test the relations
on the connected parts.
The disconnected parts carry the flavor-singlet physics.
They do not give an independent $U(1)_A$ test, however~\cite{Chiu:2026upk}.
Once the chiral group of the sector is effectively restored, the chiral
relations between singlet and nonsinglet channels give
$C^0_S-C^0_P=C^a_P-C^a_S$ for a nonsinglet $a$, and likewise
$C^0_T-C^0_X=C^a_X-C^a_T$.
The $U(1)_A$ splitting of the singlet thus equals that of the nonsinglet.
For $(S,P)$ both are given by the disconnected pseudoscalar correlator.
For two flavors, the singlet and nonsinglet $\kappa_{PS}$ are then identical, so
the singlet pair gives the same $U(1)_A$ test as the nonsinglet pair, at the
cost of the disconnected contractions, provided the test passes above $T_c$.
For $N_f\ge3$ massless flavors both splittings vanish with the chiral group.
At nonzero masses the relations between singlet and nonsinglet channels hold
only up to corrections from the quark masses, of the same order as the $U(1)_A$
splitting itself, so the singlet pair gives no independent test.
The diagonal nonsinglets $t^8$ and $t^{15}$, which set $T^{uds}$ and $T^{udsc}$
(Sec.~\ref{subsec:num_hier}), have disconnected parts
[Eqs.~(\ref{eq:su3_diag}) and (\ref{eq:su4_diag})] that cancel only for
degenerate masses.
Their connected parts are correlators of off-diagonal operators with a
degenerate valence partner, for which the symmetry relations hold exactly.
The effect of the disconnected parts on $T^{uds}$ and $T^{udsc}$ remains to be
computed.

\subsection{The largest probe over the algebra}
\label{app:algmax}

The restoration of $SU(N_f)_L\times SU(N_f)_R$ requires the degeneracy of the
partners for every element $H=\sum_a\theta^at^a$ of the Lie algebra, not only
for the generators of one basis.
With connected parts only, the operator $O^H_\Gamma=\sum_{ij}H_{ij}\,\bar
q_i\Gamma q_j$ has the correlator
\begin{equation}
C^H_\Gamma=\sum_{i,j}w_{ij}\,C^{\rm c}_{\bar q_iq_j},\qquad w_{ij}=|H_{ij}|^2,
\label{eq:CH}
\end{equation}
a sum over the flavor sectors with nonnegative weights.
In the positive normalization of Sec.~\ref{subsec:cs} the channels $S$, $P$,
$V$, $A$, $T_4$ and $X_4$ are positive, so the probe of a pair $(A,B)$ is
\begin{equation}
\kAB^H=\frac{\big|\sum_{ij}w_{ij}\Delta_{ij}\big|}{\sum_{ij}w_{ij}\Sigma_{ij}},
\qquad
\Delta_{ij}=C^{\rm c}_A-C^{\rm c}_B,\quad \Sigma_{ij}=C^{\rm c}_A+C^{\rm c}_B,
\label{eq:kappaH}
\end{equation}
with $\Delta_{ij}$ and $\Sigma_{ij}$ those of the sector $\bar q_iq_j$.
It is a weighted mean of the sector probes $|\Delta_{ij}|/\Sigma_{ij}$, so it
never exceeds the largest of them.

An off-diagonal sector $\bar q_iq_j$ is isolated by the off-diagonal generator
in rows and columns $(i,j)$, so its probe is attained exactly.
A diagonal sector $\bar q_iq_i$ cannot be isolated, since $H$ is traceless.
For a diagonal $H=\mathrm{diag}(\lambda_1,\dots,\lambda_{N_f})$ with
$\sum_j\lambda_j=0$, the Cauchy--Schwarz inequality gives
$\lambda_i^2\le\frac{N_f-1}{N_f}\sum_j\lambda_j^2$.
The bound is reached only for $\lambda\propto(1,\dots,1,-(N_f-1))$ with flavor
$i$ in the last place, which is the last diagonal generator of
Eq.~(\ref{eq:tdiag}).
The splittings grow with the quark mass, and the mass order gives the heaviest
flavor its largest possible weight: $\bar ss$ in $t^8$ with weight $2/3$, and
$\bar cc$ in $t^{15}$ with weight $3/4$.
Another diagonal element can give a larger probe when the heaviest correlator is
small at the separation of the probe, as for $\bar cc$ at $zT=1$ and
$T\le308$~MeV.
There, however, the off-diagonal sector $\bar sc$ is larger still.
The connected part of the singlet gives every flavor the weight $1/N_f$, which
is smaller for $N_f\ge3$.
For example, if the light splitting is negligible,
\begin{equation}
\frac{\kappa^8}{\kappa^0}
=\frac{4\Sigma_{\bar ud}+2\Sigma_{\bar ss}}{\Sigma_{\bar ud}+2\Sigma_{\bar ss}}
\ge1,
\label{eq:k8k0}
\end{equation}
and the data give $\kappa^8_{T_4X_4}/\kappa^0_{T_4X_4}=2.0$ at $zT=1$ from $192$
to $513$~MeV.
This is why the connected parts of the singlets do not set $T_1$ in
Table~\ref{tab:hier}.
The disconnected parts are not included (see above).

For the connected parts, the maximum over all $H$ is therefore the larger of
the largest off-diagonal probe and the probe of the last diagonal generator in
the mass order.
We have checked this numerically.
For $\kappa_{VA}$, $\kappa_{T_4X_4}$ and $\kappa_{PS}$ at $zT=1$ and at every
temperature, the maximum over all $H$ of $SU(3)$ and $SU(4)$ agrees with the
maximum over the off-diagonal generators, $t^8$ and $t^{15}$ to within
$3\times10^{-4}$.
In another order the generators of Eq.~(\ref{eq:tdiag}) miss this maximum.
For $(c,s,d,u)$ they give $\bar cc$ the weights $1/2$, $1/6$ and $1/12$ in
$t^3$, $t^8$ and $t^{15}$, and at $513$~MeV the largest $\kappa_{T_4X_4}$ over
them is $0.049$, against $0.100$ for $t^{15}$ in the mass order.

\section{Free-fermion values of the chiral-spin probes}
\label{app:free}

Here we derive the noninteracting values quoted in Eq.~(\ref{eq:free_values})
for the meson $\bar q_1\Gamma q_2$ with arbitrary quark masses $m_1,m_2$.
We treat both the thermal ($t$) and the screening ($z$) correlators.
The analytic values are those of the continuum theory in infinite spatial
volume.
The corresponding values on the $32^3\times N_t$ lattices of
Sec.~\ref{sec:numerics}, computed with free domain-wall quarks, are given in
Appendix~\ref{app:freelat}.

For the partners of an exact symmetry, the free correlators are degenerate at
every separation, and $\RAB=1$.
For the partners of an emergent symmetry such as $SU(2)_{CS}$, the free
correlators differ, with $C_{V_k}\ne C_{T_k}$ at every separation except the
thermal midpoint, where all channels coincide (Appendix~\ref{app:thermal}).
With the midpoint reference of Sec.~\ref{subsec:common}, every $\RAB$ is unity
at tree level, so the bare and renormalized probes coincide at this order.
We work throughout with the bare free-fermion correlators.
At zero coupling, the only ingredients are the Dirac-trace coefficients and one
overall factor common to every channel.
This factor cancels in the ratios $\kappa_{CS}$ and $\kappa_{fade}$, so the
values obtained below are pure numbers.

\subsection{Thermal probe: $\kappa_{fade}(t)=2/3$}
\label{app:thermal}

At the fermionic Matsubara frequency $\omega_n=(2n+1)\pi T$, the free Euclidean
propagator is
\begin{equation}
S(i\omega_n,\vec p)=\frac{1}{i\slashed p+m}
=\frac{m-i(\gamma_4\omega_n+\vec\gamma\!\cdot\!\vec p)}{\omega_n^2+E^2},
\qquad E=\sqrt{\vec p^{\,2}+m^2}.
\end{equation}
We Fourier transform to $S(\tau,\vec p)=T\sum_n
e^{-i\omega_n\tau}S(i\omega_n,\vec p)$, and the $m$ and
$-i\vec\gamma\!\cdot\!\vec p$ terms share the sum
\begin{equation}
h_0(\tau)=T\sum_n\frac{e^{-i\omega_n\tau}}{\omega_n^2+E^2}
=\frac{1}{2E}\big[(1-n_F)e^{-E\tau}-n_F e^{E\tau}\big],\qquad
n_F=\frac{1}{e^{\beta E}+1}.
\end{equation}
The $\gamma_4$ term carries an extra $\omega_n$, and since $-i\omega_n
e^{-i\omega_n\tau}=\partial_\tau e^{-i\omega_n\tau}$ its coefficient is
$h_1=\partial_\tau h_0=-\tfrac12[(1-n_F)e^{-E\tau}+n_F e^{E\tau}]$.
Thus
\begin{equation}
S(\tau,\vec p)=\gamma_4\,h_1+h_0\,(m-i\vec\gamma\!\cdot\!\vec p).
\end{equation}

The sum defining $h_0$ is elementary, and we evaluate it by contour integration.
Since $\omega_n^2+E^2$ is even in $\omega_n$ the sum is real, so the phase may
be taken as $e^{+i\omega_n\tau}$.
The partial fractions are
$(\omega_n^2+E^2)^{-1}=\tfrac1{2E}[(E-i\omega_n)^{-1}+(E+i\omega_n)^{-1}]$, and
they give $h_0=\tfrac1{2E}[P_{-E}-P_E]$ with the single-mode sum
$P_E(\tau)=T\sum_n e^{+i\omega_n\tau}/(i\omega_n-E)$.
Write the summand as $F(z)=e^{z\tau}/(z-E)$ at $z=i\omega_n$, and use the Fermi
weight $n_F(z)=[e^{\beta z}+1]^{-1}$, whose poles are the Matsubara frequencies
with residue $-T$.
Then $P_E=-\sum_n\mathrm{Res}_{i\omega_n}[n_F F]$, which is the integral of $n_F
F$ over a contour that hugs the imaginary axis and encircles the Matsubara
poles.
We deform that contour outward to a circle of radius $|z|\to\infty$, which
sweeps it across the pole at $z=E$ off the axis, so the circle encloses the
Matsubara tower together with $z=E$.
For $0<\tau<\beta$ the product $n_F(z)F(z)$ decays in every direction.
As $\mathrm{Re}\,z\to-\infty$, $e^{z\tau}\to0$, and as
$\mathrm{Re}\,z\to+\infty$, $n_F(z)e^{z\tau}\sim e^{-(\beta-\tau)z}\to0$.
Near the axis $F\sim1/z$, and the unsuppressed window shrinks as $1/|z|$.
The large-circle integral therefore vanishes, and the residues at all its poles
sum to zero.
This gives $P_E=\mathrm{Res}_{z=E}[n_F F]=n_F\,e^{E\tau}$.
The partner is the same sum at $E\to-E$, and by the crossing relation
$n_F(-E)=1-n_F$ we get $P_{-E}=n_F(-E)e^{-E\tau}=(1-n_F)e^{-E\tau}$.
Hence
$h_0=\tfrac1{2E}[P_{-E}-P_E]=\tfrac1{2E}[(1-n_F)e^{-E\tau}-n_F\,e^{E\tau}]$.

The weight $n_F$ is the Fermi--Dirac occupation of a mode of energy $E$, and it
carries all the temperature dependence.
The forward piece $e^{-E\tau}$ carries the probability $1-n_F$ that the level is
empty, and the backward piece $e^{+E\tau}$ carries the occupation $n_F$ of the
medium.
At $T\to0$, $n_F\to0$ and $h_0\to e^{-E\tau}/2E$, the vacuum propagator.

The zero-momentum meson correlator needs $S(-\tau,\vec p)$.
The Kubo--Martin--Schwinger condition~\cite{Kubo:1957mj,Martin:1959jp} gives
antiperiodicity in Euclidean time for fermions, $S(-\tau)=-S(\beta-\tau)$, and
the identities $(1-n_F)e^{-\beta E}=n_F$ and $n_F e^{\beta E}=1-n_F$ imply
$h_0(\beta-\tau)=-h_0(\tau)$ and $h_1(\beta-\tau)=h_1(\tau)$.
Hence
\begin{equation}
S(-\tau,\vec p)=-\gamma_4\,h_1+h_0\,(m-i\vec\gamma\!\cdot\!\vec p).
\end{equation}
Only the $\gamma_4 h_1$ term flips sign.
In the meson $\bar q_1\Gamma q_2$ the forward line carries mass $m_1$ and the
backward line mass $m_2$.
Let $h_0^{(i)},h_1^{(i)}$ be built from $E_i=\sqrt{\vec p^{\,2}+m_i^2}$ and
$n_F^{(i)}$ at energy $E_i$.
The two lines combine into the products $h_0^{(1)}h_0^{(2)}$ and
$h_1^{(1)}h_1^{(2)}$ at the same $\tau$, with no cross terms.

With $\bar\Gamma=\gamma_4\Gamma^\dagger\gamma_4$ the correlator is
\begin{equation}
C_\Gamma(\tau)=-N_c\!\int\!\frac{d^3p}{(2\pi)^3}\,
\mathrm{tr}[\Gamma\,S_1(\tau)\,\bar\Gamma\,S_2(-\tau)].
\end{equation}
After the angular average $\langle p_ip_j\rangle=\tfrac13\vec
p^{\,2}\delta_{ij}$ it involves three integrals,
\begin{equation}
J_{pp}=N_c\!\!\int_{\vec p}\!\vec p^{\,2}h_0^{(1)}h_0^{(2)},\qquad
J_{mm}=N_c\!\!\int_{\vec p}\!m_1m_2\, h_0^{(1)}h_0^{(2)},\qquad
J_{hh}=N_c\!\!\int_{\vec p}\!h_1^{(1)}h_1^{(2)},
\end{equation}
with $\int_{\vec p}\equiv\int d^3p/(2\pi)^3$.
The Dirac traces give
\begin{alignat}{3}
&C_S=4J_{pp}-4J_{mm}+4J_{hh}, &\qquad &C_P=4J_{pp}+4J_{mm}+4J_{hh}, \nonumber\\
&C_V=\tfrac43 J_{pp}+4J_{mm}+4J_{hh}, &\qquad &C_A=\tfrac43
J_{pp}-4J_{mm}+4J_{hh}, \\
&C_T=-\tfrac43 J_{pp}+4J_{mm}+4J_{hh}, &\qquad &C_X=-\tfrac43
J_{pp}-4J_{mm}+4J_{hh}, \nonumber
\end{alignat}
for $\Gamma=\mathbf
1,\gamma_5,\gamma_k,\gamma_k\gamma_5,\gamma_4\gamma_k,\gamma_4\gamma_k\gamma_5$.
%(We checked these six traces numerically, with hermitian Euclidean $\gamma_\mu$
%and $\gamma_5=\gamma_1\gamma_2\gamma_3\gamma_4$.)
The scalar and pseudoscalar collect all three spatial directions, giving
$4J_{pp}$, while each of $V,A,T,X$ collects one, giving $\pm\tfrac43 J_{pp}$.
The mass enters only through the product $J_{mm}\propto m_1m_2$, which comes
from a scalar insertion on each line.

The average over the chirally completed multiplet is $M_1=\tfrac14(V+A+T+X)$.
In $V+A$ and in $T+X$ the antisymmetric $\pm4J_{mm}$ cancels, and the $\tfrac43
J_{pp}$ pieces cancel between the two pairs.
So
\begin{equation}
M_1=4J_{hh},\qquad M_0=\tfrac12(S+P)=4J_{pp}+4J_{hh},\qquad M_0-M_1=4J_{pp},
\end{equation}
with no $J_{mm}$.
The $SU(2)_{CS}$-breaking split is likewise mass free, $V-T=A-X=\tfrac83
J_{pp}$, and hence
\begin{equation}
\kappa_{fade}(t)=\frac{|V-T|}{|M_0-M_1|}=\frac{8/3\,J_{pp}}{4J_{pp}}=\frac23 ,
\end{equation}
\emph{exactly}, for any quark masses $m_1,m_2$.
Every $J_{mm}$ cancels in $M_0-M_1$ and in the numerator, and the remaining
$J_{pp}$ cancels in the ratio, so the value $2/3$ is independent of the masses,
of $T$, and of the cutoff.
The chiral split $T-X=8J_{mm}$ is a $U(1)_A$ splitting, and it is not part of
the $SU(2)_{CS}$ probe.

The intrinsic-splitting probe is $\kappa_{CS}(t)=|V-T|/(|V|+|T|)=\tfrac13
J_{pp}/(J_{mm}+J_{hh})$, since $V,T>0$.
Define $a_i=(1-n_F^{(i)})e^{-E_i\tau}$ and $b_i=n_F^{(i)}e^{E_i\tau}$.
Then
$E_1E_2\,h_0^{(1)}h_0^{(2)}=\tfrac14(a_1-b_1)(a_2-b_2)\le%
\tfrac14(a_1+b_1)(a_2+b_2)=h_1^{(1)}h_1^{(2)}$,
and also $E_1E_2\ge\vec p^{\,2}+m_1m_2$.
Both $h_0^{(i)}$ change sign at $\tau=\beta/2$, so $h_0^{(1)}h_0^{(2)}\ge0$.
Hence $J_{pp}+J_{mm}\le J_{hh}$ and
\begin{equation}
0\le\kappa_{CS}(t)\le\tfrac13 .
\end{equation}
The bound is approached as $\tau\to0$, and the probe vanishes at the midpoint
$\tau=\beta/2$, where $h_0^{(1)}=h_0^{(2)}=0$ by KMS and all channels are
degenerate.

The individual channels do depend on the masses, through $\mp4J_{mm}$ and
through $E_i$ in $J_{pp}$ and $J_{hh}$, but these dependences cancel in
$\kappa_{fade}(t)$.

At $\tau=\beta/2$ the part of the propagator proportional to $h_0$ vanishes for
every momentum and mass, so the propagator is pure $\gamma_4$.
With $\bar\Gamma=\gamma_4\Gamma^\dagger\gamma_4$, the trace is
$\mathrm{tr}[\Gamma\gamma_4\bar\Gamma\gamma_4]=\mathrm{tr}[\Gamma\Gamma^%
\dagger]=4$
for every $\Gamma$ of Table~\ref{tab:bilinear}, and the correlator is therefore
$4J_{hh}$ in every channel.
The midpoint is thus the channel-blind point of the free theory, where every
$\RAB$, exact or emergent, is unity at tree level.

\subsection{Screening probe: $\kappa_{fade}(z)=2/3$}
\label{app:screen}

The same steps apply to propagation along $z$, with the roles of the time and
$z$ directions exchanged.
Let $\slashed p=\gamma_1 k_1+\gamma_2 k_2+\gamma_3 p_z+\gamma_4\omega_n$ with
$\omega_n=(2n+1)\pi T$.
The propagator is $S=(m-i\slashed p)/(p_z^2+\tilde E^2)$ with $\tilde
E=\sqrt{\omega_n^2+k_\perp^2+m^2}$, and we Fourier transform in $p_z$.
For $z>0$,
$\int\!\tfrac{dp_z}{2\pi}\,e^{ip_z z}/(p_z^2+\tilde E^2)=e^{-\tilde E z}/2\tilde
E$
and $\int\!\tfrac{dp_z}{2\pi}\,p_z\,e^{ip_z z}/(p_z^2+\tilde
E^2)=\tfrac{i}{2}e^{-\tilde E z}$.
The $-i\gamma_3 p_z$ term then becomes $+\tilde E\gamma_3$, and
\begin{equation}
S(z)=\frac{e^{-\tilde E z}}{2\tilde E}\,N_+,\qquad
N_+=-i(\gamma_4\omega_n+\gamma_1 k_1+\gamma_2 k_2)+m+\tilde E\gamma_3 .
\end{equation}
The return line reverses $z$, and closing the contour the other way flips the
sign of the $\tilde E\gamma_3$ term,
\begin{equation}
S(-z)=\frac{e^{-\tilde E z}}{2\tilde E}\,N_-,\qquad
N_-=-i(\gamma_4\omega_n+\gamma_1 k_1+\gamma_2 k_2)+m-\tilde E\gamma_3 .
\end{equation}
This mirrors the sign flip of $\gamma_4h_1$ in the thermal case.

The meson $\bar q_1\Gamma q_2$ carries a mass on each line, with $\tilde
E_i=\sqrt{\omega_n^2+k_\perp^2+m_i^2}$, and the matrices $N_\pm^{(i)}$ are built
from $m_i,\tilde E_i$.
With the transverse average $\langle k_ik_j\rangle=\tfrac12
k_\perp^2\delta_{ij}$,
\begin{equation}
C_\Gamma(z)\propto\sum_n\!\int\! d^2k_\perp\,
\frac{e^{-(\tilde E_1+\tilde E_2) z}}{\tilde E_1\tilde E_2}\,
\mathrm{tr}[\Gamma\,N_+^{(1)}\,\bar\Gamma\,N_-^{(2)}],\qquad
\mathcal K=\sum_n\!\int\! d^2k_\perp\,
\frac{e^{-(\tilde E_1+\tilde E_2) z}}{\tilde E_1\tilde E_2}\,k_\perp^2 .
\end{equation}
In the chiral limit the chiral partners are degenerate, $V_1=A_1$ and $T_4=X_4$.
Take $\tilde E_1=\tilde E_2=\tilde E$ and leave implicit the common weight
$\sum_n\!\int d^2k_\perp\,(e^{-2\tilde E z}/\tilde E^2)$.
Then
\begin{equation}
M_0\propto 8\tilde E^2,\qquad
V_1=A_1\propto 4\tilde E^2+4\omega_n^2,\qquad
T_4=X_4\propto 8\omega_n^2 ,
\end{equation}
with $\tilde E^2=\omega_n^2+k_\perp^2$, while the longitudinal $\gamma_3$
channel vanishes in the chiral limit.
The channels of $M_4$ have, inside the same weight, $V_4=A_4\propto-8k_\perp^2$
and $T_k=X_k\propto-4k_\perp^2$ ($k=1,2$).
%, as we verified numerically.
With $\bar\Gamma=\gamma_4\Gamma^\dagger\gamma_4$ the traces of these channels
are negative.
For each of them
$\gamma_3\Gamma^\dagger\gamma_3=-\gamma_4\Gamma^\dagger\gamma_4$, so the sign
fixed by reflection positivity along $z$ makes them positive.
For the channels of $M_0$ and $M_2$ the two conjugations coincide.
The traces of $M_0\propto8\tilde E^2$ and of the members of $M_2$ are positive,
and they do not vanish at $k_\perp=0$.
This fixes the large-$z$ behavior~\cite{Chiu:2023hnm}.
At large $z$ the lowest Matsubara mode $|\omega_n|=\pi T$ dominates, with
$\tilde E\simeq\pi T+k_\perp^2/(2\pi T)$.
The transverse integral gives $\int d^2k_\perp\,e^{-k_\perp^2 z/\pi T}\propto
1/z$ for the channels of $M_0$ and $M_2$, while the channels of $M_4$ have
numerators that carry $k_\perp^2$ and get one more power of $1/z$:
\begin{align}
C_{S,P,V_1,A_1,T_4,X_4}(z)\;\to\;c_0\,\frac{e^{-2\pi Tz}}{z},\qquad
C_{V_4,A_4,T_1,X_1}(z)\;\to\;c_4\,\frac{e^{-2\pi Tz}}{z^2}.
\label{eq:M4_asym}
\end{align}
The two families have the same screening mass $2\pi T$ but different prefactors,
so $M_4$ never merges with $M_0$ and $M_2$ in the noninteracting
theory~\cite{Chiu:2023hnm}.
The multiplet $M_4$ therefore plays no role in the fade.
Equation~(\ref{eq:M4_asym}) is a concrete instance of the point made in
Sec.~\ref{subsec:cs}, because a comparison of screening masses alone would
declare all ten channels degenerate.
The amplitudes, which the renormalized correlators retain, show that they are
not.

The reason is an asymptotic symmetry.
For $|\omega_n|=\pi T$ and $k_\perp=0$ the free propagator along $z$ reduces to
$S(z)\simeq-\tfrac{i}{2}e^{-\pi T|z|}\gamma_4(1+s\,\Omega)$, with $\Omega\equiv
i\gamma_4\gamma_3$, $\Omega^2=1$, and $s=\pm1$ the product of the signs of
$\omega_n$ and $z$.
The corresponding part of the action,
$\bar\psi\gamma_4(\omega_n-i\Omega\partial_z)\psi$, is invariant under $\psi\to
U\psi$, $\bar\psi\to\bar\psi\gamma_4U^\dagger\gamma_4$ for every $U$ that
commutes with $\Omega$.
This is the group $U(2N_f)_+\times U(2N_f)_-$ acting on the eigenspaces
$\Omega=\pm1$, and it contains $U(1)_A$ and $SU(2)_{CS}$ with the transverse
generators $\{\gamma_k,-i\gamma_5\gamma_k,\gamma_5\}$, $k=1,2$.
The bilinears $\bar\psi\Gamma\psi$ with $\Gamma\gamma_4$ anticommuting with
$\Omega$, namely $S$, $P$, $V_1$, $V_2$, $A_1$, $A_2$, $T_4$ and $X_4$, form one
off-diagonal multiplet, and the projectors of the quark and the antiquark give
them a common leading term.
The bilinears with $\Gamma\gamma_4$ commuting with $\Omega$, namely the channels
of $M_4$ together with $V_3$ and $A_3$, lie in the block-diagonal
representation, and their leading term vanishes, since
$\mathrm{tr}[(1+s\Omega)(1-s\Omega)]=0$.
No transformation of the group relates the two representations, so $M_4$ cannot
merge with $M_0$ and $M_2$.
The transverse kinetic term $\bar\psi\gamma_\perp\cdot\partial_\perp\psi$ breaks
the symmetry by $k_\perp^2/(\pi T)^2\sim1/(zT)$, which is the relative size of
both $|M_0-M_2|$ and the intrinsic splittings $V-T_4=A-X_4$ inside $M_2$, all
proportional to the same $\mathcal K$ below.

With the chirally completed average $M_2=\tfrac14(V+A+T_4+X_4)$, the
$SU(2)_{CS}$-breaking splits are mass free:
\begin{equation}
V-T_4=A-X_4=4\mathcal K,\quad M_0-M_2=6\mathcal K,\quad
\kappa_{fade}(z)=\frac{|A-X_4|}{|M_0-M_2|}=\frac{4}{6}=\frac23 ,
\end{equation}
independent of the cutoff.
Away from the chiral limit, the channels carry an antisymmetric $\pm4m_1m_2$ and
a common $4(\omega_n^2+\tilde E_1\tilde E_2)$, which all cancel in $M_0-M_2$,
$V-T_4$ and $A-X_4$.
They leave $6\mathcal K$ and $4\mathcal K$ with no mass.
%, as we confirmed numerically for unequal masses.
So $\kappa_{fade}(z)=2/3$ \emph{exactly} for any masses $m_1,m_2$.
The chiral split $T_4-X_4=8m_1m_2$ inside the weight is a $U(1)_A$ splitting,
and it is not part of the probe.
The intrinsic-splitting probe is
\begin{equation}
\kappa_{CS}(z)=\frac{|A-X_4|}{|A|+|X_4|}
=\frac{\sum_n\!\int d^2k_\perp\,W\,k_\perp^2}
       {\sum_n\!\int d^2k_\perp\,W\,
        (2\omega_n^2-k_\perp^2+2\tilde E_1\tilde E_2-2m_1m_2)},
\label{eq:kcs_z}
\end{equation}
with $W=e^{-(\tilde E_1+\tilde E_2)z}/(\tilde E_1\tilde E_2)$.
In the chiral limit the denominator reduces to $4\omega_n^2+k_\perp^2$, and the
ratio approaches $1/3$ as $z\to0$, because the sum over $n$ and the integral
over $k_\perp$ become isotropic in $(\omega_n,\vec k_\perp)$, so $\langle
k_\perp^2\rangle=2\langle\omega_n^2\rangle$.
The bound $\kappa_{CS}(z)\le1/3$ is not proven analytically here, but we checked
it numerically in the chiral limit for $0.02\le zT\le3$, where the maximum is
$0.32$, reached at the smallest $z$.
Unlike the temporal probe, the spatial probe does not vanish at a finite
separation.
In infinite volume there is no middle slice in $z$.
On a finite lattice the middle slice $z=N_s/2$ exists, and for antiperiodic
boundary conditions in $z$ the free-quark correlators of all channels have the
same value there, as at the thermal midpoint.
%, which we checked numerically with the image sum of the continuum propagators.
The average over periodic and antiperiodic boundary conditions removes this
equality.
For lattice free quarks at $N_t=16$ and $z=16=N_s/2$, the channels $S$, $V$,
$T_4$ and $V_4$ are $1.94$, $1.80$, $1.67$ and $0.26$ in units of $10^{-6}$
(Supplemental Material~\cite{Suppl}).
At large $z$ the integral projects onto the lowest mode $|\omega_n|=\pi T$ with
$k_\perp\to0$, so the numerator vanishes while the denominator keeps
$4\omega_n^2$, and the probe $\kappa_{CS}(z)$ decays as a slow tail.

\subsection{Interpretation}

The thermal and screening results above give the free-quark values of
Eq.~(\ref{eq:free_values}), $\kappa_{fade}=2/3$ and $0\le\kappa_{CS}\le1/3$,
for any quark masses.
Both $\kappa_{fade}$ values are the pure number $2/3$, fixed by the Dirac-trace
coefficients alone, because the common dynamical integral cancels in the ratio.
Every $m_1m_2$ term cancels in $M_0-M_1=4J_{pp}$ and $M_0-M_2=6\mathcal K$ and
is absent from $V-T$ and $A-X_4$.
So $\kappa_{fade}=2/3$ \emph{exactly}, for any quark masses $m_1,m_2$ and
independent of $T$ and of the lattice spacing.

Free quarks of any mass therefore sit at the endpoint of the fade,
$\kappa_{fade}=2/3$ for both the thermal and the screening probes.
There the chirally completed multiplet $\{V,A,T,X\}$ is split by a fixed
fraction of the $M_0$--$M_1$ gap, so the symmetry is absent, even though
$\kappa_{CS}\to0$ at the midpoint.
The absence of chiral-spin symmetry in the free theory is thus diagnosed by
$\kappa_{fade}$ and not by $\kappa_{CS}$, and a small $\kappa_{CS}$ is necessary
but not sufficient.

The noninteracting values $\kappa_{CS}\le1/3$ and $\kappa_{fade}=2/3$ are the
high-temperature endpoints of the bare QCD probes.
The renormalized probes match them only up to $O(\alpha_s(T_r))$ corrections
from $\RAB\neq1$, and they are evaluated only for $T\le T_r$, since above
$T_r$ the
renormalized $\kappa_{fade}$ has a pole (Sec.~\ref{subsec:num_cs}).
They set the reference scale for the emergence and fading of chiral-spin
symmetry in QCD.

\subsection{Numerical values of lattice free fermions}
\label{app:freelat}

The midpoint degeneracy is exact on the lattice as well, for any quark mass.
We tested this on free optimal domain-wall quarks, with all gauge links set to
the identity and every other parameter as in Sec.~\ref{sec:numerics}.
The lattice was $32^3\times16$, and we used the zero-momentum correlators
normalized as in Appendix~\ref{app:tables}, with every channel positive.
At the light quark mass, the six channels $(S,P,V,A,T,X)$ are
$(2.80,2.80,2.07,2.07,1.34,1.34)\times10^{-2}$ at $t=1$ and
$(6.18,6.18,4.55,4.55,2.93,2.93)\times10^{-4}$ at $t=4$.
Hence $C_V/C_T=1.554$ and $C_V/C_S=0.737$ at $tT=1/4$.
At the midpoint $t=8$ all six equal $1.2962\times10^{-4}$, so
$C_V/C_T=C_V/C_S=1.0000$ there, while at $t=7$ the ratios are already $1.035$
and $0.967$.
At the charm mass, the six channels equal $6.6826\times10^{-7}$ at $t=8$, yet
$V\neq A$ and $S\neq P$ at every other $t$, with $C_V/C_A=2.29$ and
$C_S/C_P=0.53$ at $t=4$.
The same holds at every $N_t$ of Table~\ref{tab:ensembles} (Supplemental
Material~\cite{Suppl}).
A reference at $tT=1/4$ would instead build a tree-level splitting of $55\%$ and
$26\%$ into the $(V,T)$ and $(V,S)$ pairs.
The measured $R_{VT}=1.105$ and $R_{VS}=0.846$ of QCD at $T_r=770$~MeV are
therefore pure interaction effects, with the size stated in
Sec.~\ref{subsec:common}.
The spread between the two references in Table~\ref{tab:scheme} is the
corresponding uncertainty in any comparison of the renormalized probes with the
free values.

Table~\ref{tab:free_lat} gives the lattice free-fermion values of $\kappa_{CS}$
in the $\bar ud$ sector for every $N_t$ of Table~\ref{tab:ensembles}, together
with the continuum formula of this appendix evaluated with the discrete
transverse momenta $\vec k_\perp=2\pi\vec n/32$ of the lattice box.
Two effects separate the lattice values from the continuum values $0.1416$,
$0.0747$ and $0.0384$ at $zT=1/2$, $1$ and $2$.

The first effect is the transverse volume.
At $N_t=16$ the box has $L_\perp T=2$, and the numerator of $\kappa_{CS}$
carries $k_\perp^2$, so the first transverse mode $k_\perp=2\pi/32=\pi T$ cuts
it.
At $z=16$ this mode is suppressed by $e^{-k_\perp^2z/\pi T}=e^{-\pi}\simeq0.04$.
This halves $\kappa_{CS}(zT=1)$ to $0.037$ in the continuum formula and $0.039$
on the lattice.
The effect is gone by $N_t=8$, where the continuum formula in the $L_\perp=32$
box is within $0.5\%$ of its infinite-volume value.

The second effect is the cutoff.
From $N_t=12$ downward, the lattice value exceeds the continuum formula in the
same box, by $5$, $8$, $12$, $21$ and $33\%$ for $N_t=12$, $10$, $8$, $6$ and
$4$.
The excess grows with $aT=1/N_t$ down to $N_t=4$.
At $N_t=2$ the lattice free-fermion value is $0.086$, $16\%$ above the continuum
value.

In every row, $\kappa_{fade}=0.6667$, the midpoint ratios $R_{VT}$ and $R_{VS}$
are unity to six digits, and the probes $\kappa_{T_4X_4}$ and $\kappa_{VA}$ are
below $2\times10^{-5}$.
The other mass combinations differ from $\bar ud$ by less than $0.005$ in
$\kappa_{CS}(zT=1)$, with the exceptions $\bar sc$ ($+0.005$ to $+0.015$) and
$\bar cc$ ($0.236$ at $N_t=16$ down to $0.131$ at $N_t=4$).
These lattice values, not the continuum ones, are the tree-level comparators for
the measured probes at the same $N_t$ in Sec.~\ref{sec:numerics}.

\begin{table}[tbp]
\caption{Lattice free-fermion values of $\kappa_{CS}$ in the $\bar ud$ sector on
$32^3\times N_t$. The quarks are optimal domain-wall quarks with unit gauge
links and the boundary conditions of Sec.~\ref{sec:numerics}, with $\RAB=1$. The
column ``cont.\ $L_\perp=32$'' is the continuum chiral-limit formula at $zT=1$
with the discrete transverse momenta $k_x,k_y=2\pi n/32$ of the $32^3$ box. In
infinite volume it gives $0.1416$, $0.0747$ and $0.0384$ at $zT=1/2$, $1$ and
$2$. The last two columns are the thermal $\kappa_{CS}(t)$ and the ratios
$R_{VT}$, $R_{VS}$ at $tT=1/4$. At $N_t=10$ and $6$ they are taken at $tT=1/5$
and $1/6$.}
\label{tab:free_lat}
\begin{ruledtabular}
\begin{tabular}{cccccccc}
 & & \multicolumn{3}{c}{$\kappa_{CS}(z)$} & & $\kappa_{CS}(t)$ & \\
$N_t$ & $T$ [MeV] & $zT=\tfrac12$ & $zT=1$ & $zT=2$ & cont.\ $L_\perp=32$ &
$tT=\tfrac14$ & $R_{VT}$, $R_{VS}$ \\
\hline
16 & 192  & 0.1274 & 0.0388 & --     & 0.0367 & 0.2169 & 1.554, 0.737 \\
12 & 257  & 0.1470 & 0.0674 & --     & 0.0641 & 0.2177 & 1.557, 0.737 \\
10 & 308  & 0.1527 & 0.0773 & --     & 0.0718 & 0.2473 & 1.657, 0.716 \\
8  & 385  & 0.1578 & 0.0832 & 0.0396 & 0.0743 & 0.1969 & 1.490, 0.753 \\
6  & 513  & 0.1595 & 0.0902 & 0.0455 & 0.0747 & 0.1748 & 1.424, 0.771 \\
4  & 770  & 0.1412 & 0.0996 & 0.0538 & 0.0747 & 0.1038 & 1.232, 0.842 \\
2  & 1539 & 0.0751 & 0.0864 & 0.0685 & 0.0746 & --     & -- \\
\end{tabular}
\end{ruledtabular}
\end{table}

%======================================================================
\section{Tables of correlators}
\label{app:tables}
%======================================================================

The complete bare correlators are provided as Supplemental
Material~\cite{Suppl}, and all probes of Sec.~\ref{sec:numerics} are built from
them.
They cover thermal and screening correlators of all six flavor sectors and of
the connected parts of $t^8$, $t^{15}$ and the $SU(3)$ and $SU(4)$ singlets on
the seven ensembles, in all channels and at all separations.
This appendix reproduces the subset that fixes the scheme and enters the tables
of the main text.
Table~\ref{tab:Ct_ref} gives the thermal correlators of the $\bar ud$ sector on
the two reference ensembles, $N_t=6$ and $4$, and their midpoint rows are the
$C_\Gamma(t_r,T_r)$ of Eq.~(\ref{eq:R_sym}).
Tables~\ref{tab:Cz_sec_a}--\ref{tab:Cz_sec_c} give the screening correlators of
the six channels of $M_0$ and $M_2$ (Sec.~\ref{subsec:cs}), for all six sectors
on all seven ensembles, including $N_t=2$, at $zT=1/2$ and $1$, and at $zT=2$
for $N_t\le8$.
These separations enter Tables~\ref{tab:sectors}--\ref{tab:cswin_eps} and
Fig.~\ref{fig:sectors}.
Tables~\ref{tab:Cz_gen_a} and \ref{tab:Cz_gen_b} give the same for the
connected parts of $t^8$, $t^{15}$ and the two singlets, which enter
Table~\ref{tab:hier}.
With Eqs.~(\ref{eq:kappa_RAB}), (\ref{eq:kappa_CS}), (\ref{eq:kappa_fade}),
(\ref{eq:M1M2}) and (\ref{eq:R_sym}), the tables reproduce the main-text values
of $\kappa_{CS}$ and $\kappa_{fade}$ at $zT=1$ within rounding, and those of
$\kappa_{T_4X_4}$ where it exceeds the rounding.

All correlators carry the factor $1/2$ of Eq.~(\ref{eq:offdiag}) relative to the
connected correlator of the bilinear $\bar q_1\Gamma q_2$, and it cancels in
every ratio.
For $q_1\neq q_2$ they are the correlators of the off-diagonal operators
$O^a_\Gamma$ of Eq.~(\ref{eq:Oa_def}), and for $\bar ss$ and $\bar cc$ they are
connected parts only (Appendix~\ref{app:flavor}).
The generator correlators of Tables~\ref{tab:Cz_gen_a} and \ref{tab:Cz_gen_b}
carry the same normalization.
The thermal correlators are averaged over $t$ and $N_t-t$ and over the three
spatial components of $V$, $A$, $T$ and $X$, and the screening correlators are
averaged over $z$ and $32-z$ and over the two transverse components of $V$ and
$A$.
The statistical uncertainties come from the jackknife method, with the bin size
at which the statistical error saturates.

The entries are quoted to the precision set by their errors, which suffices to
reconstruct $\kappa_{CS}$ and $\kappa_{fade}$ within their quoted errors, since
their numerators are $10$--$50\%$ of the correlators.
It does not suffice for exact-partner tests below about $10^{-3}$, such as
$\kappa_{VA}$ and, above $192$~MeV, $\kappa_{T_4X_4}$ of the $\bar ud$ sector.
These tests are consistent with zero (Sec.~\ref{subsec:num_subgroup}), and the
tables show the exact partners agreeing within errors.
The partner columns $C_V$, $C_A$ and $C_{T_4}$, $C_{X_4}$ are nevertheless
listed separately, so that the degeneracy of the exact partners and the
splitting of the emergent ones can be compared side by side.

\begin{table*}[tbp]
\caption{Thermal correlators $C_\Gamma(t)$ of Eq.~(\ref{eq:corr_def}) of the
$\bar ud$ sector on the two reference ensembles. They are $32^3\times6$
($T=513$~MeV) and $32^3\times4$ ($T=770$~MeV) at $\beta=6.20$. The correlators
are averaged forward and backward, and include the factor $1/2$ of the flavor
normalization $\tr_F(t^at^b)=\delta^{ab}/2$. The reference values
$C_\Gamma(t_r,T_r)$ of Eq.~(\ref{eq:R_sym}) are the midpoint rows, $t=3$ of
$N_t=6$ and $t=2$ of $N_t=4$.}
\label{tab:Ct_ref}
\setlength{\tabcolsep}{4pt}\centering\footnotesize
\begin{adjustbox}{max width=\linewidth}
\begin{tabular}{|ccc|cccccc|}
\hline
$N_t$ & $T$ & $t/a$ & $C_{P}(t)$ & $C_{S}(t)$ & $C_{V}(t)$ & $C_{A}(t)$ &
$C_{T}(t)$ & $C_{X}(t)$\\
\hline
6 & 513 & 1 & $1.9080(40)\times10^{-2}$ & $1.9080(40)\times10^{-2}$ &
$1.4916(27)\times10^{-2}$ & $1.4916(27)\times10^{-2}$ &
$1.1441(17)\times10^{-2}$ & $1.1441(17)\times10^{-2}$\\
 &  & 2 & $2.928(12)\times10^{-3}$ & $2.928(12)\times10^{-3}$ &
 $2.0048(91)\times10^{-3}$ & $2.0048(91)\times10^{-3}$ &
 $1.4975(83)\times10^{-3}$ & $1.4975(83)\times10^{-3}$\\
 &  & 3 & $1.5218(84)\times10^{-3}$ & $1.5217(84)\times10^{-3}$ &
 $1.0257(62)\times10^{-3}$ & $1.0257(62)\times10^{-3}$ &
 $8.594(68)\times10^{-4}$ & $8.594(68)\times10^{-4}$\\
\hline
4 & 770 & 1 & $1.9167(30)\times10^{-2}$ & $1.9167(30)\times10^{-2}$ &
$1.5832(25)\times10^{-2}$ & $1.5832(25)\times10^{-2}$ &
$1.2980(21)\times10^{-2}$ & $1.2980(21)\times10^{-2}$\\
 &  & 2 & $4.924(12)\times10^{-3}$ & $4.924(12)\times10^{-3}$ &
 $4.165(16)\times10^{-3}$ & $4.165(16)\times10^{-3}$ & $3.771(17)\times10^{-3}$
 & $3.771(17)\times10^{-3}$\\
\hline
\end{tabular}\end{adjustbox}
\end{table*}

\begin{table*}[tbp]
\caption{Screening correlators $C_\Gamma(z)$ of the $\bar ud$ and $\bar us$
sectors in the channels of $M_0$ and $M_2$ (Sec.~\ref{subsec:cs}). The
separations are $z=N_t/2$, $N_t$ and $2N_t$ ($zT=1/2$, $1$, $2$). They are used
in Tables~\ref{tab:sectors}--\ref{tab:cswin_eps}. The lattices are $32^3\times
N_t$ at $\beta=6.20$, and the factor $1/2$ of the flavor normalization is
included. Separations beyond $z=16$ are not available. The correlators at all
separations and in all channels are given in the Supplemental
Material~\cite{Suppl}.}
\label{tab:Cz_sec_a}
\setlength{\tabcolsep}{5pt}\centering\scriptsize\renewcommand{%
\arraystretch}{0.92}
\begin{adjustbox}{max width=\linewidth}
\begin{tabular}{|cccc|cccccc|}
\hline
sector & $N_t$ & $T$ & $z/a$ & $C_{S}(z)$ & $C_{P}(z)$ & $C_{V}(z)$ & $C_{A}(z)$
& $C_{T_4}(z)$ & $C_{X_4}(z)$\\
\hline
$\bar ud$ & 16 & 192 & 8 & $1.373(63)\times10^{-4}$ & $1.405(76)\times10^{-4}$ &
$1.698(36)\times10^{-5}$ & $1.697(36)\times10^{-5}$ & $7.40(25)\times10^{-6}$ &
$7.37(25)\times10^{-6}$\\
 & & & 16 & $3.59(61)\times10^{-5}$ & $3.76(64)\times10^{-5}$ &
 $1.387(66)\times10^{-6}$ & $1.387(66)\times10^{-6}$ & $4.18(67)\times10^{-7}$ &
 $4.16(67)\times10^{-7}$\\
\hline
 & 12 & 257 & 6 & $1.628(49)\times10^{-4}$ & $1.629(49)\times10^{-4}$ &
 $3.898(46)\times10^{-5}$ & $3.897(46)\times10^{-5}$ & $2.051(25)\times10^{-5}$
 & $2.050(25)\times10^{-5}$\\
 & & & 12 & $1.27(11)\times10^{-5}$ & $1.27(11)\times10^{-5}$ &
 $1.412(79)\times10^{-6}$ & $1.411(79)\times10^{-6}$ & $6.61(40)\times10^{-7}$ &
 $6.61(40)\times10^{-7}$\\
\hline
 & 10 & 308 & 5 & $1.907(18)\times10^{-4}$ & $1.909(18)\times10^{-4}$ &
 $6.614(50)\times10^{-5}$ & $6.614(50)\times10^{-5}$ & $3.822(29)\times10^{-5}$
 & $3.822(29)\times10^{-5}$\\
 & & & 10 & $9.82(18)\times10^{-6}$ & $9.83(18)\times10^{-6}$ &
 $1.974(30)\times10^{-6}$ & $1.974(30)\times10^{-6}$ & $1.092(17)\times10^{-6}$
 & $1.092(17)\times10^{-6}$\\
\hline
 & 8 & 385 & 4 & $3.024(27)\times10^{-4}$ & $3.024(27)\times10^{-4}$ &
 $1.344(11)\times10^{-4}$ & $1.344(11)\times10^{-4}$ & $8.316(69)\times10^{-5}$
 & $8.316(69)\times10^{-5}$\\
 & & & 8 & $1.327(29)\times10^{-5}$ & $1.328(29)\times10^{-5}$ &
 $3.900(64)\times10^{-6}$ & $3.900(64)\times10^{-6}$ & $2.431(38)\times10^{-6}$
 & $2.431(38)\times10^{-6}$\\
 & & & 16 & $8.73(41)\times10^{-8}$ & $8.74(41)\times10^{-8}$ &
 $1.242(47)\times10^{-8}$ & $1.242(47)\times10^{-8}$ & $6.68(21)\times10^{-9}$ &
 $6.68(21)\times10^{-9}$\\
\hline
 & 6 & 513 & 3 & $6.028(36)\times10^{-4}$ & $6.029(37)\times10^{-4}$ &
 $3.470(21)\times10^{-4}$ & $3.470(21)\times10^{-4}$ & $2.324(14)\times10^{-4}$
 & $2.324(14)\times10^{-4}$\\
 & & & 6 & $2.246(30)\times10^{-5}$ & $2.246(30)\times10^{-5}$ &
 $9.55(10)\times10^{-6}$ & $9.55(10)\times10^{-6}$ & $6.489(69)\times10^{-6}$ &
 $6.489(69)\times10^{-6}$\\
 & & & 12 & $5.81(16)\times10^{-8}$ & $5.82(16)\times10^{-8}$ &
 $1.509(36)\times10^{-8}$ & $1.509(36)\times10^{-8}$ & $9.52(24)\times10^{-9}$ &
 $9.52(24)\times10^{-9}$\\
\hline
 & 4 & 770 & 2 & $1.8777(51)\times10^{-3}$ & $1.8777(51)\times10^{-3}$ &
 $1.3809(37)\times10^{-3}$ & $1.3808(37)\times10^{-3}$ &
 $1.0368(26)\times10^{-3}$ & $1.0368(26)\times10^{-3}$\\
 & & & 4 & $6.607(32)\times10^{-5}$ & $6.607(32)\times10^{-5}$ &
 $4.159(22)\times10^{-5}$ & $4.159(22)\times10^{-5}$ & $3.105(16)\times10^{-5}$
 & $3.105(16)\times10^{-5}$\\
 & & & 8 & $1.887(18)\times10^{-7}$ & $1.887(18)\times10^{-7}$ &
 $8.90(10)\times10^{-8}$ & $8.90(10)\times10^{-8}$ & $6.480(78)\times10^{-8}$ &
 $6.480(78)\times10^{-8}$\\
\hline
 & 2 & 1539 & 1 & $1.2718(18)\times10^{-2}$ & $1.2718(18)\times10^{-2}$ &
 $1.1326(15)\times10^{-2}$ & $1.1326(15)\times10^{-2}$ &
 $1.0018(12)\times10^{-2}$ & $1.0018(12)\times10^{-2}$\\
 & & & 2 & $6.789(19)\times10^{-4}$ & $6.789(19)\times10^{-4}$ &
 $5.835(16)\times10^{-4}$ & $5.835(16)\times10^{-4}$ & $5.039(13)\times10^{-4}$
 & $5.039(13)\times10^{-4}$\\
 & & & 4 & $5.026(24)\times10^{-6}$ & $5.026(24)\times10^{-6}$ &
 $4.076(20)\times10^{-6}$ & $4.076(20)\times10^{-6}$ & $3.491(17)\times10^{-6}$
 & $3.491(17)\times10^{-6}$\\
\hline
$\bar us$ & 16 & 192 & 8 & $1.133(36)\times10^{-4}$ & $1.162(39)\times10^{-4}$ &
$1.578(32)\times10^{-5}$ & $1.562(30)\times10^{-5}$ & $7.11(22)\times10^{-6}$ &
$7.03(22)\times10^{-6}$\\
 & & & 16 & $2.24(17)\times10^{-5}$ & $2.34(18)\times10^{-5}$ &
 $1.147(48)\times10^{-6}$ & $1.126(48)\times10^{-6}$ & $4.06(36)\times10^{-7}$ &
 $4.00(36)\times10^{-7}$\\
\hline
 & 12 & 257 & 6 & $1.575(46)\times10^{-4}$ & $1.579(46)\times10^{-4}$ &
 $3.827(44)\times10^{-5}$ & $3.822(44)\times10^{-5}$ & $2.022(24)\times10^{-5}$
 & $2.019(24)\times10^{-5}$\\
 & & & 12 & $1.173(97)\times10^{-5}$ & $1.178(97)\times10^{-5}$ &
 $1.345(73)\times10^{-6}$ & $1.342(73)\times10^{-6}$ & $6.36(38)\times10^{-7}$ &
 $6.35(38)\times10^{-7}$\\
\hline
 & 10 & 308 & 5 & $1.884(18)\times10^{-4}$ & $1.886(18)\times10^{-4}$ &
 $6.558(49)\times10^{-5}$ & $6.555(49)\times10^{-5}$ & $3.794(29)\times10^{-5}$
 & $3.792(29)\times10^{-5}$\\
 & & & 10 & $9.54(16)\times10^{-6}$ & $9.57(17)\times10^{-6}$ &
 $1.934(29)\times10^{-6}$ & $1.932(29)\times10^{-6}$ & $1.073(17)\times10^{-6}$
 & $1.072(17)\times10^{-6}$\\
\hline
 & 8 & 385 & 4 & $3.008(26)\times10^{-4}$ & $3.009(26)\times10^{-4}$ &
 $1.339(11)\times10^{-4}$ & $1.339(11)\times10^{-4}$ & $8.285(69)\times10^{-5}$
 & $8.282(69)\times10^{-5}$\\
 & & & 8 & $1.311(29)\times10^{-5}$ & $1.312(29)\times10^{-5}$ &
 $3.860(63)\times10^{-6}$ & $3.859(63)\times10^{-6}$ & $2.408(38)\times10^{-6}$
 & $2.407(38)\times10^{-6}$\\
 & & & 16 & $8.51(40)\times10^{-8}$ & $8.52(40)\times10^{-8}$ &
 $1.215(45)\times10^{-8}$ & $1.215(45)\times10^{-8}$ & $6.55(20)\times10^{-9}$ &
 $6.55(20)\times10^{-9}$\\
\hline
 & 6 & 513 & 3 & $6.016(36)\times10^{-4}$ & $6.017(36)\times10^{-4}$ &
 $3.464(20)\times10^{-4}$ & $3.464(20)\times10^{-4}$ & $2.320(14)\times10^{-4}$
 & $2.320(14)\times10^{-4}$\\
 & & & 6 & $2.235(30)\times10^{-5}$ & $2.236(30)\times10^{-5}$ &
 $9.51(10)\times10^{-6}$ & $9.51(10)\times10^{-6}$ & $6.462(68)\times10^{-6}$ &
 $6.461(68)\times10^{-6}$\\
 & & & 12 & $5.75(16)\times10^{-8}$ & $5.76(16)\times10^{-8}$ &
 $1.495(35)\times10^{-8}$ & $1.495(35)\times10^{-8}$ & $9.44(23)\times10^{-9}$ &
 $9.44(23)\times10^{-9}$\\
\hline
 & 4 & 770 & 2 & $1.8766(51)\times10^{-3}$ & $1.8766(51)\times10^{-3}$ &
 $1.3801(37)\times10^{-3}$ & $1.3801(37)\times10^{-3}$ &
 $1.0362(26)\times10^{-3}$ & $1.0362(26)\times10^{-3}$\\
 & & & 4 & $6.598(32)\times10^{-5}$ & $6.598(32)\times10^{-5}$ &
 $4.153(22)\times10^{-5}$ & $4.153(22)\times10^{-5}$ & $3.100(16)\times10^{-5}$
 & $3.100(16)\times10^{-5}$\\
 & & & 8 & $1.882(18)\times10^{-7}$ & $1.882(18)\times10^{-7}$ &
 $8.87(10)\times10^{-8}$ & $8.87(10)\times10^{-8}$ & $6.461(78)\times10^{-8}$ &
 $6.461(78)\times10^{-8}$\\
\hline
 & 2 & 1539 & 1 & $1.2716(18)\times10^{-2}$ & $1.2717(18)\times10^{-2}$ &
 $1.1326(15)\times10^{-2}$ & $1.1326(15)\times10^{-2}$ &
 $1.0018(12)\times10^{-2}$ & $1.0018(12)\times10^{-2}$\\
 & & & 2 & $6.788(19)\times10^{-4}$ & $6.788(19)\times10^{-4}$ &
 $5.834(16)\times10^{-4}$ & $5.834(16)\times10^{-4}$ & $5.038(13)\times10^{-4}$
 & $5.038(13)\times10^{-4}$\\
 & & & 4 & $5.024(24)\times10^{-6}$ & $5.024(24)\times10^{-6}$ &
 $4.075(20)\times10^{-6}$ & $4.075(20)\times10^{-6}$ & $3.489(17)\times10^{-6}$
 & $3.489(17)\times10^{-6}$\\
\hline
\end{tabular}\end{adjustbox}
\end{table*}

\begin{table*}[tbp]
\caption{Screening correlators $C_\Gamma(z)$ of the $\bar ss$ and $\bar uc$
sectors in the channels of $M_0$ and $M_2$ (Sec.~\ref{subsec:cs}). The
separations are $z=N_t/2$, $N_t$ and $2N_t$ ($zT=1/2$, $1$, $2$). They are used
in Tables~\ref{tab:sectors}--\ref{tab:cswin_eps}. The lattices are $32^3\times
N_t$ at $\beta=6.20$, and the factor $1/2$ of the flavor normalization is
included. Separations beyond $z=16$ are not available. The correlators at all
separations and in all channels are given in the Supplemental
Material~\cite{Suppl}.}
\label{tab:Cz_sec_b}
\setlength{\tabcolsep}{5pt}\centering\scriptsize\renewcommand{%
\arraystretch}{0.92}
\begin{adjustbox}{max width=\linewidth}
\begin{tabular}{|cccc|cccccc|}
\hline
sector & $N_t$ & $T$ & $z/a$ & $C_{S}(z)$ & $C_{P}(z)$ & $C_{V}(z)$ & $C_{A}(z)$
& $C_{T_4}(z)$ & $C_{X_4}(z)$\\
\hline
$\bar ss$ & 16 & 192 & 8 & $8.68(27)\times10^{-5}$ & $1.127(37)\times10^{-4}$ &
$1.588(33)\times10^{-5}$ & $1.350(27)\times10^{-5}$ & $7.45(20)\times10^{-6}$ &
$6.12(21)\times10^{-6}$\\
 & & & 16 & $1.305(54)\times10^{-5}$ & $2.02(12)\times10^{-5}$ &
 $1.096(48)\times10^{-6}$ & $8.24(43)\times10^{-7}$ & $4.24(31)\times10^{-7}$ &
 $3.18(27)\times10^{-7}$\\
\hline
 & 12 & 257 & 6 & $1.492(40)\times10^{-4}$ & $1.565(45)\times10^{-4}$ &
 $3.820(44)\times10^{-5}$ & $3.689(41)\times10^{-5}$ & $2.030(24)\times10^{-5}$
 & $1.950(23)\times10^{-5}$\\
 & & & 12 & $1.048(83)\times10^{-5}$ & $1.136(91)\times10^{-5}$ &
 $1.318(70)\times10^{-6}$ & $1.242(66)\times10^{-6}$ & $6.29(36)\times10^{-7}$ &
 $5.94(35)\times10^{-7}$\\
\hline
 & 10 & 308 & 5 & $1.848(17)\times10^{-4}$ & $1.878(18)\times10^{-4}$ &
 $6.544(49)\times10^{-5}$ & $6.457(48)\times10^{-5}$ & $3.796(28)\times10^{-5}$
 & $3.734(28)\times10^{-5}$\\
 & & & 10 & $9.17(15)\times10^{-6}$ & $9.42(17)\times10^{-6}$ &
 $1.913(29)\times10^{-6}$ & $1.872(28)\times10^{-6}$ & $1.065(16)\times10^{-6}$
 & $1.041(16)\times10^{-6}$\\
\hline
 & 8 & 385 & 4 & $2.983(26)\times10^{-4}$ & $3.002(26)\times10^{-4}$ &
 $1.337(11)\times10^{-4}$ & $1.329(11)\times10^{-4}$ & $8.282(69)\times10^{-5}$
 & $8.221(68)\times10^{-5}$\\
 & & & 8 & $1.289(28)\times10^{-5}$ & $1.302(28)\times10^{-5}$ &
 $3.837(63)\times10^{-6}$ & $3.803(62)\times10^{-6}$ & $2.397(38)\times10^{-6}$
 & $2.373(37)\times10^{-6}$\\
 & & & 16 & $8.23(38)\times10^{-8}$ & $8.37(39)\times10^{-8}$ &
 $1.196(45)\times10^{-8}$ & $1.182(44)\times10^{-8}$ & $6.46(20)\times10^{-9}$ &
 $6.38(20)\times10^{-9}$\\
\hline
 & 6 & 513 & 3 & $5.997(36)\times10^{-4}$ & $6.011(36)\times10^{-4}$ &
 $3.462(20)\times10^{-4}$ & $3.454(20)\times10^{-4}$ & $2.320(14)\times10^{-4}$
 & $2.313(14)\times10^{-4}$\\
 & & & 6 & $2.221(29)\times10^{-5}$ & $2.229(30)\times10^{-5}$ &
 $9.48(10)\times10^{-6}$ & $9.45(10)\times10^{-6}$ & $6.446(68)\times10^{-6}$ &
 $6.423(68)\times10^{-6}$\\
 & & & 12 & $5.69(16)\times10^{-8}$ & $5.72(16)\times10^{-8}$ &
 $1.484(35)\times10^{-8}$ & $1.478(35)\times10^{-8}$ & $9.37(23)\times10^{-9}$ &
 $9.33(23)\times10^{-9}$\\
\hline
 & 4 & 770 & 2 & $1.8749(51)\times10^{-3}$ & $1.8761(51)\times10^{-3}$ &
 $1.3798(37)\times10^{-3}$ & $1.3789(37)\times10^{-3}$ &
 $1.0361(26)\times10^{-3}$ & $1.0353(26)\times10^{-3}$\\
 & & & 4 & $6.585(32)\times10^{-5}$ & $6.592(32)\times10^{-5}$ &
 $4.149(22)\times10^{-5}$ & $4.145(22)\times10^{-5}$ & $3.097(16)\times10^{-5}$
 & $3.094(16)\times10^{-5}$\\
 & & & 8 & $1.875(17)\times10^{-7}$ & $1.877(18)\times10^{-7}$ &
 $8.85(10)\times10^{-8}$ & $8.84(10)\times10^{-8}$ & $6.445(78)\times10^{-8}$ &
 $6.439(78)\times10^{-8}$\\
\hline
 & 2 & 1539 & 1 & $1.2715(18)\times10^{-2}$ & $1.2716(18)\times10^{-2}$ &
 $1.1325(15)\times10^{-2}$ & $1.1324(15)\times10^{-2}$ &
 $1.0018(12)\times10^{-2}$ & $1.0016(12)\times10^{-2}$\\
 & & & 2 & $6.786(19)\times10^{-4}$ & $6.787(19)\times10^{-4}$ &
 $5.833(16)\times10^{-4}$ & $5.832(16)\times10^{-4}$ & $5.038(13)\times10^{-4}$
 & $5.037(13)\times10^{-4}$\\
 & & & 4 & $5.022(24)\times10^{-6}$ & $5.023(24)\times10^{-6}$ &
 $4.074(20)\times10^{-6}$ & $4.073(20)\times10^{-6}$ & $3.488(17)\times10^{-6}$
 & $3.488(17)\times10^{-6}$\\
\hline
$\bar uc$ & 16 & 192 & 8 & $7.01(13)\times10^{-6}$ & $7.19(14)\times10^{-6}$ &
$2.089(30)\times10^{-6}$ & $2.039(28)\times10^{-6}$ & $1.284(20)\times10^{-6}$ &
$1.249(20)\times10^{-6}$\\
 & & & 16 & $5.88(19)\times10^{-8}$ & $6.16(19)\times10^{-8}$ &
 $1.040(37)\times10^{-8}$ & $9.94(35)\times10^{-9}$ & $6.47(30)\times10^{-9}$ &
 $6.21(29)\times10^{-9}$\\
\hline
 & 12 & 257 & 6 & $3.061(34)\times10^{-5}$ & $3.083(34)\times10^{-5}$ &
 $1.0888(88)\times10^{-5}$ & $1.0818(88)\times10^{-5}$ &
 $6.711(61)\times10^{-6}$ & $6.657(60)\times10^{-6}$\\
 & & & 12 & $3.12(10)\times10^{-7}$ & $3.16(10)\times10^{-7}$ &
 $6.94(20)\times10^{-8}$ & $6.86(19)\times10^{-8}$ & $4.34(14)\times10^{-8}$ &
 $4.29(14)\times10^{-8}$\\
\hline
 & 10 & 308 & 5 & $6.596(41)\times10^{-5}$ & $6.620(41)\times10^{-5}$ &
 $2.763(16)\times10^{-5}$ & $2.753(16)\times10^{-5}$ & $1.730(11)\times10^{-5}$
 & $1.723(11)\times10^{-5}$\\
 & & & 10 & $9.17(11)\times10^{-7}$ & $9.23(11)\times10^{-7}$ &
 $2.544(29)\times10^{-7}$ & $2.530(29)\times10^{-7}$ & $1.643(18)\times10^{-7}$
 & $1.632(18)\times10^{-7}$\\
\hline
 & 8 & 385 & 4 & $1.596(10)\times10^{-4}$ & $1.599(10)\times10^{-4}$ &
 $7.781(56)\times10^{-5}$ & $7.767(56)\times10^{-5}$ & $5.000(36)\times10^{-5}$
 & $4.989(36)\times10^{-5}$\\
 & & & 8 & $3.123(45)\times10^{-6}$ & $3.134(45)\times10^{-6}$ &
 $1.058(13)\times10^{-6}$ & $1.055(13)\times10^{-6}$ & $7.067(91)\times10^{-7}$
 & $7.044(90)\times10^{-7}$\\
 & & & 16 & $4.33(13)\times10^{-9}$ & $4.35(13)\times10^{-9}$ &
 $7.98(22)\times10^{-10}$ & $7.95(22)\times10^{-10}$ & $4.94(12)\times10^{-10}$
 & $4.92(12)\times10^{-10}$\\
\hline
 & 6 & 513 & 3 & $4.434(24)\times10^{-4}$ & $4.438(24)\times10^{-4}$ &
 $2.635(15)\times10^{-4}$ & $2.633(15)\times10^{-4}$ & $1.791(10)\times10^{-4}$
 & $1.789(10)\times10^{-4}$\\
 & & & 6 & $1.098(12)\times10^{-5}$ & $1.099(12)\times10^{-5}$ &
 $4.841(45)\times10^{-6}$ & $4.835(45)\times10^{-6}$ & $3.347(32)\times10^{-6}$
 & $3.342(32)\times10^{-6}$\\
 & & & 12 & $1.318(28)\times10^{-8}$ & $1.321(28)\times10^{-8}$ &
 $3.612(68)\times10^{-9}$ & $3.606(68)\times10^{-9}$ & $2.365(49)\times10^{-9}$
 & $2.361(49)\times10^{-9}$\\
\hline
 & 4 & 770 & 2 & $1.6908(43)\times10^{-3}$ & $1.6912(43)\times10^{-3}$ &
 $1.2505(32)\times10^{-3}$ & $1.2501(32)\times10^{-3}$ &
 $9.442(23)\times10^{-4}$ & $9.439(23)\times10^{-4}$\\
 & & & 4 & $5.171(23)\times10^{-5}$ & $5.174(23)\times10^{-5}$ &
 $3.251(16)\times10^{-5}$ & $3.250(16)\times10^{-5}$ & $2.423(12)\times10^{-5}$
 & $2.423(12)\times10^{-5}$\\
 & & & 8 & $1.1412(97)\times10^{-7}$ & $1.1418(97)\times10^{-7}$ &
 $5.293(58)\times10^{-8}$ & $5.291(58)\times10^{-8}$ & $3.837(43)\times10^{-8}$
 & $3.836(43)\times10^{-8}$\\
\hline
 & 2 & 1539 & 1 & $1.2525(17)\times10^{-2}$ & $1.2526(17)\times10^{-2}$ &
 $1.1169(14)\times10^{-2}$ & $1.1169(14)\times10^{-2}$ &
 $9.897(12)\times10^{-3}$ & $9.896(12)\times10^{-3}$\\
 & & & 2 & $6.547(18)\times10^{-4}$ & $6.548(18)\times10^{-4}$ &
 $5.624(15)\times10^{-4}$ & $5.624(15)\times10^{-4}$ & $4.858(13)\times10^{-4}$
 & $4.858(13)\times10^{-4}$\\
 & & & 4 & $4.727(22)\times10^{-6}$ & $4.727(22)\times10^{-6}$ &
 $3.820(18)\times10^{-6}$ & $3.819(18)\times10^{-6}$ & $3.264(16)\times10^{-6}$
 & $3.264(16)\times10^{-6}$\\
\hline
\end{tabular}\end{adjustbox}
\end{table*}

\begin{table*}[tbp]
\caption{Screening correlators $C_\Gamma(z)$ of the $\bar sc$ and $\bar cc$
sectors in the channels of $M_0$ and $M_2$ (Sec.~\ref{subsec:cs}). The
separations are $z=N_t/2$, $N_t$ and $2N_t$ ($zT=1/2$, $1$, $2$). They are used
in Tables~\ref{tab:sectors}--\ref{tab:cswin_eps}. The lattices are $32^3\times
N_t$ at $\beta=6.20$, and the factor $1/2$ of the flavor normalization is
included. Separations beyond $z=16$ are not available. The correlators at all
separations and in all channels are given in the Supplemental
Material~\cite{Suppl}.}
\label{tab:Cz_sec_c}
\setlength{\tabcolsep}{5pt}\centering\scriptsize\renewcommand{%
\arraystretch}{0.92}
\begin{adjustbox}{max width=\linewidth}
\begin{tabular}{|cccc|cccccc|}
\hline
sector & $N_t$ & $T$ & $z/a$ & $C_{S}(z)$ & $C_{P}(z)$ & $C_{V}(z)$ & $C_{A}(z)$
& $C_{T_4}(z)$ & $C_{X_4}(z)$\\
\hline
$\bar sc$ & 16 & 192 & 8 & $5.42(10)\times10^{-6}$ & $8.33(16)\times10^{-6}$ &
$2.428(36)\times10^{-6}$ & $1.594(28)\times10^{-6}$ & $1.558(21)\times10^{-6}$ &
$9.24(24)\times10^{-7}$\\
 & & & 16 & $3.50(17)\times10^{-8}$ & $7.62(24)\times10^{-8}$ &
 $1.291(39)\times10^{-8}$ & $6.05(28)\times10^{-9}$ & $8.29(30)\times10^{-9}$ &
 $3.65(24)\times10^{-9}$\\
\hline
 & 12 & 257 & 6 & $2.745(26)\times10^{-5}$ & $3.315(40)\times10^{-5}$ &
 $1.1688(94)\times10^{-5}$ & $9.781(81)\times10^{-6}$ & $7.350(66)\times10^{-6}$
 & $5.884(55)\times10^{-6}$\\
 & & & 12 & $2.545(74)\times10^{-7}$ & $3.52(12)\times10^{-7}$ &
 $7.67(21)\times10^{-8}$ & $5.71(16)\times10^{-8}$ & $4.87(15)\times10^{-8}$ &
 $3.53(11)\times10^{-8}$\\
\hline
 & 10 & 308 & 5 & $6.229(38)\times10^{-5}$ & $6.890(43)\times10^{-5}$ &
 $2.875(17)\times10^{-5}$ & $2.606(16)\times10^{-5}$ & $1.823(12)\times10^{-5}$
 & $1.610(11)\times10^{-5}$\\
 & & & 10 & $8.282(96)\times10^{-7}$ & $9.79(11)\times10^{-7}$ &
 $2.689(30)\times10^{-7}$ & $2.304(27)\times10^{-7}$ & $1.752(19)\times10^{-7}$
 & $1.474(17)\times10^{-7}$\\
\hline
 & 8 & 385 & 4 & $1.5478(97)\times10^{-4}$ & $1.635(11)\times10^{-4}$ &
 $7.957(57)\times10^{-5}$ & $7.538(53)\times10^{-5}$ & $5.152(38)\times10^{-5}$
 & $4.805(35)\times10^{-5}$\\
 & & & 8 & $2.963(42)\times10^{-6}$ & $3.233(47)\times10^{-6}$ &
 $1.090(14)\times10^{-6}$ & $1.005(12)\times10^{-6}$ & $7.319(94)\times10^{-7}$
 & $6.674(85)\times10^{-7}$\\
 & & & 16 & $3.98(12)\times10^{-9}$ & $4.52(13)\times10^{-9}$ &
 $8.23(23)\times10^{-10}$ & $7.39(20)\times10^{-10}$ & $5.11(12)\times10^{-10}$
 & $4.56(11)\times10^{-10}$\\
\hline
 & 6 & 513 & 3 & $4.374(23)\times10^{-4}$ & $4.483(24)\times10^{-4}$ &
 $2.662(15)\times10^{-4}$ & $2.598(14)\times10^{-4}$ & $1.815(10)\times10^{-4}$
 & $1.760(10)\times10^{-4}$\\
 & & & 6 & $1.072(11)\times10^{-5}$ & $1.115(12)\times10^{-5}$ &
 $4.904(45)\times10^{-6}$ & $4.733(43)\times10^{-6}$ & $3.398(32)\times10^{-6}$
 & $3.265(31)\times10^{-6}$\\
 & & & 12 & $1.273(27)\times10^{-8}$ & $1.343(28)\times10^{-8}$ &
 $3.655(69)\times10^{-9}$ & $3.500(65)\times10^{-9}$ & $2.395(50)\times10^{-9}$
 & $2.292(47)\times10^{-9}$\\
\hline
 & 4 & 770 & 2 & $1.6831(43)\times10^{-3}$ & $1.6970(43)\times10^{-3}$ &
 $1.2548(32)\times10^{-3}$ & $1.2445(31)\times10^{-3}$ &
 $9.480(23)\times10^{-4}$ & $9.391(23)\times10^{-4}$\\
 & & & 4 & $5.133(23)\times10^{-5}$ & $5.198(24)\times10^{-5}$ &
 $3.264(16)\times10^{-5}$ & $3.228(16)\times10^{-5}$ & $2.434(12)\times10^{-5}$
 & $2.406(12)\times10^{-5}$\\
 & & & 8 & $1.1288(96)\times10^{-7}$ & $1.1480(98)\times10^{-7}$ &
 $5.311(58)\times10^{-8}$ & $5.243(57)\times10^{-8}$ & $3.847(43)\times10^{-8}$
 & $3.804(43)\times10^{-8}$\\
\hline
 & 2 & 1539 & 1 & $1.2513(17)\times10^{-2}$ & $1.2536(17)\times10^{-2}$ &
 $1.1178(14)\times10^{-2}$ & $1.1158(14)\times10^{-2}$ &
 $9.904(12)\times10^{-3}$ & $9.887(12)\times10^{-3}$\\
 & & & 2 & $6.539(18)\times10^{-4}$ & $6.553(18)\times10^{-4}$ &
 $5.627(15)\times10^{-4}$ & $5.619(15)\times10^{-4}$ & $4.859(13)\times10^{-4}$
 & $4.855(13)\times10^{-4}$\\
 & & & 4 & $4.716(22)\times10^{-6}$ & $4.735(22)\times10^{-6}$ &
 $3.823(18)\times10^{-6}$ & $3.813(18)\times10^{-6}$ & $3.264(16)\times10^{-6}$
 & $3.261(16)\times10^{-6}$\\
\hline
$\bar cc$ & 16 & 192 & 8 & $5.200(80)\times10^{-7}$ & $2.064(21)\times10^{-6}$ &
$7.119(59)\times10^{-7}$ & $1.688(32)\times10^{-7}$ & $5.204(55)\times10^{-7}$ &
$7.58(22)\times10^{-8}$\\
 & & & 16 & $1.870(88)\times10^{-10}$ & $2.118(41)\times10^{-9}$ &
 $4.430(82)\times10^{-10}$ & $3.93(22)\times10^{-11}$ &
 $3.306(76)\times10^{-10}$ & $1.83(19)\times10^{-11}$\\
\hline
 & 12 & 257 & 6 & $5.431(51)\times10^{-6}$ & $1.486(12)\times10^{-5}$ &
 $5.962(40)\times10^{-6}$ & $2.146(19)\times10^{-6}$ & $4.227(31)\times10^{-6}$
 & $1.090(13)\times10^{-6}$\\
 & & & 12 & $7.29(14)\times10^{-9}$ & $3.984(71)\times10^{-8}$ &
 $1.028(15)\times10^{-8}$ & $1.978(43)\times10^{-9}$ & $7.48(12)\times10^{-9}$ &
 $1.075(32)\times10^{-9}$\\
\hline
 & 10 & 308 & 5 & $1.932(11)\times10^{-5}$ & $4.060(23)\times10^{-5}$ &
 $1.838(10)\times10^{-5}$ & $8.661(52)\times10^{-6}$ & $1.2857(77)\times10^{-5}$
 & $4.785(35)\times10^{-6}$\\
 & & & 10 & $6.544(70)\times10^{-8}$ & $2.222(19)\times10^{-7}$ &
 $6.719(67)\times10^{-8}$ & $2.061(24)\times10^{-8}$ & $4.860(43)\times10^{-8}$
 & $1.230(17)\times10^{-8}$\\
\hline
 & 8 & 385 & 4 & $7.175(40)\times10^{-5}$ & $1.1872(71)\times10^{-4}$ &
 $6.079(42)\times10^{-5}$ & $3.673(22)\times10^{-5}$ & $4.224(30)\times10^{-5}$
 & $2.192(14)\times10^{-5}$\\
 & & & 8 & $5.672(59)\times10^{-7}$ & $1.274(15)\times10^{-6}$ &
 $4.503(53)\times10^{-7}$ & $2.096(22)\times10^{-7}$ & $3.241(36)\times10^{-7}$
 & $1.335(16)\times10^{-7}$\\
 & & & 16 & $1.545(33)\times10^{-10}$ & $4.98(10)\times10^{-10}$ &
 $9.46(22)\times10^{-11}$ & $3.478(77)\times10^{-11}$ & $6.50(12)\times10^{-11}$
 & $2.228(54)\times10^{-11}$\\
\hline
 & 6 & 513 & 3 & $2.928(14)\times10^{-4}$ & $3.840(20)\times10^{-4}$ &
 $2.329(13)\times10^{-4}$ & $1.7806(94)\times10^{-4}$ &
 $1.6482(94)\times10^{-4}$ & $1.1789(63)\times10^{-4}$\\
 & & & 6 & $4.523(37)\times10^{-6}$ & $6.982(68)\times10^{-6}$ &
 $3.080(26)\times10^{-6}$ & $2.073(17)\times10^{-6}$ & $2.191(20)\times10^{-6}$
 & $1.402(12)\times10^{-6}$\\
 & & & 12 & $2.404(37)\times10^{-9}$ & $4.358(77)\times10^{-9}$ &
 $1.153(18)\times10^{-9}$ & $7.13(13)\times10^{-10}$ & $7.74(14)\times10^{-10}$
 & $4.757(83)\times10^{-10}$\\
\hline
 & 4 & 770 & 2 & $1.4511(34)\times10^{-3}$ & $1.6112(39)\times10^{-3}$ &
 $1.1964(30)\times10^{-3}$ & $1.0789(25)\times10^{-3}$ &
 $9.141(22)\times10^{-4}$ & $8.131(18)\times10^{-4}$\\
 & & & 4 & $3.769(16)\times10^{-5}$ & $4.427(19)\times10^{-5}$ &
 $2.748(13)\times10^{-5}$ & $2.388(11)\times10^{-5}$ & $2.0447(88)\times10^{-5}$
 & $1.7747(74)\times10^{-5}$\\
 & & & 8 & $6.285(49)\times10^{-8}$ & $7.840(61)\times10^{-8}$ &
 $3.466(35)\times10^{-8}$ & $2.941(30)\times10^{-8}$ & $2.463(26)\times10^{-8}$
 & $2.148(22)\times10^{-8}$\\
\hline
 & 2 & 1539 & 1 & $1.2184(16)\times10^{-2}$ & $1.2492(17)\times10^{-2}$ &
 $1.1146(14)\times10^{-2}$ & $1.0885(13)\times10^{-2}$ &
 $9.892(12)\times10^{-3}$ & $9.663(11)\times10^{-3}$\\
 & & & 2 & $6.227(17)\times10^{-4}$ & $6.406(17)\times10^{-4}$ &
 $5.477(15)\times10^{-4}$ & $5.368(14)\times10^{-4}$ & $4.714(12)\times10^{-4}$
 & $4.655(12)\times10^{-4}$\\
 & & & 4 & $4.328(20)\times10^{-6}$ & $4.574(20)\times10^{-6}$ &
 $3.643(17)\times10^{-6}$ & $3.521(17)\times10^{-6}$ & $3.075(15)\times10^{-6}$
 & $3.033(15)\times10^{-6}$\\
\hline
\end{tabular}\end{adjustbox}
\end{table*}

\begin{table*}[tbp]
\caption{Screening correlators $C_\Gamma(z)$ of the connected parts of the
diagonal
nonsinglets $t^8$ and $t^{15}$, in the channels of $M_0$ and $M_2$. In terms of
the sector correlators of Tables~\ref{tab:Cz_sec_a}--\ref{tab:Cz_sec_c}, they
are $C^8=(C_{\bar ud}+2C_{\bar ss})/3$ and $C^{15}=(2C_{\bar ud}+C_{\bar
ss}+9C_{\bar cc})/12$, with the same normalization [Eqs.~(\ref{eq:su3_diag}) and
(\ref{eq:su4_diag})]. The disconnected diagrams are not included. The
separations are $z=N_t/2$, $N_t$ and $2N_t$ ($zT=1/2$, $1$, $2$). They enter
Table~\ref{tab:hier}.}
\label{tab:Cz_gen_a}
\setlength{\tabcolsep}{5pt}\centering\scriptsize\renewcommand{%
\arraystretch}{0.92}
\begin{adjustbox}{max width=\linewidth}
\begin{tabular}{|cccc|cccccc|}
\hline
generator & $N_t$ & $T$ & $z/a$ & $C_{S}(z)$ & $C_{P}(z)$ & $C_{V}(z)$ &
$C_{A}(z)$
& $C_{T_4}(z)$ & $C_{X_4}(z)$\\
\hline
$t^8$ & 16 & 192 & 8 & $1.037(32)\times10^{-4}$ & $1.220(47)\times10^{-4}$ &
$1.624(34)\times10^{-5}$ & $1.466(29)\times10^{-5}$ & $7.43(21)\times10^{-6}$ &
$6.53(22)\times10^{-6}$\\
 & & & 16 & $2.07(21)\times10^{-5}$ & $2.60(27)\times10^{-5}$ &
$1.193(48)\times10^{-6}$ & $1.011(44)\times10^{-6}$ & $4.22(37)\times10^{-7}$ &
$3.50(35)\times10^{-7}$\\
 & 12 & 257 & 6 & $1.538(43)\times10^{-4}$ & $1.586(46)\times10^{-4}$ &
$3.846(44)\times10^{-5}$ & $3.758(43)\times10^{-5}$ & $2.037(24)\times10^{-5}$ &
$1.984(24)\times10^{-5}$\\
 & & & 12 & $1.122(91)\times10^{-5}$ & $1.180(97)\times10^{-5}$ &
$1.349(73)\times10^{-6}$ & $1.299(70)\times10^{-6}$ & $6.40(37)\times10^{-7}$ &
$6.16(37)\times10^{-7}$\\
 & 10 & 308 & 5 & $1.868(18)\times10^{-4}$ & $1.888(18)\times10^{-4}$ &
$6.567(49)\times10^{-5}$ & $6.509(48)\times10^{-5}$ & $3.805(29)\times10^{-5}$ &
$3.763(28)\times10^{-5}$\\
 & & & 10 & $9.38(15)\times10^{-6}$ & $9.56(17)\times10^{-6}$ &
$1.933(29)\times10^{-6}$ & $1.906(29)\times10^{-6}$ & $1.074(17)\times10^{-6}$ &
$1.058(16)\times10^{-6}$\\
 & 8 & 385 & 4 & $2.997(26)\times10^{-4}$ & $3.010(26)\times10^{-4}$ &
$1.340(11)\times10^{-4}$ & $1.334(11)\times10^{-4}$ & $8.294(69)\times10^{-5}$ &
$8.252(68)\times10^{-5}$\\
 & & & 8 & $1.302(28)\times10^{-5}$ & $1.311(29)\times10^{-5}$ &
$3.858(63)\times10^{-6}$ & $3.835(63)\times10^{-6}$ & $2.408(38)\times10^{-6}$ &
$2.392(38)\times10^{-6}$\\
 & & & 16 & $8.40(39)\times10^{-8}$ & $8.49(40)\times10^{-8}$ &
$1.211(45)\times10^{-8}$ & $1.202(45)\times10^{-8}$ & $6.53(20)\times10^{-9}$ &
$6.48(20)\times10^{-9}$\\
 & 6 & 513 & 3 & $6.007(36)\times10^{-4}$ & $6.017(36)\times10^{-4}$ &
$3.465(20)\times10^{-4}$ & $3.459(20)\times10^{-4}$ & $2.321(14)\times10^{-4}$ &
$2.317(14)\times10^{-4}$\\
 & & & 6 & $2.229(30)\times10^{-5}$ & $2.235(30)\times10^{-5}$ &
$9.50(10)\times10^{-6}$ & $9.48(10)\times10^{-6}$ & $6.460(68)\times10^{-6}$ &
$6.445(68)\times10^{-6}$\\
 & & & 12 & $5.73(16)\times10^{-8}$ & $5.75(16)\times10^{-8}$ &
$1.492(35)\times10^{-8}$ & $1.488(35)\times10^{-8}$ & $9.42(23)\times10^{-9}$ &
$9.40(23)\times10^{-9}$\\
 & 4 & 770 & 2 & $1.8758(51)\times10^{-3}$ & $1.8767(51)\times10^{-3}$ &
$1.3801(37)\times10^{-3}$ & $1.3795(37)\times10^{-3}$ &
$1.0363(26)\times10^{-3}$ & $1.0358(26)\times10^{-3}$\\
 & & & 4 & $6.592(32)\times10^{-5}$ & $6.597(32)\times10^{-5}$ &
$4.152(22)\times10^{-5}$ & $4.150(22)\times10^{-5}$ & $3.100(16)\times10^{-5}$ &
$3.098(16)\times10^{-5}$\\
 & & & 8 & $1.879(18)\times10^{-7}$ & $1.881(18)\times10^{-7}$ &
$8.86(10)\times10^{-8}$ & $8.86(10)\times10^{-8}$ & $6.456(78)\times10^{-8}$ &
$6.453(78)\times10^{-8}$\\
 & 2 & 1539 & 1 & $1.2716(18)\times10^{-2}$ & $1.2717(18)\times10^{-2}$ &
$1.1326(15)\times10^{-2}$ & $1.1325(15)\times10^{-2}$ &
$1.0018(12)\times10^{-2}$ & $1.0017(12)\times10^{-2}$\\
 & & & 2 & $6.787(19)\times10^{-4}$ & $6.788(19)\times10^{-4}$ &
$5.834(16)\times10^{-4}$ & $5.833(16)\times10^{-4}$ & $5.038(13)\times10^{-4}$ &
$5.038(13)\times10^{-4}$\\
 & & & 4 & $5.023(24)\times10^{-6}$ & $5.024(24)\times10^{-6}$ &
$4.075(20)\times10^{-6}$ & $4.074(20)\times10^{-6}$ & $3.489(17)\times10^{-6}$ &
$3.489(17)\times10^{-6}$\\
\hline
$t^{15}$ & 16 & 192 & 8 & $3.05(12)\times10^{-5}$ & $3.44(15)\times10^{-5}$ &
$4.687(90)\times10^{-6}$ & $4.080(79)\times10^{-6}$ & $2.244(60)\times10^{-6}$ &
$1.795(60)\times10^{-6}$\\
 & & & 16 & $7.1(10)\times10^{-6}$ & $8.0(11)\times10^{-6}$ &
$3.23(14)\times10^{-7}$ & $3.00(13)\times10^{-7}$ & $1.05(13)\times10^{-7}$ &
$9.6(13)\times10^{-8}$\\
 & 12 & 257 & 6 & $4.36(11)\times10^{-5}$ & $5.13(13)\times10^{-5}$ &
$1.415(14)\times10^{-5}$ & $1.118(12)\times10^{-5}$ & $8.280(80)\times10^{-6}$ &
$5.860(65)\times10^{-6}$\\
 & & & 12 & $2.99(25)\times10^{-6}$ & $3.09(26)\times10^{-6}$ &
$3.53(19)\times10^{-7}$ & $3.40(19)\times10^{-7}$ & $1.682(97)\times10^{-7}$ &
$1.604(96)\times10^{-7}$\\
 & 10 & 308 & 5 & $6.168(47)\times10^{-5}$ & $7.791(57)\times10^{-5}$ &
$3.026(18)\times10^{-5}$ & $2.290(14)\times10^{-5}$ & $1.918(13)\times10^{-5}$ &
$1.3070(91)\times10^{-5}$\\
 & & & 10 & $2.449(40)\times10^{-6}$ & $2.591(44)\times10^{-6}$ &
$5.388(78)\times10^{-7}$ & $5.005(75)\times10^{-7}$ & $3.073(44)\times10^{-7}$ &
$2.780(42)\times10^{-7}$\\
 & 8 & 385 & 4 & $1.2907(86)\times10^{-4}$ & $1.645(12)\times10^{-4}$ &
$7.914(58)\times10^{-5}$ & $6.103(43)\times10^{-5}$ & $5.244(39)\times10^{-5}$ &
$3.715(26)\times10^{-5}$\\
 & & & 8 & $3.712(75)\times10^{-6}$ & $4.254(81)\times10^{-6}$ &
$1.307(19)\times10^{-6}$ & $1.124(17)\times10^{-6}$ & $8.48(12)\times10^{-7}$ &
$7.03(10)\times10^{-7}$\\
 & & & 16 & $2.15(10)\times10^{-8}$ & $2.19(10)\times10^{-8}$ &
$3.14(12)\times10^{-9}$ & $3.08(11)\times10^{-9}$ & $1.700(52)\times10^{-9}$ &
$1.662(52)\times10^{-9}$\\
 & 6 & 513 & 3 & $3.700(19)\times10^{-4}$ & $4.386(24)\times10^{-4}$ &
$2.614(15)\times10^{-4}$ & $2.202(12)\times10^{-4}$ & $1.817(10)\times10^{-4}$ &
$1.4643(81)\times10^{-4}$\\
 & & & 6 & $8.986(98)\times10^{-6}$ & $1.084(12)\times10^{-5}$ &
$4.692(45)\times10^{-6}$ & $3.934(37)\times10^{-6}$ & $3.262(32)\times10^{-6}$ &
$2.668(25)\times10^{-6}$\\
 & & & 12 & $1.623(43)\times10^{-8}$ & $1.773(46)\times10^{-8}$ &
$4.62(10)\times10^{-9}$ & $4.281(96)\times10^{-9}$ & $2.949(69)\times10^{-9}$ &
$2.722(64)\times10^{-9}$\\
 & 4 & 770 & 2 & $1.5575(38)\times10^{-3}$ & $1.6777(42)\times10^{-3}$ &
$1.2424(31)\times10^{-3}$ & $1.1543(28)\times10^{-3}$ & $9.447(23)\times10^{-4}$
& $8.689(20)\times10^{-4}$\\
 & & & 4 & $4.476(20)\times10^{-5}$ & $4.971(22)\times10^{-5}$ &
$3.100(15)\times10^{-5}$ & $2.830(13)\times10^{-5}$ & $2.309(11)\times10^{-5}$ &
$2.1063(92)\times10^{-5}$\\
 & & & 8 & $9.421(80)\times10^{-8}$ & $1.0590(89)\times10^{-7}$ &
$4.820(52)\times10^{-8}$ & $4.425(48)\times10^{-8}$ & $3.464(39)\times10^{-8}$ &
$3.228(36)\times10^{-8}$\\
 & 2 & 1539 & 1 & $1.2317(17)\times10^{-2}$ & $1.2548(17)\times10^{-2}$ &
$1.1191(14)\times10^{-2}$ & $1.0995(14)\times10^{-2}$ & $9.923(12)\times10^{-3}$
& $9.752(11)\times10^{-3}$\\
 & & & 2 & $6.368(17)\times10^{-4}$ & $6.502(18)\times10^{-4}$ &
$5.566(15)\times10^{-4}$ & $5.485(15)\times10^{-4}$ & $4.795(12)\times10^{-4}$ &
$4.751(12)\times10^{-4}$\\
 & & & 4 & $4.502(21)\times10^{-6}$ & $4.687(21)\times10^{-6}$ &
$3.751(18)\times10^{-6}$ & $3.659(18)\times10^{-6}$ & $3.179(15)\times10^{-6}$ &
$3.147(15)\times10^{-6}$\\
\hline
\end{tabular}\end{adjustbox}
\end{table*}

\begin{table*}[tbp]
\caption{Screening correlators $C_\Gamma(z)$ of the connected parts of the
$SU(3)$ and
$SU(4)$ singlets $t^0_{uds}$ and $t^0_{udsc}$, as in Table~\ref{tab:Cz_gen_a}.
In terms of the sector correlators, they are $(2C_{\bar ud}+C_{\bar ss})/3$ and
$(2C_{\bar ud}+C_{\bar ss}+C_{\bar cc})/4$. They enter the $T_1$ columns of
Table~\ref{tab:hier}.}
\label{tab:Cz_gen_b}
\setlength{\tabcolsep}{5pt}\centering\scriptsize\renewcommand{%
\arraystretch}{0.92}
\begin{adjustbox}{max width=\linewidth}
\begin{tabular}{|cccc|cccccc|}
\hline
generator & $N_t$ & $T$ & $z/a$ & $C_{S}(z)$ & $C_{P}(z)$ & $C_{V}(z)$ &
$C_{A}(z)$
& $C_{T_4}(z)$ & $C_{X_4}(z)$\\
\hline
$t^0_{uds}$ & 16 & 192 & 8 & $1.205(46)\times10^{-4}$ & $1.312(60)\times10^{-4}$
& $1.661(35)\times10^{-5}$ & $1.581(32)\times10^{-5}$ & $7.42(23)\times10^{-6}$
& $6.95(23)\times10^{-6}$\\
 & & & 16 & $2.83(41)\times10^{-5}$ & $3.18(46)\times10^{-5}$ &
$1.290(57)\times10^{-6}$ & $1.199(54)\times10^{-6}$ & $4.20(51)\times10^{-7}$ &
$3.83(51)\times10^{-7}$\\
 & 12 & 257 & 6 & $1.583(46)\times10^{-4}$ & $1.608(48)\times10^{-4}$ &
$3.872(45)\times10^{-5}$ & $3.828(44)\times10^{-5}$ & $2.044(25)\times10^{-5}$ &
$2.017(24)\times10^{-5}$\\
 & & & 12 & $1.195(100)\times10^{-5}$ & $1.23(10)\times10^{-5}$ &
$1.380(76)\times10^{-6}$ & $1.355(74)\times10^{-6}$ & $6.50(39)\times10^{-7}$ &
$6.39(38)\times10^{-7}$\\
 & 10 & 308 & 5 & $1.887(18)\times10^{-4}$ & $1.898(18)\times10^{-4}$ &
$6.591(49)\times10^{-5}$ & $6.561(49)\times10^{-5}$ & $3.813(29)\times10^{-5}$ &
$3.793(29)\times10^{-5}$\\
 & & & 10 & $9.60(16)\times10^{-6}$ & $9.70(17)\times10^{-6}$ &
$1.954(30)\times10^{-6}$ & $1.940(29)\times10^{-6}$ & $1.083(17)\times10^{-6}$ &
$1.075(17)\times10^{-6}$\\
 & 8 & 385 & 4 & $3.010(26)\times10^{-4}$ & $3.017(27)\times10^{-4}$ &
$1.342(11)\times10^{-4}$ & $1.339(11)\times10^{-4}$ & $8.305(69)\times10^{-5}$ &
$8.284(69)\times10^{-5}$\\
 & & & 8 & $1.315(29)\times10^{-5}$ & $1.319(29)\times10^{-5}$ &
$3.879(63)\times10^{-6}$ & $3.868(63)\times10^{-6}$ & $2.419(38)\times10^{-6}$ &
$2.411(38)\times10^{-6}$\\
 & & & 16 & $8.57(40)\times10^{-8}$ & $8.62(40)\times10^{-8}$ &
$1.227(46)\times10^{-8}$ & $1.222(46)\times10^{-8}$ & $6.61(21)\times10^{-9}$ &
$6.58(20)\times10^{-9}$\\
 & 6 & 513 & 3 & $6.018(36)\times10^{-4}$ & $6.023(36)\times10^{-4}$ &
$3.467(20)\times10^{-4}$ & $3.465(20)\times10^{-4}$ & $2.323(14)\times10^{-4}$ &
$2.320(14)\times10^{-4}$\\
 & & & 6 & $2.237(30)\times10^{-5}$ & $2.241(30)\times10^{-5}$ &
$9.53(10)\times10^{-6}$ & $9.52(10)\times10^{-6}$ & $6.474(68)\times10^{-6}$ &
$6.467(68)\times10^{-6}$\\
 & & & 12 & $5.77(16)\times10^{-8}$ & $5.79(16)\times10^{-8}$ &
$1.501(36)\times10^{-8}$ & $1.499(36)\times10^{-8}$ & $9.47(23)\times10^{-9}$ &
$9.46(23)\times10^{-9}$\\
 & 4 & 770 & 2 & $1.8768(51)\times10^{-3}$ & $1.8772(51)\times10^{-3}$ &
$1.3805(37)\times10^{-3}$ & $1.3802(37)\times10^{-3}$ &
$1.0365(26)\times10^{-3}$ & $1.0363(26)\times10^{-3}$\\
 & & & 4 & $6.600(32)\times10^{-5}$ & $6.602(32)\times10^{-5}$ &
$4.156(22)\times10^{-5}$ & $4.154(22)\times10^{-5}$ & $3.102(16)\times10^{-5}$ &
$3.101(16)\times10^{-5}$\\
 & & & 8 & $1.883(18)\times10^{-7}$ & $1.884(18)\times10^{-7}$ &
$8.88(10)\times10^{-8}$ & $8.88(10)\times10^{-8}$ & $6.468(78)\times10^{-8}$ &
$6.466(78)\times10^{-8}$\\
 & 2 & 1539 & 1 & $1.2717(18)\times10^{-2}$ & $1.2717(18)\times10^{-2}$ &
$1.1326(15)\times10^{-2}$ & $1.1326(15)\times10^{-2}$ &
$1.0018(12)\times10^{-2}$ & $1.0018(12)\times10^{-2}$\\
 & & & 2 & $6.788(19)\times10^{-4}$ & $6.789(19)\times10^{-4}$ &
$5.834(16)\times10^{-4}$ & $5.834(16)\times10^{-4}$ & $5.039(13)\times10^{-4}$ &
$5.039(13)\times10^{-4}$\\
 & & & 4 & $5.024(24)\times10^{-6}$ & $5.025(24)\times10^{-6}$ &
$4.076(20)\times10^{-6}$ & $4.075(20)\times10^{-6}$ & $3.490(17)\times10^{-6}$ &
$3.490(17)\times10^{-6}$\\
\hline
$t^0_{udsc}$ & 16 & 192 & 8 & $9.05(35)\times10^{-5}$ & $9.89(45)\times10^{-5}$
& $1.264(26)\times10^{-5}$ & $1.190(24)\times10^{-5}$ & $5.69(17)\times10^{-6}$
& $5.23(18)\times10^{-6}$\\
 & & & 16 & $2.12(31)\times10^{-5}$ & $2.38(34)\times10^{-5}$ &
$9.67(42)\times10^{-7}$ & $8.99(40)\times10^{-7}$ & $3.15(39)\times10^{-7}$ &
$2.87(38)\times10^{-7}$\\
 & 12 & 257 & 6 & $1.201(35)\times10^{-4}$ & $1.243(36)\times10^{-4}$ &
$3.053(35)\times10^{-5}$ & $2.925(33)\times10^{-5}$ & $1.639(19)\times10^{-5}$ &
$1.540(18)\times10^{-5}$\\
 & & & 12 & $8.96(75)\times10^{-6}$ & $9.20(77)\times10^{-6}$ &
$1.038(57)\times10^{-6}$ & $1.017(56)\times10^{-6}$ & $4.90(29)\times10^{-7}$ &
$4.79(29)\times10^{-7}$\\
 & 10 & 308 & 5 & $1.464(13)\times10^{-4}$ & $1.525(14)\times10^{-4}$ &
$5.402(36)\times10^{-5}$ & $5.138(37)\times10^{-5}$ & $3.181(23)\times10^{-5}$ &
$2.964(22)\times10^{-5}$\\
 & & & 10 & $7.22(12)\times10^{-6}$ & $7.33(13)\times10^{-6}$ &
$1.482(22)\times10^{-6}$ & $1.460(22)\times10^{-6}$ & $8.25(13)\times10^{-7}$ &
$8.10(12)\times10^{-7}$\\
 & 8 & 385 & 4 & $2.437(20)\times10^{-4}$ & $2.560(22)\times10^{-4}$ &
$1.1584(93)\times10^{-4}$ & $1.0963(88)\times10^{-4}$ & $7.285(59)\times10^{-5}$
& $6.761(54)\times10^{-5}$\\
 & & & 8 & $1.000(22)\times10^{-5}$ & $1.021(22)\times10^{-5}$ &
$3.022(49)\times10^{-6}$ & $2.953(48)\times10^{-6}$ & $1.895(29)\times10^{-6}$ &
$1.842(29)\times10^{-6}$\\
 & & & 16 & $6.43(30)\times10^{-8}$ & $6.48(30)\times10^{-8}$ &
$9.22(34)\times10^{-9}$ & $9.17(34)\times10^{-9}$ & $4.97(15)\times10^{-9}$ &
$4.94(15)\times10^{-9}$\\
 & 6 & 513 & 3 & $5.245(31)\times10^{-4}$ & $5.477(33)\times10^{-4}$ &
$3.183(18)\times10^{-4}$ & $3.044(18)\times10^{-4}$ & $2.154(13)\times10^{-4}$ &
$2.035(12)\times10^{-4}$\\
 & & & 6 & $1.791(23)\times10^{-5}$ & $1.855(24)\times10^{-5}$ &
$7.916(83)\times10^{-6}$ & $7.656(81)\times10^{-6}$ & $5.404(56)\times10^{-6}$ &
$5.201(54)\times10^{-6}$\\
 & & & 12 & $4.39(12)\times10^{-8}$ & $4.45(12)\times10^{-8}$ &
$1.154(27)\times10^{-8}$ & $1.142(27)\times10^{-8}$ & $7.30(18)\times10^{-9}$ &
$7.21(18)\times10^{-9}$\\
 & 4 & 770 & 2 & $1.7704(46)\times10^{-3}$ & $1.8107(48)\times10^{-3}$ &
$1.3345(35)\times10^{-3}$ & $1.3049(34)\times10^{-3}$ &
$1.0059(25)\times10^{-3}$ & $9.805(24)\times10^{-4}$\\
 & & & 4 & $5.892(28)\times10^{-5}$ & $6.058(29)\times10^{-5}$ &
$3.804(19)\times10^{-5}$ & $3.713(19)\times10^{-5}$ & $2.838(14)\times10^{-5}$ &
$2.770(14)\times10^{-5}$\\
 & & & 8 & $1.569(14)\times10^{-7}$ & $1.609(15)\times10^{-7}$ &
$7.526(87)\times10^{-8}$ & $7.393(85)\times10^{-8}$ & $5.467(65)\times10^{-8}$ &
$5.387(64)\times10^{-8}$\\
 & 2 & 1539 & 1 & $1.2583(17)\times10^{-2}$ & $1.2661(18)\times10^{-2}$ &
$1.1281(15)\times10^{-2}$ & $1.1215(14)\times10^{-2}$ & $9.986(12)\times10^{-3}$
& $9.929(12)\times10^{-3}$\\
 & & & 2 & $6.648(18)\times10^{-4}$ & $6.693(18)\times10^{-4}$ &
$5.745(16)\times10^{-4}$ & $5.718(15)\times10^{-4}$ & $4.958(13)\times10^{-4}$ &
$4.943(13)\times10^{-4}$\\
 & & & 4 & $4.850(23)\times10^{-6}$ & $4.912(23)\times10^{-6}$ &
$3.967(19)\times10^{-6}$ & $3.937(19)\times10^{-6}$ & $3.386(17)\times10^{-6}$ &
$3.376(17)\times10^{-6}$\\
\hline
\end{tabular}\end{adjustbox}
\end{table*}


\clearpage
\begin{thebibliography}{99}

%\cite{Chiu:2024jyz}
\bibitem{Chiu:2024jyz}
T.~W.~Chiu,
``Symmetries of spatial correlators of light and heavy mesons in high
temperature lattice QCD'',
\href{https://doi.org/10.1103/PhysRevD.110.014502}{\blue Phys.\ Rev.\ D
\textbf{110}, 014502 (2024)};
Erratum: \href{https://doi.org/10.1103/gnhb-zmhn}{\blue Phys.\ Rev.\ D
\textbf{114}, 079903 (2026)}
[arXiv:2404.15932 [hep-lat]].

%\cite{Chiu:2025sym}
\bibitem{Chiu:2025sym}
T.~W.~Chiu,
``Symmetries in high-temperature lattice QCD with $(u,d,s,c,b)$ optimal
domain-wall quarks'',
\href{https://doi.org/10.3390/sym17050700}{\blue Symmetry \textbf{17}, 700
(2025)}
[arXiv:2411.16705 [hep-lat]].

%\cite{Chiu:2026sxy}
\bibitem{Chiu:2026sxy}
T.~W.~Chiu, T.~H.~Hsieh,
``RG-invariant symmetry ratio for QCD: A study of $U(1)_A$ and chiral symmetry
restoration'',
Phys.\ Rev.\ D, in press (2026)
[arXiv:2602.14811 [hep-lat]].

%\cite{Chiu:2026upk}
\bibitem{Chiu:2026upk}
T.~W.~Chiu,
``Renormalization of meson susceptibilities and RG-invariant symmetry ratios in
QCD'',
arXiv:2607.17816 [hep-lat].

%\cite{Ginsparg:1981bj}
\bibitem{Ginsparg:1981bj}
P.~H.~Ginsparg, K.~G.~Wilson,
``A Remnant of Chiral Symmetry on the Lattice'',
\href{https://doi.org/10.1103/PhysRevD.25.2649}{\blue Phys.\ Rev.\ D
\textbf{25}, 2649 (1982)}.

%\cite{Luscher:1998pqa}
\bibitem{Luscher:1998pqa}
M.~L\"uscher,
``Exact chiral symmetry on the lattice and the Ginsparg-Wilson relation'',
\href{https://doi.org/10.1016/S0370-2693(98)00423-7}{\blue Phys.\ Lett.\ B
\textbf{428}, 342 (1998)}
[hep-lat/9802011].

%\cite{Neuberger:1997fp}
\bibitem{Neuberger:1997fp}
H.~Neuberger,
``Exactly massless quarks on the lattice'',
\href{https://doi.org/10.1016/S0370-2693(97)01368-3}{\blue Phys.\ Lett.\ B
\textbf{417}, 141 (1998)}
[hep-lat/9707022].

%\cite{Kaplan:1992bt}
\bibitem{Kaplan:1992bt}
D.~B.~Kaplan,
``A Method for simulating chiral fermions on the lattice'',
\href{https://doi.org/10.1016/0370-2693(92)91112-M}{\blue Phys.\ Lett.\ B
\textbf{288}, 342 (1992)}
[hep-lat/9206013].

%\cite{Shamir:1993zy}
\bibitem{Shamir:1993zy}
Y.~Shamir,
``Chiral fermions from lattice boundaries'',
\href{https://doi.org/10.1016/0550-3213(93)90162-I}{\blue Nucl.\ Phys.\ B
\textbf{406}, 90 (1993)}
[hep-lat/9303005].

%\cite{Borici:1999zw}
\bibitem{Borici:1999zw}
A.~Borici,
``Truncated overlap fermions'',
\href{https://doi.org/10.1016/S0920-5632(00)91802-4}{\blue Nucl.\ Phys.\ B
Proc.\ Suppl.\ \textbf{83}, 771 (2000)}
[hep-lat/9909057].

%\cite{Chiu:2002ir}
\bibitem{Chiu:2002ir}
T.~W.~Chiu,
``Optimal domain wall fermions'',
\href{https://doi.org/10.1103/PhysRevLett.90.071601}{\blue Phys.\ Rev.\ Lett.\
\textbf{90}, 071601 (2003)}
[hep-lat/0209153].

%\cite{Brower:2012vk}
\bibitem{Brower:2012vk}
R.~C.~Brower, H.~Neff, and K.~Orginos,
``The M\"obius domain wall fermion algorithm'',
\href{https://doi.org/10.1016/j.cpc.2017.01.024}{\blue Comput.\ Phys.\ Commun.\
\textbf{220}, 1 (2017)}
[arXiv:1206.5214 [hep-lat]].

%\cite{Bazavov:2011nk}
\bibitem{Bazavov:2011nk}
A.~Bazavov \textit{et al.} [HotQCD Collaboration],
``The chiral and deconfinement aspects of the QCD transition'',
\href{https://doi.org/10.1103/PhysRevD.85.054503}{\blue Phys.\ Rev.\ D
\textbf{85}, 054503 (2012)}
[arXiv:1111.1710 [hep-lat]].

%\cite{HotQCD:2019xnw}
\bibitem{HotQCD:2019xnw}
H.~T.~Ding \textit{et al.} [HotQCD Collaboration],
``Chiral phase transition temperature in (2+1)-flavor QCD'',
\href{https://doi.org/10.1103/PhysRevLett.123.062002}{\blue Phys.\ Rev.\ Lett.\
\textbf{123}, 062002 (2019)}
[arXiv:1903.04801 [hep-lat]].

%\cite{Dick:2015twa}
\bibitem{Dick:2015twa}
V.~Dick, F.~Karsch, E.~Laermann, S.~Mukherjee, S.~Sharma,
``Microscopic origin of $U_A(1)$ symmetry violation in the high temperature
phase of QCD'',
\href{https://doi.org/10.1103/PhysRevD.91.094504}{\blue Phys.\ Rev.\ D
\textbf{91}, 094504 (2015)}
[arXiv:1502.06190 [hep-lat]].

%\cite{Brandt:2016daq}
\bibitem{Brandt:2016daq}
B.~B.~Brandt, A.~Francis, H.~B.~Meyer, O.~Philipsen, D.~Robaina, H.~Wittig,
``On the strength of the $U_A(1)$ anomaly at the chiral phase transition in
$N_f=2$ QCD'',
\href{https://doi.org/10.1007/JHEP12(2016)158}{\blue JHEP \textbf{12}, 158
(2016)}
[arXiv:1608.06882 [hep-lat]].

%\cite{Aarts:2026}
\bibitem{Aarts:2026}
G.~Aarts, C.~Allton, R.~Bignell, B.~J\"ager, S.~Kim, J.~I.~Skullerud and
A.~Smecca,
``On the effective restoration of $U(1)_A$ symmetry at finite temperature'',
arXiv:2604.11916 [hep-lat].

%\cite{Kotov:2021rah}
\bibitem{Kotov:2021rah}
A.~Y.~Kotov, M.~P.~Lombardo, A.~Trunin,
``QCD transition at the physical point, and its scaling window from twisted mass
Wilson fermions'',
\href{https://doi.org/10.1016/j.physletb.2021.136749}{\blue Phys.\ Lett.\ B
\textbf{823}, 136749 (2021)}
[arXiv:2105.09842 [hep-lat]].

%\cite{Glozman:2014mka}
\bibitem{Glozman:2014mka}
L.~Y.~Glozman,
``SU(4) symmetry of the dynamical QCD string and genesis of hadron spectra'',
\href{https://doi.org/10.1140/epja/i2015-15027-x}{\blue Eur.\ Phys.\ J.\ A
\textbf{51}, 27 (2015)}
[arXiv:1407.2798 [hep-ph]].

%\cite{Rohrhofer:2019qwq}
\bibitem{Rohrhofer:2019qwq}
C.~Rohrhofer, Y.~Aoki, G.~Cossu, H.~Fukaya, C.~Gattringer, L.~Y.~Glozman,
S.~Hashimoto,
C.~B.~Lang, S.~Prelovsek,
``Symmetries of spatial meson correlators in high temperature QCD'',
\href{https://doi.org/10.1103/PhysRevD.100.014502}{\blue Phys.\ Rev.\ D
\textbf{100}, 014502 (2019)}
[arXiv:1902.03191 [hep-lat]].

%\cite{Rohrhofer:2019qal}
\bibitem{Rohrhofer:2019qal}
C.~Rohrhofer, Y.~Aoki, L.~Y.~Glozman, S.~Hashimoto,
``Chiral-spin symmetry of the meson spectral function above $T_c$'',
\href{https://doi.org/10.1016/j.physletb.2020.135245}{\blue Phys.\ Lett.\ B
\textbf{802}, 135245 (2020)}
[arXiv:1909.00927 [hep-lat]].

%\cite{Chiu:2023hnm}
\bibitem{Chiu:2023hnm}
T.~W.~Chiu,
``Symmetries of meson correlators in high-temperature QCD with physical
(u/d,s,c) domain-wall quarks'',
\href{https://doi.org/10.1103/PhysRevD.107.114501}{\blue Phys.\ Rev.\ D
\textbf{107}, 114501 (2023)}
[arXiv:2302.06073 [hep-lat]].

%\cite{Chimirri:2024tns}
\bibitem{Chimirri:2024tns}
L.~Chimirri, P.~Fritzsch, J.~Heitger, F.~Joswig, M.~Panero, C.~Pena, D.~Preti,
``Non-perturbative renormalisation and improvement of non-singlet tensor
currents in $N_f=3$ QCD'',
\href{https://doi.org/10.1007/JHEP07(2024)089}{\blue JHEP \textbf{07}, 089
(2024)}
[arXiv:2309.04314 [hep-lat]].

%\cite{Karsten:1981gd}
\bibitem{Karsten:1981gd}
L.~H.~Karsten,
``Lattice fermions in Euclidean space-time'',
\href{https://doi.org/10.1016/0370-2693(81)90133-X}{\blue Phys.\ Lett.\ B
\textbf{104}, 315 (1981)}.

%\cite{Wilczek:1987kw}
\bibitem{Wilczek:1987kw}
F.~Wilczek,
``Lattice fermions'',
\href{https://doi.org/10.1103/PhysRevLett.59.2397}{\blue Phys.\ Rev.\ Lett.\
\textbf{59}, 2397 (1987)}.

%\cite{Creutz:2007af}
\bibitem{Creutz:2007af}
M.~Creutz,
``Four-dimensional graphene and chiral fermions'',
\href{https://doi.org/10.1088/1126-6708/2008/04/017}{\blue JHEP \textbf{04}, 017
(2008)}
[arXiv:0712.1201 [hep-lat]].

%\cite{Borici:2007kz}
\bibitem{Borici:2007kz}
A.~Bori\c{c}i,
``Creutz fermions on an orthogonal lattice'',
\href{https://doi.org/10.1103/PhysRevD.78.074504}{\blue Phys.\ Rev.\ D
\textbf{78}, 074504 (2008)}
[arXiv:0712.4401 [hep-lat]].

%\cite{Glozman:2015qva}
\bibitem{Glozman:2015qva}
L.~Y.~Glozman, M.~Pak,
``Exploring a new SU(4) symmetry of meson interpolators'',
\href{https://doi.org/10.1103/PhysRevD.92.016001}{\blue Phys.\ Rev.\ D
\textbf{92}, 016001 (2015)}
[arXiv:1504.02323 [hep-lat]].

%\cite{Wilson:1975id}
\bibitem{Wilson:1975id}
K.~G.~Wilson,
``Quarks and strings on a lattice'',
in \textit{New Phenomena in Subnuclear Physics}, edited by A.~Zichichi (Plenum,
New York, 1977), p.~69,
\href{https://doi.org/10.1007/978-1-4613-4208-3_6}{\blue
doi:10.1007/978-1-4613-4208-3\_6}.

%\cite{Sheikholeslami:1985ij}
\bibitem{Sheikholeslami:1985ij}
B.~Sheikholeslami, R.~Wohlert,
``Improved Continuum Limit Lattice Action for QCD with Wilson Fermions'',
\href{https://doi.org/10.1016/0550-3213(85)90002-1}{\blue Nucl.\ Phys.\ B
\textbf{259}, 572 (1985)}.

%\cite{Frezzotti:2000nk}
\bibitem{Frezzotti:2000nk}
R.~Frezzotti, P.~A.~Grassi, S.~Sint, P.~Weisz,
``Lattice QCD with a chirally twisted mass term'',
\href{https://doi.org/10.1088/1126-6708/2001/08/058}{\blue JHEP \textbf{08}, 058
(2001)}
[hep-lat/0101001].

%\cite{Frezzotti:2003ni}
\bibitem{Frezzotti:2003ni}
R.~Frezzotti, G.~C.~Rossi,
``Chirally improving Wilson fermions. 1. O(a) improvement'',
\href{https://doi.org/10.1088/1126-6708/2004/08/007}{\blue JHEP \textbf{08}, 007
(2004)}
[hep-lat/0306014].

%\cite{Kogut:1974ag}
\bibitem{Kogut:1974ag}
J.~B.~Kogut, L.~Susskind,
``Hamiltonian formulation of Wilson's lattice gauge theories'',
\href{https://doi.org/10.1103/PhysRevD.11.395}{\blue Phys.\ Rev.\ D \textbf{11},
395 (1975)}.

%\cite{Susskind:1976jm}
\bibitem{Susskind:1976jm}
L.~Susskind,
``Lattice fermions'',
\href{https://doi.org/10.1103/PhysRevD.16.3031}{\blue Phys.\ Rev.\ D
\textbf{16}, 3031 (1977)}.

%\cite{Matsui:1986dk}
\bibitem{Matsui:1986dk}
T.~Matsui and H.~Satz,
``$J/\psi$ Suppression by Quark-Gluon Plasma Formation'',
\href{https://doi.org/10.1016/0370-2693(86)91404-8}{\blue Phys.\ Lett.\ B
\textbf{178}, 416 (1986)}.

%\cite{Chiu:inprep}
\bibitem{Chiu:inprep}
T.~W.~Chiu \textit{et al.} (in preparation).

%\cite{Kubo:1957mj}
\bibitem{Kubo:1957mj}
R.~Kubo,
``Statistical mechanical theory of irreversible processes. 1. General theory and
simple applications in magnetic and conduction problems'',
\href{https://doi.org/10.1143/JPSJ.12.570}{\blue J.\ Phys.\ Soc.\ Jpn.\
\textbf{12}, 570 (1957)}.

%\cite{Martin:1959jp}
\bibitem{Martin:1959jp}
P.~C.~Martin and J.~S.~Schwinger,
``Theory of many particle systems. 1'',
\href{https://doi.org/10.1103/PhysRev.115.1342}{\blue Phys.\ Rev.\ \textbf{115},
1342 (1959)}.

%\cite{Suppl}
\bibitem{Suppl}
See Supplemental Material at [URL will be inserted by publisher], provided on
arXiv as the ancillary file \texttt{supplemental\_tables.pdf}, for the tables of
the thermal and screening correlators of all six flavor sectors and of the
connected parts of $t^8$, $t^{15}$ and the $SU(3)$ and $SU(4)$ singlets on the
seven ensembles, in all channels and at all separations.

\end{thebibliography}
\end{document}